\documentclass[journal=jacsat,manuscript=article]{achemso}

\usepackage{chemformula} 
\usepackage[T1]{fontenc} 

\usepackage[utf8]{inputenc}
\usepackage[version=4]{mhchem}
\usepackage{graphicx}
\usepackage{amsmath}
\usepackage{float}
\usepackage{comment}
\usepackage[justification=centerfirst]{subfig}

\usepackage{caption}
\usepackage{multirow}
\usepackage{amssymb}

\usepackage{xr}
\newcommand{\neuron}{d-CN} 

\newcommand{\one}{($i$) }
\newcommand{\two}{($ii$) }
\newcommand{\three}{($iii$) }

\newcommand{\ecoli}{{\em E.coli } }

\newcommand{\figref}[1]{Fig.~\ref{#1}}
\newcommand{\tabref}[1]{Table~\ref{#1}}

\usepackage{listings}
 
\usepackage{todonotes} 

\definecolor{dkblue}{rgb}{0,0,0.6}
\definecolor{dkred}{rgb}{0.6,0,0.0}
\definecolor{dkgreen}{rgb}{0,0.6,0}
\definecolor{gray}{rgb}{0.5,0.5,0.5}
\lstdefinestyle{dsd}
{ aboveskip=3mm,
belowskip=3mm,  showstringspaces=false, columns=flexible, basicstyle={\small\ttfamily}, language=Caml, numbers=left, numberstyle=\tiny\color{gray}, stepnumber=1, numbersep=5pt, keywordstyle=\color{blue}, commentstyle=\color{dkgreen}, morecomment=[l][\color{dkgreen}]{//}, stringstyle=\color{dkred}, breaklines=true, breakatwhitespace=true, tabsize=4, emph={directive,simulator,inference,deterministic,parameters,plot_settings,sweeps,plot,time,simulation,rendering,units,compilation,def,dom,init}, emphstyle=\color{blue} }
\author{Jakub Fil}
\email{jakub.fil@manchester.ac.uk}
\affiliation[University of Manchester]
{APT Group, School of Computer Science, The University of Manchester, Manchester, M13 9PL, UK}
\author{Neil Dalchau}
\affiliation[Microsoft Research]
{Microsoft Research, Cambridge, CB1 2FB, UK}
\author{Dominique Chu}
\affiliation[University of Kent]
{CEMS, School of Computing, University of Kent, Canterbury, CT2 7NF, UK}
\title[Programming a molecular neuron]{ Programming molecular systems to emulate a learning spiking neuron}
\abbreviations{CN, c-CN, d-CN, SN, CRN, DSD, TC, FB}
\begin{document}

\begin{tocentry}

\includegraphics[width=1\textwidth]{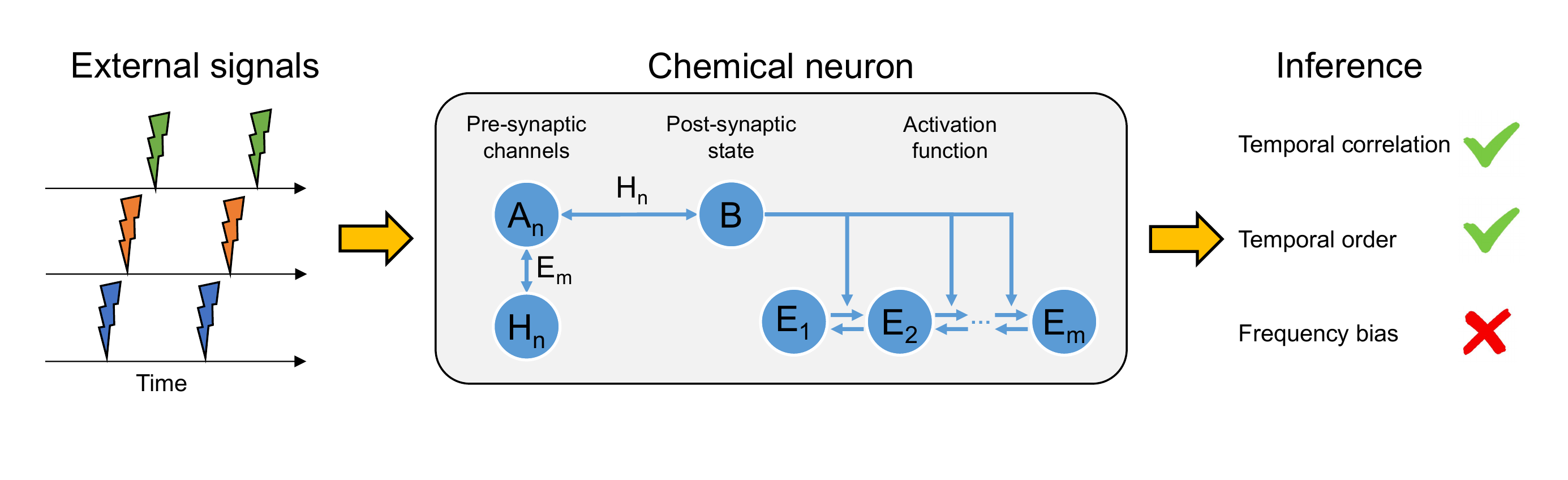}

\end{tocentry}

\begin{abstract} 
 Hebbian theory seeks to explain how the neurons in the brain adapt to stimuli, to enable learning. An interesting feature of Hebbian learning is that it is an unsupervised method and as such, does not require feedback, making it suitable in contexts where systems have to learn autonomously. This paper explores how molecular systems can be designed to show such proto-intelligent behaviours, and proposes the first chemical reaction network (CRN) that can exhibit autonomous Hebbian learning across arbitrarily many input channels. The system emulates a spiking neuron, and we demonstrate that it can learn statistical biases of incoming inputs. The basic CRN is a minimal, thermodynamically plausible set of micro-reversible chemical equations that can be analysed with respect to their energy requirements. However, to explore how such chemical systems might be engineered de novo, we also propose an extended version based on enzyme-driven compartmentalised reactions. Finally, we also show how a purely DNA system, built upon the paradigm of DNA strand displacement, can realise neuronal dynamics. Our analysis provides a compelling blueprint for exploring autonomous learning in biological settings, bringing us closer to realising real synthetic biological intelligence.  
\end{abstract}
%

\subsection{Keywords}
\label{sec:keywords}
Hebbian learning, Spiking neurons, DNA strand displacement, Autonomous learning, Biochemical intelligence

\section{Introduction}
\label{sec:intro}

While intelligent behaviours are usually associated with higher organisms that have a nervous system, adaptive and proto-intelligent behaviours are well documented in unicellular organisms. Examples include sensing \citep{government1, government2, alon2019introduction}, chemotaxis \citep{Yi_2000, Hoffer_2001} or diauxic growth  \citep{my1, my2, my3}.   This begs the question whether it is possible to rationally build molecular systems that show proto-intelligent behaviours and can  be used  as machines to monitor or control their chemical environment at a microscopic scale.  Systems of this type  could find applications  in areas such as  drug delivery, bioprocessing or biofabrication.   
\par
As a step in this direction, we will probe how artificial intelligence can be realised in molecular systems. More specifically, we will show how to realise \emph{artificial neurons}, as they are widely used in computer science as components of neural networks \citep{aggravate}. Individual artificial neurons are simple machines, but nevertheless show a remarkable ability to learn from observation. For the purpose of this article, we will consider a particular type of neuron,  a  {\em spiking neuron} (SN).   SNs are widely used  in machine learning  \citep{deep_snn, afshar_snn} and it is well known that they have significant learning capabilities \citep{Fil_2020, Gutig2016} including principal component analysis \citep{oja_1982}, recognition of handwriting \citep{diehl_digits} or classification of  fighter-planes \citep{van_schaik_planes}. There are a number of  different models of  SNs in the literature. Commonly a SN  has  an internal state, usually represented by  a positive real number.  The internal  state may decay, which means that it reduces over time with some rate. The internal state  variable  increases when the SN receives a stimulus (an input {\em spike}) via one of its $N$ input channels. Importantly, these input channels are weighted.  The higher the weight, the more the internal state variable increases following an input spike through this channel. This weighting is crucial for the behaviours of the neuron. Consequently, ``learning,'' in the context of neural networks, normally means adjusting the weights. 
\par
There have been numerous attempts to build neurons in chemical systems. The earliest dates back to the 1980s by Okamoto and collaborators \citep{Okamoto_1988} who showed that certain biochemical systems implement the McCulloch-Pitts neuronic equations. Later, a mathematical description of a neuron  has been proposed \citep{Hjelmfelt_1991}, but this system had no ability to learn. Banda {\em et al.} \citep{chemperceptron}  used an artificial chemistry  to emulate an artificial neuron and a fully fledged feed-forward neural network \citep{chemperceptron2} which could solve the XOR problem. Their model requires regular interventions by outside operators, however. Beside these simulation studies, there have also been attempts to implement learning {\em in vivo} \citep{assmodel,assmodel2,biochemimp,forcement}, but again, these systems are not autonomous: they rely on iterative measurement and manipulation protocols, which limit their practical deployment as computing machines within a molecular environment.
\par
An attractive concept of learning that avoids the  need to monitor the molecular neurons is {\em Hebbian learning}.  This concept originated from  neuroscience but is now widely used in artificial intelligence to train neural networks. The  basic idea of Hebbian learning is that the connection between  neurons that fire at the same time is strengthened. This update scheme is attractive, because unlike many other learning algorithms, it does not require evaluating an objective function, which would be difficult to achieve in general with chemical networks. 
\par
To illustrate the basic idea of Hebbian learning --- or associative learning, as it is often called when there are only two input channels --- consider a neuron with two inputs $A_1$ and $A_2$. Let the weights associated with the inputs be set such that (an output firing of) the neuron is triggered whenever $A_1$ fires, but not when $A_2$  fires. Assume now that  $A_2$ fires usually  at around the same time as $A_1$.  Then its weights will be strengthened by the Hebbian rule because of the coincidence of $A_1$ and $A_2$. Eventually, the weights of the second channel will have  increased  sufficiently such that  firing of $A_2$ on its own will be sufficient to trigger an output. 
\par
Molecular models of Hebbian learning have been proposed before.
A biochemical model of associative learning was proposed by Fernando and co-workers \citep{Fernando2009}. Their model is fully autonomous, but it is also inflexible.  Association is learned after just a single coincidence, and hence the model is unable to detect statistical correlations robustly. Moreover, the system cannot forget the association between the inputs. McGregor {\em et al.} \citep{mcgregor_cn} introduced an improved design with systems that were found by evolutionary processes. A biochemically more plausible system was proposed by Sol\'{e} and co-workers \citep{solebiochemimp}, but this system is also limited to learning two coinciding inputs and relies on an explicit operator manipulation in order to forget past associations. 
\par
In this article  we will propose a fully autonomous chemical  artificial neuron, henceforth referred to as {\em CN}, that goes beyond the state of the art in that  it can  learn statistical relations between an arbitrary number of inputs. The CN is also able to forget learned associations and as such can adapt to new observations without any intervention by an external observer.  Via each of  its input channels the CN can accept  {\em boli}, that is the injection of a certain amount of chemical species, representing the input spikes of simulated neurons.  The CN will ``learn'' the statistical biases of the input boli in the sense that  the abundance of some of its  constituent species, which  play an analogous role to neuronal weights, reflect statistical biases of the boli. In particular we consider two types of biases: \one {\em Frequency biases} (FB):  one or more input channels of the CN receive boli at different rates. \two {\em Time correlations} (TC): two or more input channels are correlated in time. 
The TC task can be understood as a direct generalisation of associative learning with an arbitrary number of input channels.
\par
We will propose three different  versions of the CN. The first (basic) version will be the CN itself, which is a minimal set of chemical reactions. It is also  thermodynamically consistent  in that it comprises only micro-reversible reactions with mass-action kinetics. This first version, while compact, assumes a high degree of enzymatic multiplicity which is unlikely to be realisable.  Therefore, we shall propose a second version of the model, which is not thermodynamically explicit, but biologically plausible  in the sense that  it can be formulated in terms of known bio-chemical motifs. The  main difference between this and the previous system is that the former  is  compartmentalised. Henceforth, this compartmentalised system will be referred to as {\em c-CN}. 
\par
We also propose {\em \neuron{}}, a version of the CN that is formulated using DNA strand displacement (DSD) \citep{visual_dsd}, a type of DNA-based computing.  DSD is a  molecular computing paradigm  based  entirely on interactions of DNA strands and  Watson-Crick complementarity, and is \emph{bio-compatible}. By this we mean that DSD computers can, in principle, be injected into organisms and interact with their biochemistry \citep{amir2014universal},  and therefore have potential to be used to control molecular systems. It has been shown that DSD systems are capable of universal computation \citep{Seelig1585} and indeed that any chemical reaction network  can be emulated in DSD \citep{soloveichik2010, chen_dalchau2013}.  From a practical point of view, it is relatively easy to experimentally realise DSD systems and their behaviour can also be accurately predicted \citep{Yurke_2000, Fontana2006} using simulation software such as Visual DSD \citep{visual_dsd} or Peppercorn \citep{Badelt_2020}.  There is now also a wealth of computational methods and tools for designing DNA-based circuits \citep{Cardelli, cross_talk}. 
\par
Given these properties, there have been a number of attempts to build intelligent DSD systems. Examples include linear-threshold circuits, logic gates \citep{Seelig1585, qian2011scaling}, switches  \citep{biol_switches}, oscillators \citep{DNA_circuits}, and consensus algorithms  \citep{chen_dalchau2013}.

There were also some attempts to emulate neural networks in DSD: Qian {\em et al.} \citep{DNA_NN_cascades} proposed a Hopfield network which has the ability to complete partially shown patterns. However, because the weights connecting individual neurons  were  hard-coded into the system, the system was unable to learn. Networks of perceptron-like neurons with competitive winner-take-all  architectures have also been proposed \citep{Genot2013ScalingDD,DNA_WTA}, and shows how to use DSD reaction networks to classify patterns, such as MNIST handwritten digits \citep{mnist}. However, learning is external to these systems; weights have to be determined before building the DNA circuit and are then hard-coded into the design. 
\par
Supervised learning in DSD was proposed by Lakin and collaborators \citep{supervised}. They used a two-concentration multiplier circuit motif in order to model the gradient descent weight update rule. However, this approach requires an external observer to provide constant feedback. From the perspective of implementing artificial proto-intelligence in biochemistry, none of the above approaches can be used as a fully autonomous component of a molecular learning system, in the sense that they can operate independently of constant external maintenance.
\par

\begin{table}[ht]
 \centering
  \begin{tabular}{ll} 
   \hline
  Acronym & Definition \\
   \hline
CN &  Chemical neuron \\
c-CN &  Compartmentalised chemical neuron \\
d-CN &  DNA chemical neuron \\
FB &  Frequency biases \\
TC &  Time correlations \\
DSD &  DNA  strand displacement \\
   \hline
  \end{tabular}
  \caption{List of acronyms. }
  \label{acr_table}
\end{table}

\section{Results}

In the first part of this section we describe the micro-reversible chemical reactions that constitute the CN. Next we demonstrate that the system of reactions behaves like a spiking neuron, and we analyse the key parameters that determine the performance of the system. In the subsequent section we describe c-CN, which lends itself more easily to experimental implementation. Finally, we discuss how DNA strand displacement can be used to construct the  \neuron{}.

\subsection{The chemical neuron --- minimal model}
\label{cnsec}

\paragraph{Overview}
We model the CN as a set of micro-reversible elementary chemical reactions obeying mass-action kinetics (\tabref{reaction_table}). Micro-reversibility makes the model thermodynamically consistent. The system is best understood by thinking of each molecular species $A_i$ as an input to the system via channel $i$. 
The inputs are provided in a form of \emph{boli}, which is defined as a fixed amount of molecules introduced to the system at the time of the input. 
The weight equivalent of the $i$-th input channel of the CN is the abundance of the species $H_i$. The species $\mathcal E$ is the activated form of $E$ and plays a dual role. It is \one the learning signal, which indicates that a weight update should take place, and  \two it is the output of the CN, which could be coupled to further neurons downstream. The internal state of the CN, which acts as a memory for the system,  is represented by the abundance of the molecular species  $B$. We now proceed by discussing each reaction in \tabref{reaction_table} in turn.

\begin{figure}[ht]
\centering
\includegraphics[width=0.7\textwidth]{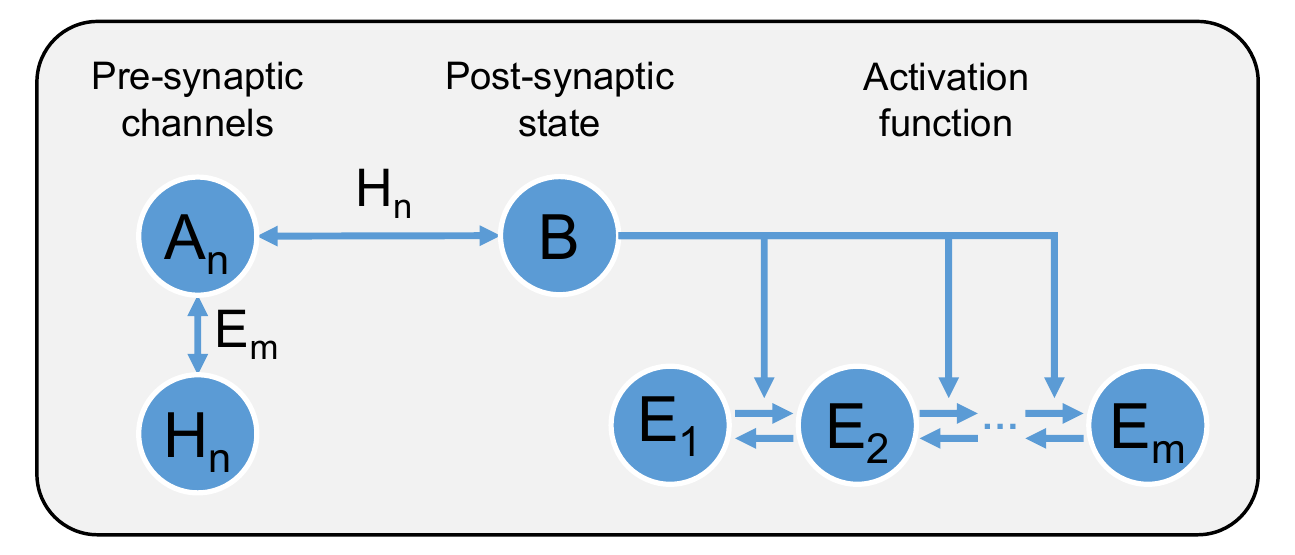}
\caption{Graphical representation of the minimal model of the CN.}
\label{CN_diagram}
\end{figure}
\begin{table}[ht]
\centering
\begin{tabular}{ll} 
 \hline
Function & Reaction(s) \\
 \hline
 \multirow{2}{*}{Input}
& \ce{I_n <=>[k_{IA}][k_{AI}] A }  \\
& \ce{A_n <=>[k_{AB}][k_{BA}] B}  \\ 
 \hline
\multirow{2}{*}{Activation function }
& \ce{B + E_{i} <=>[k^{+}][k^{-}] E_{i+1}}, \qquad $i< m-1$ \\ & \ce{B + E_{m-1} <=>[k^{+}][k^{-}_{last}] $\mathcal{E}$}   \\ 
 \hline
\multirow{2}{*}{Learning}
& \ce{A_n + $\mathcal{E}$ <=>[k_{AE}][k_{EA}] $A\mathcal{E}_{n}$ <=>[k_{EH}][k_{HE}] H_n + $\mathcal{E}$}   \\
& \ce{A_n + H_n <=>[k_{AH}][k_{HA}] AH_{n} <=>[k_{HB}][k_{BH}] B + H_n}   \\ 
\hline
\multirow{2}{*}{Leak}
& \ce{H_n ->[k_{H$\varnothing$}] $\varnothing$}   \\ 
& \ce{B ->[k_{B$\varnothing$}] $\varnothing$}   \\ 
\hline
\end{tabular}
\caption{List of chemical reactions constituting the  CN.}
\label{reaction_table}
\end{table}
%

\paragraph{Input}
We assume here that the CN has  $N$ different species of input molecules $A_1,\ldots,A_N$. These represent the $N$ input channels each of which is  associated with a corresponding weight $H_1, \ldots,H_N$. The weight molecules are the interpretable output of the neuron, in the sense that the abundance of the $H_i$ molecules will reflect statistical biases in the input. The input is always provided as an exponentially decaying bolus at a particular time  $t_i^s$, where $s$ is a label for individual spikes. Concretely, this means that at  time $t=t_i^s$ the CN is brought into contact with a reservoir consisting of $\beta$ (un-modelled) precursor molecules $I_i$ that then decay into $A_i$ molecules with a rate constant $\kappa>0$. A particular consequence of this is that the $A_i$ are not added instantaneously, but will enter the system over a certain time. This particular procedure is a model choice that has been made for convenience. Different choices are possible and would not impact on the results to be presented. The important point is that the input signal  to channel  $i$ is a  bolus of quantity $A_i$ and occurs at a  particular time $t$. This enables the system to reach a steady state provided that the input is stationary. 
\par
The basic idea of the CN is that input boli $A_i$ are converted into internal state molecules $B$. 
This reaction takes a catalysed, as well as an uncatalysed form. The uncatalysed reaction \ce{A_n <=>[k_{AB}][k_{BA}] B} is necessary in order to allow the system to learn to react in response to new stimulus, even when the weight associated with a given channel decayed to 0. 
In the case of the catalysed reaction the channel specific $H_i$ molecules play the role of the catalyst. 
Thus the speed of conversion depends on the amount of weight $H_i$. 
If at any one time there is enough of $B$ in the system then the learning signal $\mathcal E$ is created by activating $E$ molecules. Once the learning signal is present then some of the $A_i$ are converted into weight molecules, such that the weight of the particular input channel increases. This realises Hebbian learning in the sense that the coincidence of inputs $A_i$ and output $\mathcal E$ activates weight increases, following  the well known Hebbian tenet ``What fires together, wires together".  

\paragraph{Activation function}
The link between the internal state molecules $B$ and the learning signal  is often called the  {\em activation function}.
In spiking neurons, as they are used in artificial intelligence, this activation is usually a threshold function. The neuron triggers an output if the internal state crosses a threshold value. In chemical realisations, such a threshold function is difficult to realise. Throughout this contribution, our systems are parametrised such that the dynamics of the system is dominated by noise. Molecular abundances are therefore noisy.  As a consequence, the activation function has to be seen as the  {\em probability} to observe the activated form $\mathcal{E}$ as a function of the abundance of  $B$. 
\par
An ideal activation function would be a step function, but physical realisations will necessarily need to approximate the step function by a continuous function, for example a sigmoid. In the CN, this is realised as follows:  Each of the  $E$ molecules has $m$ binding sites  for  the internal state molecules  $B$. Once all $m$ binding sites are occupied, then $E$  is converted into its active form $\mathcal E$. We make the simplifying assumption that the conversion from $E$ to $\mathcal E$ is instantaneous once the last $B$ binds. Similarly, if a $B$ molecule unbinds, then the $\mathcal E$ changes immediately to $E$. In this model, the  balance between $\mathcal E$ and $E$ molecules depends on the binding and unbinding rates of $B$. We assume that there is a cooperative interaction between the $B$ molecules such that unbinding of $B$ from $\mathcal E$ is much slower than unbinding from $E$. With an appropriate choice of rate constants, this system is known to display ultra-sensitivity, i.e. the probability for the fully occupied form of the ligand chain ($\mathcal E$) to exist transitions rapidly from close to 0 to close to 1 as the concentration of ligands approaches a threshold value $\vartheta \approx k_+/k_-$. The dynamics of such systems is often approximated by the so-called Hill kinetics.  It can be shown that the maximal Hill exponent that can be achieved by such a system is $m$ \citep{Chu_2009}. This means that the chain-length  $m$, which we henceforth shall refer to as the ``non-linearity'', controls the steepness of the activation function of $\mathcal E$. In the limiting case of $m=\infty$, this will be a step function, whereby the probability to  observe $\mathcal E$ is 0 if  the abundance of $B$  is below a threshold and 1  otherwise. We are limited here to  finite values of $m>0$. In this case, the function is sigmoidal, or a saturating function in the case of $m=1$. The parameter $m$ and hence the steepness of the activation function will turn out to be a crucial factor determining the computational properties of the CN. 

\begin{figure}[ht]
\centering
\subfloat[][]{\includegraphics[width=\textwidth]{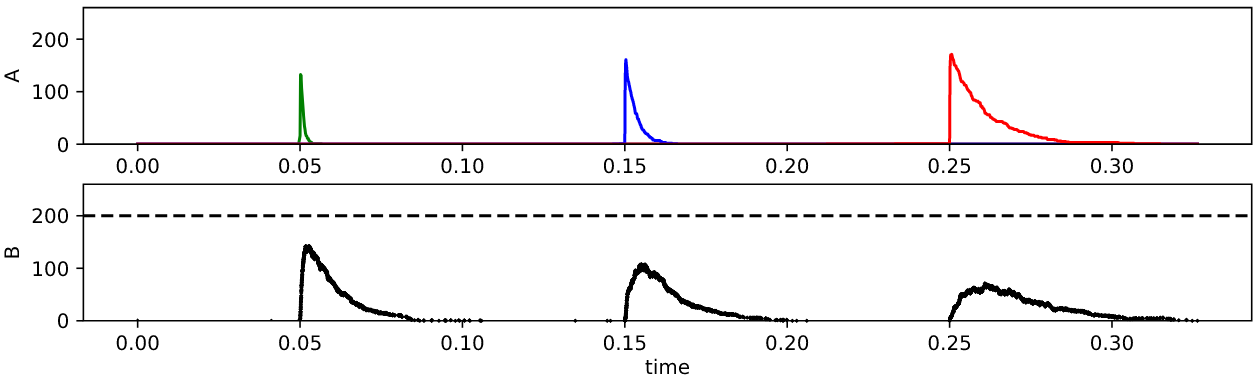}\label{a}}\\
\subfloat[][]{\includegraphics[width=\textwidth]{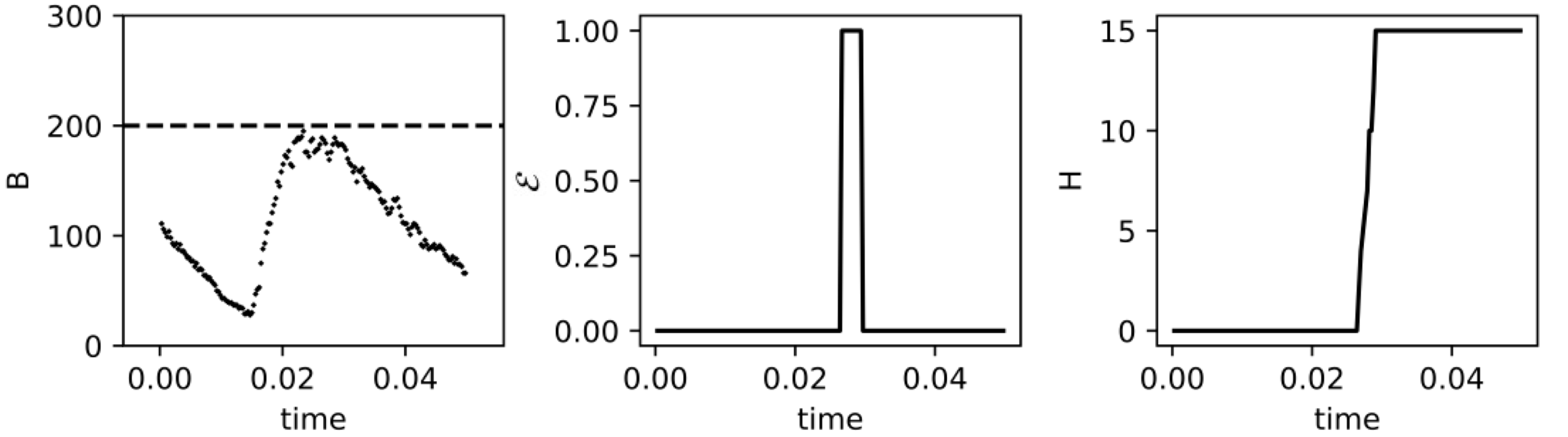}\label{b}}
\caption{
\protect\subref{a}
Example of three inputs of uniform size received from 3 different channels.  Each input shown in the second graph has a different weight associated it: $H_{\textrm{green}} = 250$, $H_{\textrm{blue}} = 50$, and $H_{\textrm{red}} = 0$.  $H$ molecules act as a catalyst in $\ce{A_{n} <=>[H_{n}] B}$ reaction, hence the change in the function of $B$ molecules over time for each of the inputs. The higher the amount of $H$, the higher is the peak of $B$ molecules caused by a particular input.  Moreover, with the increase in weights, the function of inputs also changes.  The higher the amount of $H$, the quicker its corresponding $A$ dissipates. 
\protect\subref{b}
Example simulation showing the core idea of the CN dynamics. The graphs show the internal state $B$, the learning signal $\mathcal E$ and the weight $H$ for a single channel. We assume a bolus provided at time $t=0.015$. This causes the internal state to go up and reach the threshold. A learning signal is triggered at around $t=0.03$ and consequently the weight is increased by (in this case) 15 molecules of $H$. }
\label{example_CRN_small}
\end{figure}

\paragraph{Learning}
In neural networks, ``learning'' is usually associated with the update of weights. Accordingly, in the case of the CN, learning  is the  change of abundances $H_i$.   The abundance can only increase   if two conditions are fulfilled: \one  the learning signal $\mathcal E$ is present and \two there are still input molecules $A_i$ in the system. In short, learning can only happen if input and output coincide, which is precisely the idea of Hebbian learning. For an illustrative example of how Hebbian learning works in the CN, see \figref{example_CRN_small}.

\paragraph{Leak}
Finally, we assume that the weight molecules $H_i$ and the internal state molecules $B$ decay, albeit at different rates.   This is so that the weight abundances can reach a steady state; additionally it  enables the  CN to forget past inputs and to  adapt when the statistics of the input changes. We will assume that the decay of $H_i$ is slow compared to the typical rate of input boli.

Throughout this paper we will assume that the dynamics of $A$, $B$ and $E$ are fast compared to the change in concentration of $H$. This is a crucial assumption to allow the weights to capture long-term statistics of inputs; in particular, the weights should not be influenced by high frequency noise present in the system.  Furthermore, we also assume that the lifetime of $\mathcal E$ is short. For details of the parameters used see \tabref{params_table}.

\subsubsection{Associative learning}
\label{associate}

\begin{figure}[h]
\centering
\includegraphics[width=0.8\textwidth]{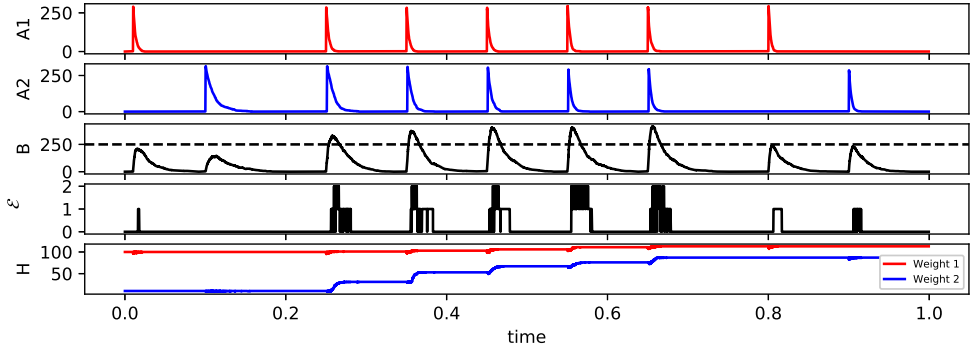}\label{association_learning}
\caption{
Associative learning in CN. The first two graphs show the inputs  $A_1$ and $A_2$. Clearly, a single $A_2$ does not lead to a sufficient increase of the internal state $B$, such that no learning signal is triggered. After a few coincidences of $A_1$ and $A_2$  the weights $H_2$ (last graph) have increased sufficiently for $A_2$ to trigger a signal in its own at time $t=0.8$.  Note the increase in weights for the second channel after each coincidence. 
} 
\label{asso}
\end{figure}

We first demonstrate that the CN is capable of associative learning (\figref{asso}). To do this, we generate a CN with $N=2$ input channels. Then,  we initialise the CN with a high weight for the first channel ($H_1=100$) and a low weight for the second channel ($H_2=0$).  Furthermore, we set the parameters of the model such that a bolus of $A_1$ is sufficient to trigger an output, but a bolus of $A_2$, corresponding to stimulating the second channel is not.
This also means that presenting simultaneously both $A_1$ and $A_2$  triggers a learning signal and increases $H_1$ and $H_2$. If  $A_1$ and $A_2$ coincide a few times, then the weights of $A_2$  have increased sufficiently so that a bolus of $A_2$ can push the internal state of the system over the threshold on its own. This demonstrates associative learning. Note that unlike some previous molecular models of associative learning (e.g. ref. \citenum{Fernando2009}), the CN requires several coincidences before it learns the association. It is thus robust against noise. 

This means that the CN can also readily unlearn the correlation if input patterns change (see fig. \ref{FB_relearn} and \ref{TC_relearn}). There are two mechanisms in the system that ensure that the neuron is able to continuously learn new input statistics. These are \one the decay of the weights, which ensures a rate of forgetting, and \two the uncatalyzed reaction $A_n$ to $B$ which allows the system to learn to react in response to new stimulus, even when the weight associated with a given channel decayed to 0.

\subsubsection{Full Hebbian learning}

We now show that the ability of  the CN to learn extends to full Hebbian learning with an arbitrary number of $N$ input channels. First we consider the FB task, where the CN should detect  input channels that fire at a higher frequency than others. To do this, we  provide random boli  to each of the $N$ input channels.  Random here means that the waiting time between two successive boli of  $A_i$ is distributed  according to an exponential distribution with parameter $1/f_i$, where $f_i$ is the frequency of the input boli to channel $i$.  The  CN should then  detect the difference in frequencies $f_i$ between input channels.  We  consider the FB task as solved if (after a transient period) the ordering of the abundances of weights reflects the input frequencies, i.e. the number of $H_i$ should be higher than the number of $H_j$ if $f_i> f_j$. Below we will show, using a number of example simulations, that the CN is indeed able to show the desired behaviour. Later on, we will probe in more detail how the response of the system depends on its parametrisation and the strength of the input signal.  	

\begin{figure}[h]
  \centering
  \subfloat[][]{\includegraphics[width=0.8\textwidth]{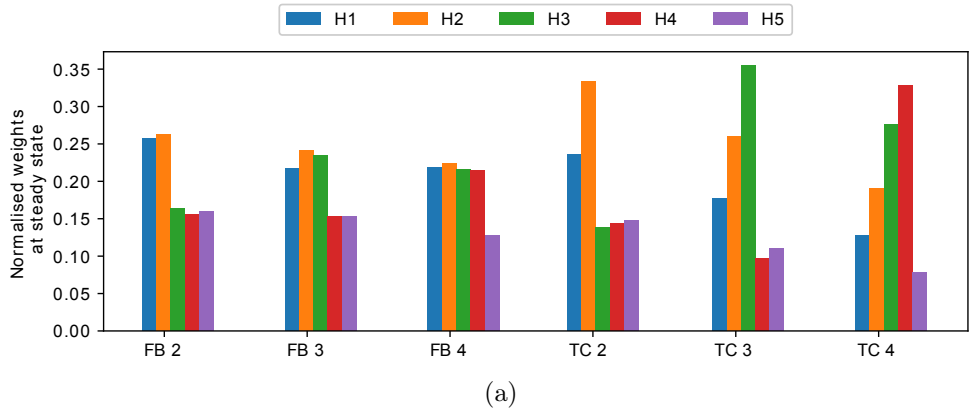}\label{tasks_fig}}\\
  \caption{
    \protect\subref{tasks_fig} Normalised weights for a variety of TC and FB tasks. The first (blue) bar refers to the first weight, the second (orange) to the weight for the second channel and so on. Each value represents the average over 300 time units of a single simulation.  Data was only collected after the weights reached the steady state (after 700 time units).  In all experiments we set the number of $E_1$ molecules at the start of the simulation to 40. The non-linearity was set  to $m=5$ for the TC, and $m=1$ for FB.
  }
  \label{bars_CN}
\end{figure}

In order to test a CN with multiple inputs ($N=5$, $m=1$), we consider 3 variants of the FB task. First, we assume that boli to the first \emph{two} input channels come at a frequency of 4 Hz whereas channels 3, 4 and 5  fire at a  frequency of  2 Hz; we call this variant FB 2. Similarly, for FB 3 and FB 4 the first 3 and 4 channels respectively fire at the higher frequency. \figref{tasks_fig} shows the steady state weights for each of the three tasks. As expected, in each of the experiments the weights of the high-frequency inputs are higher when compared to the low frequency inputs. We conclude that the CN can work as a frequency detector at least for some parametrisations.


\par
The other scenario that we will investigate is the TC task, which is the direct generalisation of the associative learning task to an arbitrary number of input channels. For this problem we assume that all input frequencies are the same, i.e. $f_i=f_j$ for all $i,j\leq N$. Instead of differences in frequency, we  allow temporal correlations between input boli of some channels. If $A_1$ and $A_2$ are temporally correlated then each bolus of $A_1$ is followed by a bolus of $A_2$ after a time period of $\delta + \xi$; with $\delta$ being a fixed number and $\xi$ is a random variable drawn from a normal distribution  with $\mu=0$ and $\sigma^2=0.0001$ for each bolus. In all simulations, the input frequency of all channels is set to 2 Hz.
\par
The CN can solve the TC task in the sense that, after a transient period, the weights indicate which channels are correlated. They also indicate the temporal order implied by the correlation, i.e. if $A_i$ tends to precede $A_j$, then the abundance of weight $H_i$ should be lower than the abundance of $H_j$. Furthermore, if $A_i$ is correlated with some other channel $k$, but $A_j$ is not, then the abundance of  $H_i$ must be greater than that of $H_j$. 
\par
In order to test whether the system is indeed able detect TC biases, we again simulated a CN with $N=5$ input channels and all  weight molecules  initialised to  $H_i=0$.  We then determined the steady state weights in four different  scenarios: there are correlations between \one  $A_1$ and $A_2$ (TC 2), \two  $A_1,A_2, A_3$ (TC 3), and \three $A_1,A_2,A_3, A_4$ (TC 4). The temporal order is always in ascending order of the index, such that in the last example, $A_1$ occurs before $A_2$, which in turn occurs before $A_3$.  We find that the behaviour of the CN is as expected (\figref{tasks_fig}). At steady state the weights reflect the correlation between input channels, including the temporal ordering, thus allowing us to conclude that, at least for some parametrisations, the CN successfully identify temporal correlations.

\subsubsection{Analysis of activation function non-linearity}

The ability of the CN to perform in the TC task depends on its ability to detect coincidences. In this section, we will now analyse in more detail how this coincidence detection depends on the non-linearity of the activation function, i.e. the parameter $m$. To do this, we consider two extreme cases: Firstly, the case of minimal non-linearity (i.e. $m=1$) and the secondly, the limiting (and hypothetical) case of maximal non-linearity (i.e. $m=\infty$). This latter case would correspond to an activation function that is a step function. While a \emph{chemical} neuron cannot realise a pure step function, considering the limiting case provides valuable insight.
\par
We consider first this latter scenario with a CN with two inputs $A_1$ and $A_2$. In this case, there will be a learning signal $\mathcal E$ in the CN if the abundance of $B$ crosses the threshold $\vartheta$. Let us now assume that the parameters are set such that a single bolus of either $A_1$ or $A_2$ is not sufficient to push the abundance of $B$ over the threshold, but a coincidence of both is. In this scenario then we have:  
\begin{itemize}
\item
A single bolus of $A_1$ will not lead to a threshold crossing. No learning signal is generated and  weights are not increased.
\item
If a bolus of $A_1$ coincides with a bolus of $A_2$ then this may lead to a crossing of the threshold of the internal state. A learning signal is generated. Weights for both input channels 1 and 2 are increased (although typically not by equal amounts).   
\end{itemize} 
Next consider an activation function tuned to the opposite extreme, i.e. $m=1$. It will still be true that both $A_1$ and $A_2$ are required to push the abundance of $B$ across the threshold. However, the learning behaviour of the CN will be different:
\begin{itemize}
\item
A single bolus of $A_1$ will not lead to a threshold crossing. A learning signal may still be generated even below the threshold because the activation function is not a strict step function.  The weight $H_1$ will increase by some amount, depending on the bolus size. 
\item
If a bolus of $A_1$ coincides with a bolus of $A_2$ then this will lead to more learning signal being  generated than in the case of  $A_1$ only.  As a result,  the  weights for both input channel 1 and 2 are increased by more than if they had occurred separately.   
\end{itemize} 
\par
These two extreme cases illustrate how the CN integrates over input. In the case of low non-linearity the weights of a channel will be a weighted sum over all input events of this channel. The weights will be  higher  for  channels  whose boli coincide often.   On the other hand, a step-like activation function will integrate only over those events where the threshold was crossed, thus specifically detect coincidences.  From this we can derive two conjectures:
\begin{itemize}
\item
The higher the non-linearity, the better the CN at detecting coincidences. Low non-linearity still allows coincidence detection, but in a much weaker form.
\item
As the bolus size increases, the CN will lose its ability to detect coincidences, especially when the bolus size is so large that a single bolus is sufficient to push the abundance of $B$ over the threshold. In this case, a single input spike can saturate the activation function, thus undermining the ability of the system to detect coincidences effectively. 
\end{itemize} 

\begin{figure}[ht]
\centering
\includegraphics[width=0.95\textwidth]{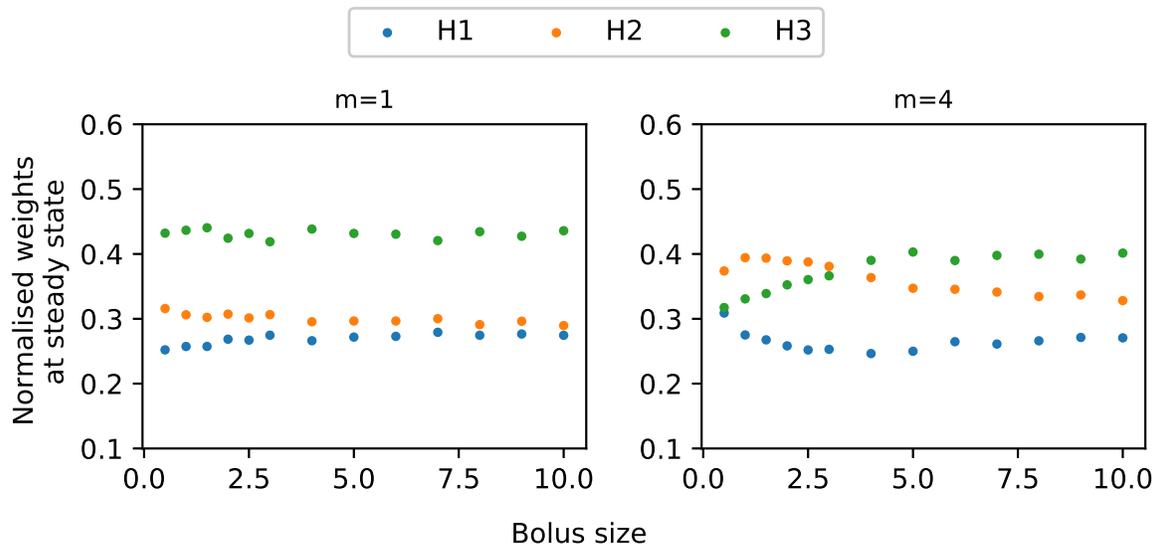}
\caption{The steady state weights as a function of bolus size for a CN with 3 inputs. Input $A_3$ (green) is provided at 4Hz, $A_1$ and $A_2$ are correlated with a $\delta= 0.0047 $ but they are only provided at 2Hz. The graph shows the normalised weights at steady state corresponding to the input channels for different bolus-sizes (here reported as a fraction of the threshold). From left to right, the bolus size increases. For $m=1$ the system detects the higher frequency of  $A_1$ as indicated by its high weight. It also differentiates between the correlated inputs, but with weaker signal. As the bolus size increases, the neuron maintains its ability to recognise FB, but can no longer detect TC, i.e. $H_1$ and $H_2$ have the same abundance. For the higher non-linearity ($m=4$) the system detects the TC ($H_2$ has a higher abundance than $H_1$). As the bolus size increases, it detects the FB, but its ability to detect TC decreases.}
\label{meanH_bolus_m1_5_mixHz}
\end{figure}

In order to check these conjectures, we simulated a version of the CN with 3 inputs, where $A_1$ and $A_2$ are correlated and $A_3$ fires at twice the frequency of $A_1$ and $A_2$. We considered the minimally non-linear case ($m=1$) and a moderate non-linearity ($m=4$), which shows the weights as a function of the bolus size (\figref{meanH_bolus_m1_5_mixHz}). The minimal non-linear CN detects both coincidences and frequency differences, but loses its ability to detect coincidences as the bolus size increases. This is consistent with the above formulated hypothesis. In contrast, for the non-linear CN and moderately low bolus-sizes the weights indicate the coincidences strongly (i.e. the weights $H_2$ are highest), and less so the FB.   As the bolus size increases the non-linear CN loses its ability to detect coincidences and becomes a frequency detector, as conjectured.

\begin{figure}[ht]
\subfloat[][]{\includegraphics[width=0.48\textwidth]{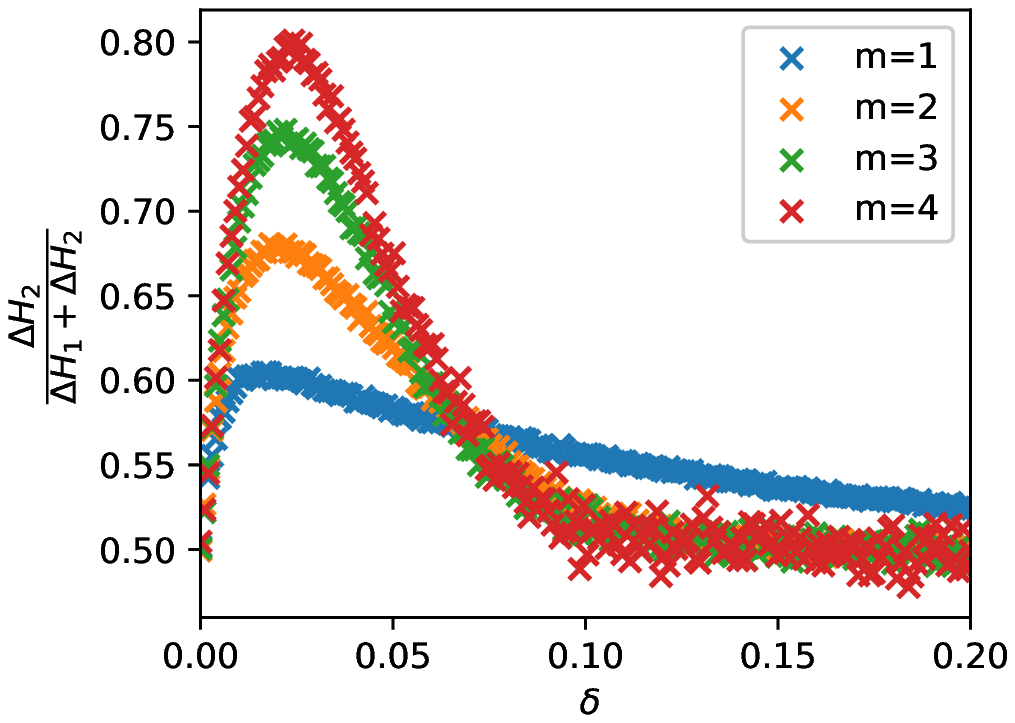}\label{aaaa}}
\subfloat[][]{\includegraphics[width=0.48\textwidth]{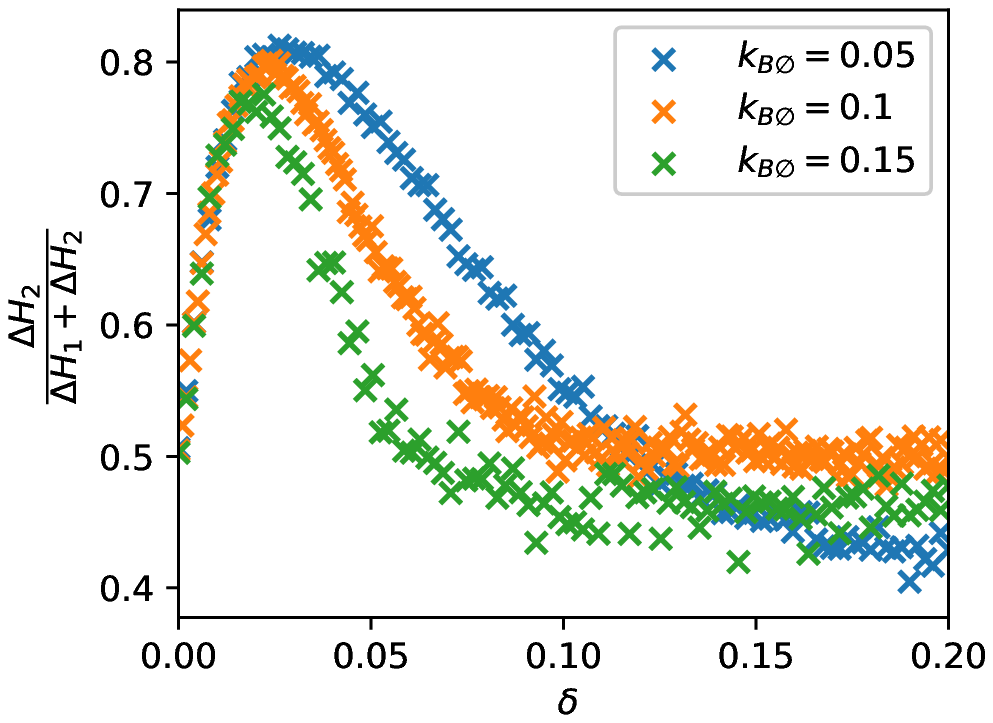}\label{long_ratio_sum} }
\caption{Differential weight increase for different non-linearities. For both graphs the  points were computed as follows:  We simulated a CN with two input channels only. We set the  initial condition  to  $H_1, H_2=0$. At time $t=0$ we provided a bolus of $A_1$ and after a time period of $\delta$ we provided the bolus $A_2$. We then continued the simulation for another $0.2$ time units. The $y$-axis records the relative increase of $H_2$  over $0.2$ time units averaged over 1000 repetitions. \protect\subref{aaaa}  We stimulate channel 1  followed by channel $2$ after a time period of $\delta$. Then we measure the amount by which weights $H_1$ and  $H_2$ were increased and record the fraction.  A value of 1 means that only the second input channel received weight accumulation. A value of $0.5$ means that the weights of both channels were updated equally. \protect\subref{long_ratio_sum}  Same, but for different removal rates of $B$. The faster the removal, the more specific the coincidence detection, i.e. inputs need to occur within a narrower window.}
\label{dist_m_long}
\end{figure}

Next, we check how the coincidence detection depends on the time-delay between the correlated signals. To do this we created a scenario where we provided two boli to the system. The first bolus $A_1$ comes at a fixed time and the second one a fixed time period $\delta$ thereafter.  We then vary the length of $\delta$ and  record the accumulation of weights $H_2$ as a fraction of the total weight accumulation. \figref{dist_m_long} shows the average weight accumulation per spike event. It confirms that the CN with the low non-linearity is less sensitive to short coincidences than the CN with higher $m$. However, it  can detect coincidences over a wider range  of  lag durations. This means that for  higher  non-linearities the differential weight update becomes more specific, but also  more limited in  its ability to detect coincidences that are far apart. In the particular case, for a $\delta> 0.1$ the CN with $m>1$ does not detect any coincidences any more, whereas the case of $m=1$ shows some differential weight update throughout.   

Next, we tested the conjecture that the TC can be solved more effectively by the CN when the non-linearity is higher.  To do this, we generated a CN with $N=5$ input channels on the TC 2 task. We then trained the CN for non-linearities $m=1,\dots, 10$. As a measure of the ability of the system to distinguish the weights, we used the index of dispersion, i.e. the standard deviation divided by the mean of the weights.
 A higher index of dispersion indicates more heterogeneity of the weights and hence a better ability of the system to discriminate between the biased and unbiased input channels. 

\begin{figure}
  \centering
  \includegraphics[width=\textwidth]{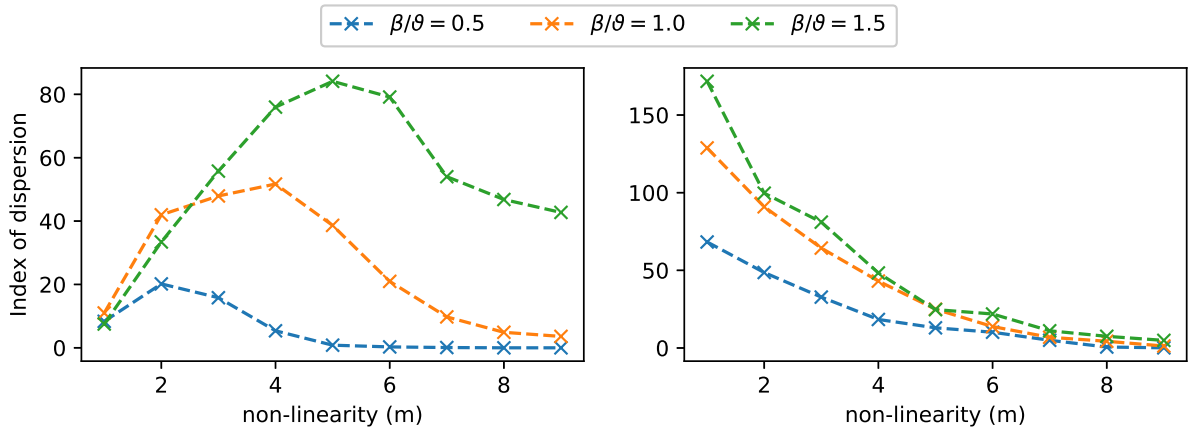}
  \caption{Index of dispersion for different bolus sizes $\beta$ expressed as a fraction of the threshold $\vartheta$. We show  TC 2 (left) and FB 2 (right). The index of dispersion measures how different the steady state weights are from one another, and hence indicates how well the CN distinguished between input channels. Completely unbiased input would give an index of dispersion of $\approx 0$. The graph shows that for the TC task, there is an optimal non-linearity. Increasing the bolus size, increases the optimal non-linearity, which is consistent with the fact that the optimum is due to resource starvation. }
  \label{iod_volume}
\end{figure}

Consistent with our hypothesis we found that the ability to distinguish temporarily correlated inputs increases with the non-linearity. However, it does so only up to a point (the optimal non-linearity), beyond which the index of dispersion reduces again (\figref{iod_volume}).  Increasing the bolus size, i.e. increasing the number of $A_i$ that are contained within a single bolus,  shifts the optimal non-linearity to the right.  This suggests that the decline in the performance of the CN for higher chain lengths is due to a resource starvation. The realisation of the sigmoidal function, i.e. the thresholding reactions in \tabref{reaction_table}, withdraws $m$ molecules of $B$  from the system. As a consequence, the CN is no longer able to represent its internal state efficiently and the activation function is distorted. If the total abundance of $B$ is high compared to $E$, then this effect is negligible. We conclude that there  is  a resource cost associated with  computing non-linearity. The higher $m$, the higher the bolus size required to faithfully realise the activation function. 
As an aside we note that, other designs for the system are also possible. For example, $B$ molecules could be used catalytically. Nevertheless, such systems would also face different trade-offs. The system presented here was one of many designs that we tested, and provided the most desirable properties for learning temporal patterns.

While the TC task requires a  non-linearity, the FB task, does not. This can be understood acknowledging that  the  FB task is fundamentally about integrating over input, which can be done naturally in chemical systems. Indeed, it can be done by systems that are much simpler than the CN. For example, the minimal system to detect FB bias is \ce{$A_{i}$ ->[d] $\varnothing$}.  For appropriately chosen values of $d$, the steady state value of $A_i$ would then reflect the input frequency. To understand this, note that the input frequency determines the  rate of increase of $A_i$. This rate, divided by the decay rate constant $d$ then determines the steady state abundance of $A_i$, such that $A_i$ trivially records its own frequency. This system is the minimal and ideal frequency detector.

The CN itself is not an ideal frequency detector because all weight updates are mediated by the internal state $B$. Hence, the weights are always convolutions over all inputs. The weights thus reflect both frequency bias and temporal correlations. In many applications this may be desired, but  sometimes it may not be. We now consider the conditions necessary to turn the CN into a pure frequency detector, i.e.~  a system that indicates only FB, but not TC. One possibility is to set the parameters such that the CN approximates the minimal system. This could be achieved by setting $k_\mathrm{BA} \ll  k_\mathrm{AB}$ and all other rate constants very high in comparison to $k_\mathrm{AB}$. The second possibility is to tune the CN such that a single bolus saturates the threshold. In this case, the strength of the learning signal does not depend on the number of boli that are active at any one time. A single bolus will trigger the maximal learning signal. This is confirmed by \figref{meanH_bolus_m1_5_mixHz}, which shows that as the bolus size increases, the system becomes increasingly unable to detect temporal correlations, but remains sensitive to frequency differences.

\subsection{c-CN: The CN with compartments}
\label{biosec}

The CN, as presented in \tabref{reaction_table}, is thermodynamically plausible and has the benefit of being easy to simulate  and analyse. However, it is biologically implausible. As written in \tabref{reaction_table} the molecular species $A_i, H_i$ and $B$ would have to be interpreted as conformations of the same molecule with  different  energy levels. Additionally, we require that these different conformations have specific enzymatic properties. Molecules with the required properties are not known currently, and it is unlikely that they will be discovered or engineered in the near future.  

As we will show now, it is possible to re-interpret the reaction network that constitutes the CN (\tabref{reaction_table}) so as to get a model whose elements are easily recognisable as common biochemical motifs. This requires only relatively minor adjustments of the reactions themselves, but a fundamental re-interpretation of what the reactions mean. 

\begin{figure}[ht]
  \centering
  \includegraphics[width=0.65\textwidth]{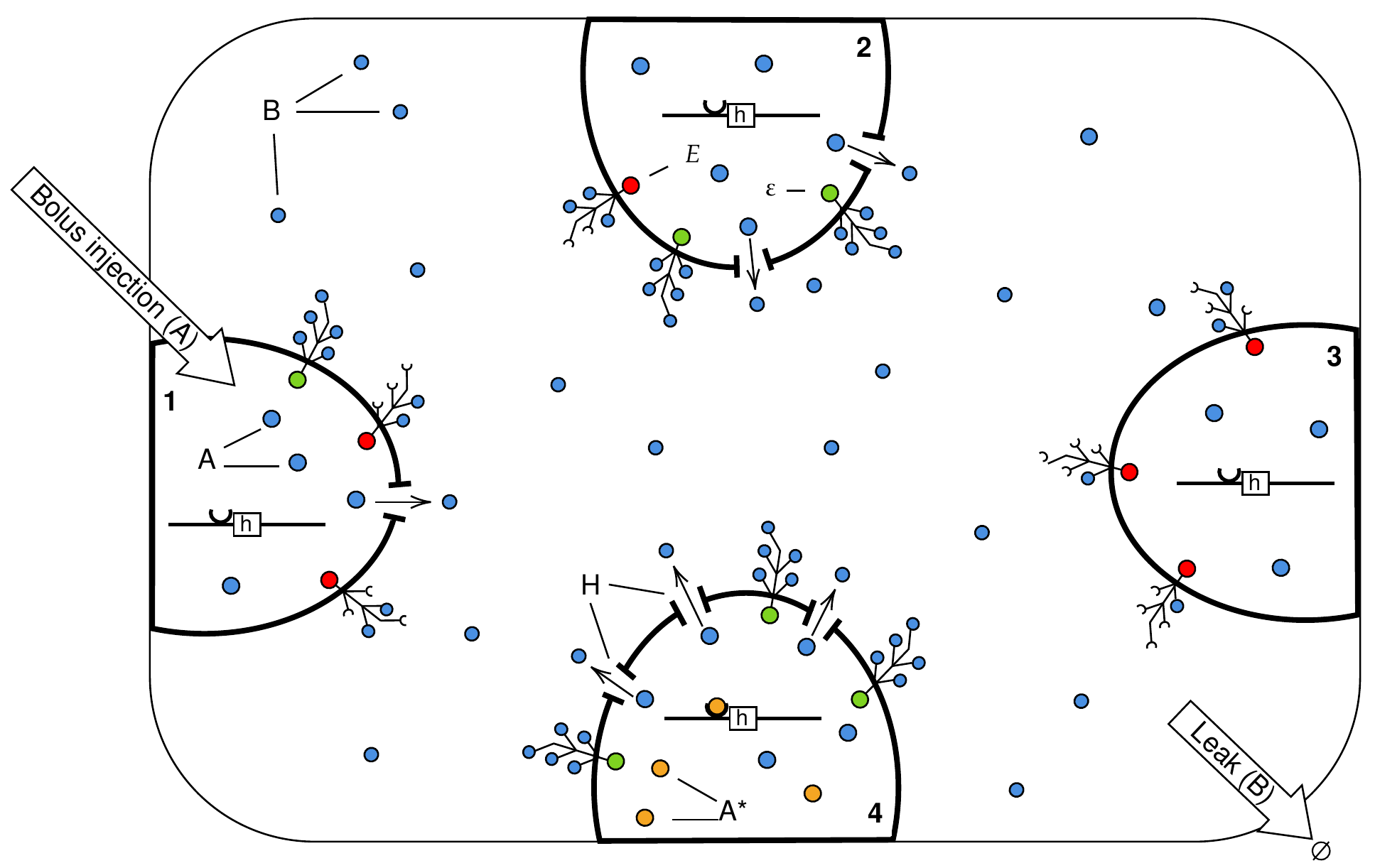}
  \caption{Graphical representation of a c-CN. $A_i$ and $A_j$ are the same molecular species but contained in different  compartments $i$ and $j$ respectively. We allow for an activated form of $A$, denoted by $A^*$, which binds to the promotor site of $h$ and activates its expression.   $H$ is an active transporter molecule for $A$. Once exported to the extracellular space, an  $A_i$ molecules becomes a molecule of $B$.  We assume that each compartment has a trans-membrane protein $E$ with $m$ extra-cellular binding sites. If all $m$ binding sites are occupied by $B$, then the internal site becomes active (indicated by green) and can catalyse the activation of $A$. }
  \label{biomodel}
\end{figure}

The main difference we introduce is that the new model is compartmentalised (\figref{biomodel}). While in the basic model the indices of $A_i$ and $H_i$ referred to different species that exist in the same volume, it should now be interpreted as the same species but living in different compartments. This means that $A_i$ and $A_j$ are the same type of molecule, but located in compartments $i$ and $j$ respectively. Similarly, $H_i$ and $H_j$ are the same species. All compartments $i$ and $j$ are themselves enveloped in a further compartment (the ``extra-cellular space''). The internal state species $B$ is the same as $A_i$ but located in the extra-cellular space. From here on, we will refer to this re-interpreted model as the c-CN. It is formally described by the reactions in \tabref{reaction_table2}.

\begin{table}[ht]
  \centering
  \begin{tabular}{ll}
   \hline
  Function & Reaction(s) \\
 \hline
\multirow{2}{*}{Input}
& \ce{I_n <=>[k_{IA}][k_{AI}] A }  \\
& \ce{A + H <=>[k_{AH}][k_{HA}] AH <=>[k_{HB}][k_{BH}] B + H}   \\
 \hline
  \multirow{2}{*}{Activation function}
  & \ce{B + E_{i} <=>[k^{+}][k^{-}] E_{i+1}}, \qquad $i< m-1$ \\
  & \ce{B + E_{m-1} <=>[k^{+}][k^{-}_{last}] $\mathcal{E}$}   \\
   \hline
  \multirow{8}{*}{Weight accumulation}
  & \ce{$\mathcal{E}$  + A <=>[k_{AE}][k_{EA}] $\mathcal{E}$A <=>[k_{A*E}][k_{EA*}] $\mathcal{E}$ + A^*} \\
  & \ce{A^* <=>[k_{A*A}][k_{AA*}] A } \\
  & \ce{$h_0$ + $A^*$   <=>[A*h][hA*]  h }  \\
  & \ce{$h_0$  ->[$k_\mathrm{leak}$] H_n + h_0 }  \\
  & \ce{$h$  ->[$k_{h}$] H_n + h }  \\
 \hline
  \multirow{2}{*}{Leak}
  & \ce{H ->[k_{H$\varnothing$}] $\varnothing$}   \\
  & \ce{B ->[k_{B$\varnothing$}] $\varnothing$}   \\
   \hline
  \end{tabular}
  \caption{List of chemical reactions constituting the  c-CN. Molecular species $A,E,\mathcal E, h_0,h$ and $H$ are compartmentalised. Each compartment has a gene $h_0$ which when activated by $A^*$ can express a transporter $H$.}
  \label{reaction_table2}
\end{table}

Input to channel $i$ is  provided by boli of the molecular species $A$ into the compartment $i$. A novelty of c-CN when compared to CN is that it has  an activated form of $A$, denoted by $A^*$.  The conversion from $A$ to $A^*$ is catalysed by the learning signal $\mathcal E$.  Also new is that each compartment contains a gene $h$ that codes for the molecule $H$ (we suppress the index indicating the compartment). Expression of the gene is activated by $A^*$ binding to the promoter site of $h$. We also allow a low leak expression by the inactivated gene (denoted as $h_0$ in \tabref{reaction_table2}). Gene activation of this type is frequently modelled using Michaelis-Menten kinetics, thus reproducing in good approximation the corresponding enzyme kinetics in the CN.
The molecules of type $H$ are now transporters for $A$.  We then interpret the conversion of $A_i$ to $B$ as export of $A$ from compartment $i$ to the extra-cellular space. The rate of export of $A$ is specific to each compartment in that it depends on the abundance of $H$ in this compartment.    Finally, we interpret the $E$ molecules as transmembrane proteins that are embedded in the membrane of each compartment. Their extra-cellular  part has $m$ binding sites for $B$ molecules which bind cooperatively. When all sites are occupied then the intra-cellular part is activated, i.e. becomes $\mathcal E$. In its activated form it can convert $A$ to $A^*$.
\par
Another difference between the two versions of the models is that the molecule $E$ is now specific to each membrane. The minimum number of copies of $E$ is thus $N$ whereas in the basic model a single copy of $E$ at time $t=0$ could be sufficient.  This has two consequences. Firstly, at any particular time the number of occupied binding sites will typically  be different across the different $N$ compartments. This is a source of additional variability. Moreover, since the number of copies of $E$ is higher than in CN, the c-CN  is more susceptible to starvation of $B$ as a result of the extra-cellular binding sites withdrawing molecules from the outer compartment.  Both of these potential problems can be overcome by tuning the model such that the abundance of $B$ molecules is high in comparison to $E$ molecules.  

\begin{figure}[ht]
\centering
\includegraphics[width=0.8\textwidth]{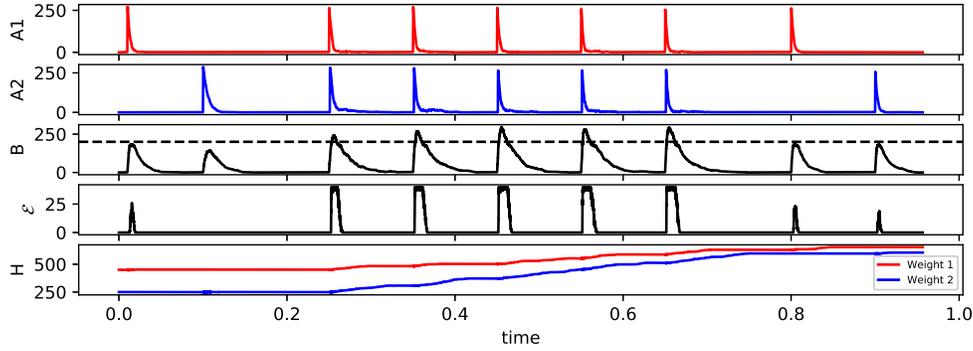}
\caption{
Same as fig. \ref{asso} but for c-CN. For the parameters used see \tabref{params_table2}. For this experiment we  approximated the ligand kinetics by a Hill function, in order to speed up the simulations.
} 
\label{asozial}
\end{figure}

\begin{figure}[h]
  \centering
  \includegraphics[width=0.8\textwidth]{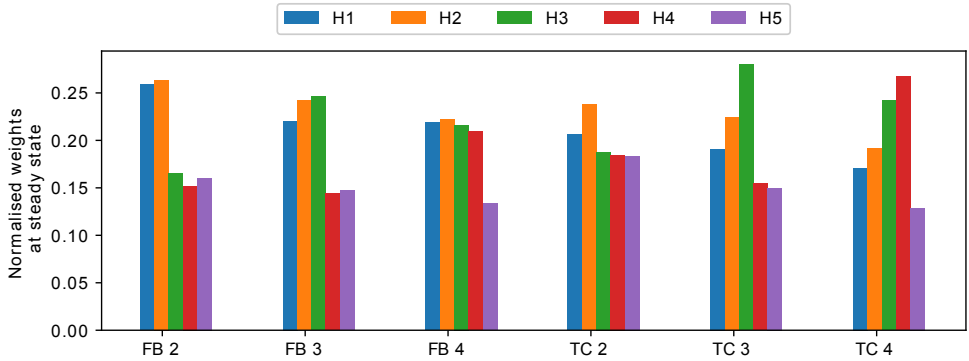}\label{biomodel_res}
  \caption{
Same as fig. \ref{bars_CN} but for c-CN. The experiments approximated the ligand dynamics by a Hill function, in order to speed up the simulations. 
  }
  \label{bars_cCN}
\end{figure}

This highlights that the difference between the basic CN and c-CN are deeper than the list of reaction suggests. Our simulations, however, confirm that the c-CN supports   associative learning (\figref{asozial}) and full Hebbian learning (\figref{bars_cCN}) just as the basic CN, provided that the parameters are set appropriately.

\subsection{\neuron{}: a neuron in DNA}
\label{dnasec}
We now show how to emulate  the chemical reaction network of  \tabref{reaction_table} using DNA strand displacement (DSD)  \citep{visual_dsd}. This is interesting, because the experimental realisation of DSD systems is straightforward and predictable when compared to bio-chemical reaction networks.
\par   
The basic idea of DNA-based computation is that double stranded DNA molecules with an overhang on one strand --- often called the {\em toehold} --- can interact with single stranded DNA that contains the Watson-Crick complement of the toehold via partial or total displacement of the existing complement.
DNA-based systems are typically analysed on two levels: sequence-level and domain-level. The former involves the  study of interactions between individual nucleotide pairs, while the latter focuses on the interactions between \emph{domains}.  Here, domains are  sequences of nucleotides of varied length.  There are two types of domains which are differentiated by their length. Short domains or \emph{toeholds} are between 4 to 10 nucleotides, and are assumed to be able to bind and unbind from complementary strands.  Long domains, or {\em  recognition} domains, are at least 20 nucleotides in length, and assumed to bind irreversibly.  DSD is a domain-level mechanism for performing computational tasks with DNA via two basic operations: toehold mediated branch migration and strand displacement. 

\subsubsection{Implementing the \neuron{} using two-domain DNA strand displacement}

In order to emulate the chemical neuron in DNA, we will focus here on two-domain strand displacement  \citep{Cardelli, chen_dalchau2013}, where each molecular species comprises a toehold and a long domain only. These species can interact with double-stranded gates which facilitate the computation.  Restricting computation to two-domain strands helps to protect against unexpected interactions between single stranded species, which can occur with more complex molecules.  Also, as all double-stranded structures are stable, and can only change once a single-stranded component has bound, there is no possibility for gate complexes to polymerise and interact with each other.
\par
Here, we will be using the standard syntax of the Visual DSD programming language \citep{visual_dsd} to describe the species present in our system.  We denote double-stranded molecules as \texttt{\string[r\string]}, where its upper strand \texttt{<r>} is connected to a complementary lower strand \texttt{\string{r*\string}}.  Each of the reactants and products in our system is an upper single-stranded molecule composed of a short toehold domain (annotated with a prefix \texttt{t} and an identifier \texttt{\^}) and a corresponding long domain: \texttt{<tr\textasciicircum\  r>}.
We  will refer to a short domain of a two-domain DSD strand $A_n$ as \texttt{ta} and its corresponding long domain as \texttt{an}, where $n$ is a channel index.  Note that the toehold is not specific to the species index $n$, and therefore  the recognition of each input and weight strand is dependent on their long domains, rather than their toeholds.  We will use the same convention for all other channel specific two-domain species.  For the detailed description of the nucleotide structure and binding rates; see Table \ref{tab:strand_table_dsd_new} and \ref{nucleotide_table_dsd_new} in the SI.
The four main two-domain strands that enable communication between different modules of the \neuron{} are shown in \tabref{species_table_dsd}.  
\par
While there is a theoretical guarantee that any chemical reaction network can be mimicked by a DSD circuit \citep{chen_dalchau2013}, it is often difficult to find circuits. However, there are now a number of general design motifs with known behaviours in the literature. Here, we will make extensive use of the two-domain scheme, which introduces a {\em Join-Fork} motif to mimic a chemical reaction.
While the abstract chemical system remains broadly similar to the CN model, there are some crucial differences, see \tabref{reaction_table_dsd}.
The general strategy we take to convert the CN to DSD is to translate each of the catalytic reactions in \tabref{reaction_table_dsd} into a Join-Fork gate \citep{Cardelli, chen_dalchau2013}. Subsequently, we will simulate the gates acting in concert. 

\begin{table}[ht]
 \centering
  \begin{tabular}{ll} 
   \hline
  Function & Reaction \\
   \hline
  Signal integration &  \ce{Fsi_{n} + A_{n} <=>  A_{n} + B}\\  
  Weight accumulation &  \ce{ A_{n} + E <=> E + H_{n} }\\
  Signal modulation &  \ce{ A_{n} + H_{n} <=> H_{n} + B } \\ 
   \hline
    & \ce{ B + E_{0} <=> E_{1}} \\ 
   Activation function  & ... \\ 
    & \ce{ B + $E_{m-1}$ <=> $\mathcal E$} \\ 
   \hline
  \end{tabular}
  \caption{List of reactions that constitute the d-CN.}
  \label{reaction_table_dsd}
\end{table}

We first explain how we use the Join-Fork gates.   For each reaction, a \emph{Join} gate is  able to  bind the \emph{reactants} and produces a translator strand.  Then, the translator activates a \emph{Fork} gate, which in turn releases the reaction \emph{products}. Additional energy must be supplied to completely release all products from Fork gates, as the translator strand will only displace the first product.  Appropriately designed \emph{helper} strands are therefore placed in the  solution to release subsequent products.  After the first product has unbound, an exposed toehold is left, which can  lead to unwanted side-effects.  To address this, we follow \citep{chen_dalchau2013} and extend the original design from \citep{Cardelli} by incorporating an additional long domain on the left-hand side of the Fork gate, which upon binding an appropriate auxiliary molecule, \emph{seals} the gate to prevent rebinding of its outputs. Here, we extend all Join gates in an equivalent way to prevent rebinding of the translator strand. This addition allows us to avoid interactions of the double-stranded complexes with waste molecules. 

\begin{figure}[ht]
  \subfloat[model_SI][Signal integration: \ce{Fsi_{n} + A_n -> A_n + B}.]{
      \includegraphics[width=0.48\textwidth]{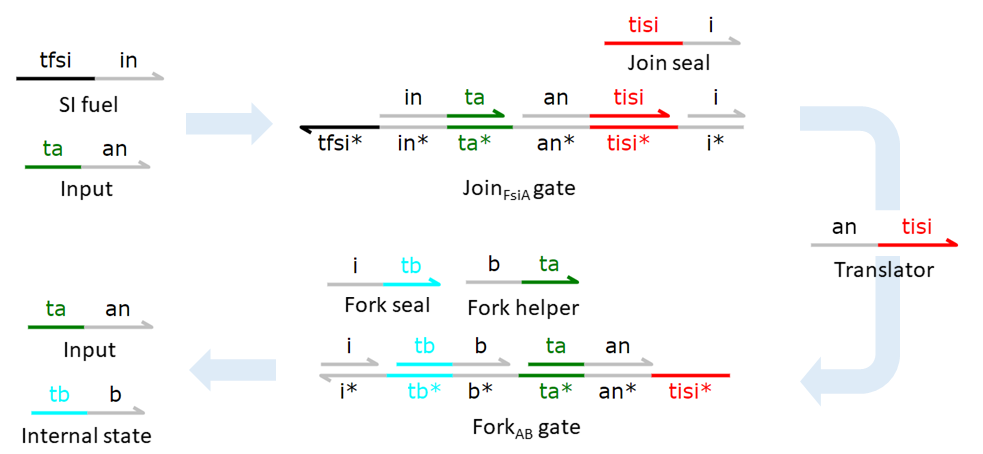}
         \label{model_SI}
  }
  \subfloat[model_SM][Signal modulation: \ce{ A_{n} + H_{n} -> H_{n} + B}. ]{
      \includegraphics[width=0.48\textwidth]{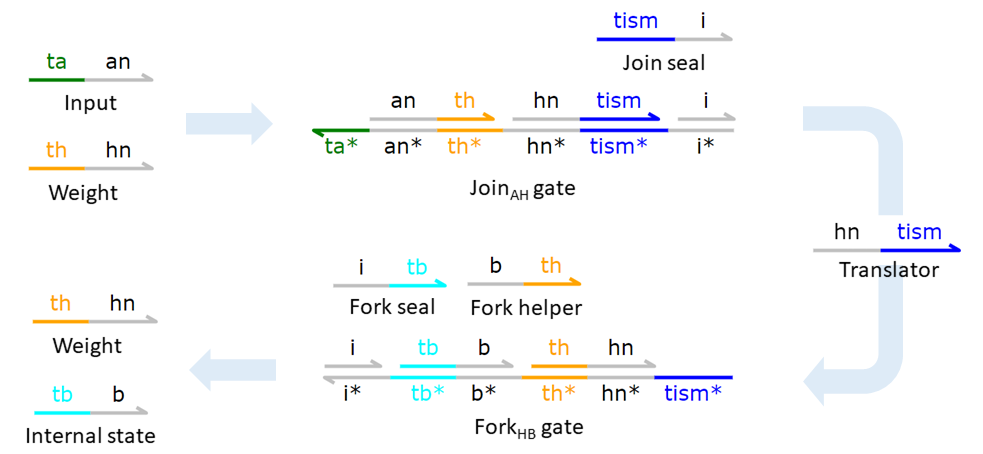}
             \label{model_SM}
  }
 \caption{Mapping the CRN neuron to a DNA neuron. We use a two domain Join-Fork gate to emulate each of the  catalytic reactions in the CN (\tabref{reaction_table_dsd}). In each case, a Join gate binds the two reactants in sequence, first displacing a waste molecule, and secondly displacing a translator molecule, which triggers the  corresponding Fork gate to release strands representing the reaction \emph{products}. The translator displaces the first product, and then a Fork helper displaces the second product. Both Join and Fork gates can be \emph{sealed} upon binding of an appropriate auxiliary strand (labelled Join seal and Fork seal), which displaces the final incumbent bound \texttt{\string<i\string>} strand.}
\label{dsd_model}
\end{figure}

In our design, binding of the translator immediately releases an $A_n$ (\texttt{\string<ta\string^ an\string>}) strand, the first of the reaction products. 
The second product, $B$ (\texttt{\string<tb\string^ b\string>}), is released upon binding of a \emph{Fork helper} strand \texttt{\string<b ta\string^\string>}.
Finally, the Fork$_\text{AB}$ gate is sealed upon binding of the \emph{Fork seal} strand \texttt{\string<i tb\string^\string>}.
The pair of Join and Fork gates together consume 1 molecule for each of the reactants, and produce 1 molecule for  each of the products, ensuring equivalent stoichiometry to the abstract reaction.

In order to illustrate the mapping from the CN to DSD, we describe now in detail the reaction \ce{Fsi_n + A_n  -> A_n + B} (\figref{model_SI}), which serves as a representative of all 3 catalytic reactions in the \neuron{}. A Join$_\text{AFsi}$ gate is defined by a structure that enables the binding of $Fsi_n$ and $A_n$; the gate is  only active  if both input species  are present.  First, $Fsi_n$ binds and displaces the incumbent bound \texttt{\string<in ta\string^\string>} molecule, exposing the \texttt{ta\string^} toehold.  This enables the binding of $A_n$ (\texttt{\string<ta\string^ an\string>}), which then displaces the \texttt{\string<an tisi\string^\string>} translator strand, signalling that the reactants have been received and that the overall reaction can fire.  The Join$_\text{FsiA}$ gate is then sealed by the binding of \texttt{\string<tisi\string^ i\string>}, preventing rebinding of the translator, and producing a further waste molecule \texttt{\string<i\string>}.  The  Fork$_\text{AB}$ gate is designed in such a way that upon triggering by the translator strand of the corresponding Join gate, it is able to release both product molecules. 

\begin{table}[ht]
\centering
\begin{tabular}{lcc} 
 \hline
Name & Signal & DSD Species \\
 \hline
Input & $A_{n}$ & \texttt{<ta\textasciicircum\ an>} \\ 
Weights & $H_{n}$ & \texttt{<th\textasciicircum\ hn>}\\
Internal state & $B$ &  \texttt{<tb\textasciicircum\ b>} \\ 
Learning signal & $\mathcal E$ & \texttt{<b tem\textasciicircum\ b>}  \\ 
Signal integration fuel & $Fsi_{n}$ & \texttt{<tfsi\textasciicircum\ fsin>}  \\ 
 \hline
\end{tabular}
\caption{List of key DNA strands which facilitate learning.}
\label{species_table_dsd}
\end{table}

\subsubsection{Controlling the activation function non-linearity with extended polymers}

The only reaction which takes a different form than a combination of Join and Fork gates is the activation function. We first describe the simplest case of an activation function with minimal non-linearity, i.e. $m=1$. In this case it takes the form:  \texttt{\string{tb\string^\string*\string}\string[b te0\string^\string]\string:\string[b te1\string^\string]\string<b\string>}, or graphically: \includegraphics[height=\baselineskip]{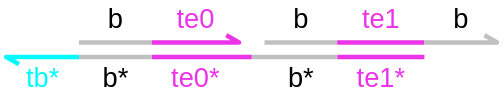}.
$B$ molecules can bind to this compound; in doing so they expose the   \texttt{te0} short domain which allows for binding of $E_{0}$.
When $E_{0}$ binds to the complex, it displaces a long domain \texttt{b} and releases the learning signal $\mathcal E$, which in the case of $m=1$ is represented by three-domain species: \texttt{\string<b te1\string^ b\string>}.
\par
\begin{figure}[ht]
  \centering \includegraphics[height=0.5\textheight]{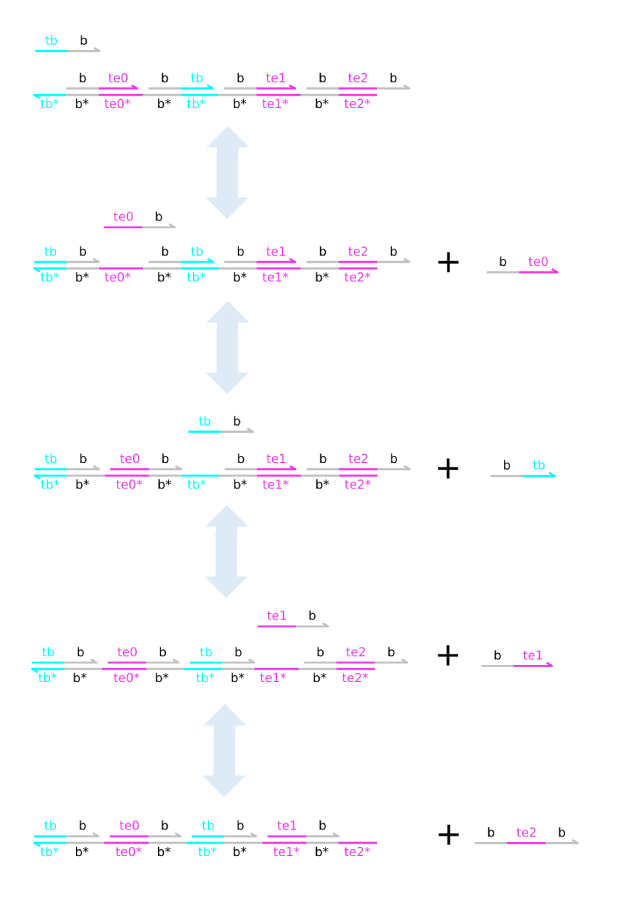}
\caption{The activation function for $m=2$ is modelled  as a long polymer which accommodates for binding of $B$ and subsequent $E$ molecules to its surface. These two species can bind to the polymer in an alternate manner. First the binding of $B$ frees up a \texttt{te0} toehold, next the binding of $E_{1}$ frees up a \texttt{tb} toehold etc. Altogether, this process consumes $B$  molecules.  At the end of the process a three-domain learning signal molecule $\mathcal E$ is produced.  In the case of $m=2$ this molecule takes the  following form: \texttt{\string<b te2\string^ b\string>.} This mechanism can also run backwards to produce $B$ molecules.  }
\label{model_AF}
\end{figure}

This system can now be generalised to arbitrary integer values of $m$, by extending the polymer with additional segments to accommodate for binding of more $B$ and $E_{m}$ molecules (\figref{model_AF}). We use segments of the form  \texttt{\string[b tb\string^\string]\string:\string[b tek\string^\string]}, where $k$ is the index of the $k$-th extra  segment in the complex. 
Each new segments should be added before the last fragment which contains $\mathcal E$: \texttt{\string[b te1\string^\string]\string<b\string>}.
In the case of  $m=2$ the activation function then is \texttt{\string{tb\string^\string*\string}\string[b te0\string^\string]\string:\string[b tb\string^\string]\string:\string[b te1\string^\string]\string:\string[b te2\string^\string]\string<b\string>} or graphically: \includegraphics[height=\baselineskip]{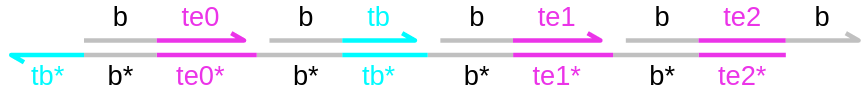}. 

\begin{figure}[ht]
\centering
    \includegraphics[width=0.48\textwidth]{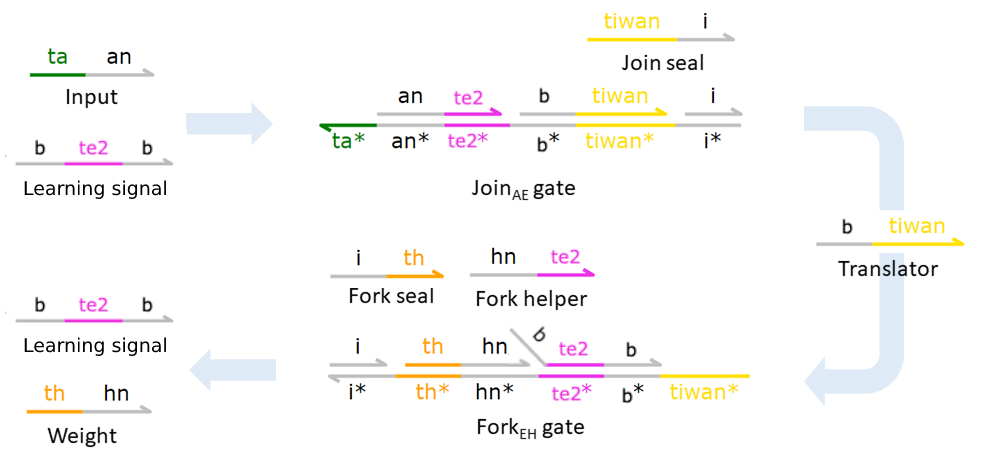}
\caption{Weight accumulation ($m=2$): \ce{ A_{n} + $\mathcal E$ -> $\mathcal E$ + H_{n}}.  The catalytic reaction  that realises the weight accumulation function   is a Join gate and a modified variant of the Fork gate.  In this variation, the first reactant of the Join gate and the first product of the Fork gate is a three-domain species $\mathcal E$, which represents the learning signal.  The initial Fork gate complex now has a long domain \texttt{b} branching out of the double-stranded structure. This modification is needed to ensure complementarity with the tunable activation function.  }
\label{model_WA}
\end{figure}

The weight accumulation function is distinguished from standard gates in that the first reactant of the Join gate, i.e.   \texttt{\string<b te2  b\string>} representing the learning signal $\mathcal E$ and the first product of the Fork gate are both three domain species.  The initial form of the Fork gate complex has a long domain \texttt{b} branching out of the double-stranded structure (\figref{model_WA}).  This modification is necessary in order to allow for $\mathcal E$ to catalyse the reaction.

An alternative way to implement this mechanism could be the use of a multi-step cascade of gates. 
This approach, however, would necessitate the use of additional toehold definitions, thus limiting the number of input channels that could be simulated. 

\subsubsection{Computational Complexity}

Extending the \neuron{} to accommodate additional input channels requires the user to define a single new toehold domain definition \texttt{tiwan}, which is responsible for weight accumulation in each of the  $N$ channels. Moreover, there are six toehold domains that remain the same regardless of the number of input channels (\texttt{ta}, \texttt{th}, \texttt{tb}, \texttt{tfsi},  \texttt{tism}, \texttt{tisi}).  Therefore, the system with $N=3$ input channels requires 9 toehold definitions ($6+N$).
Additionally, depending on the length of the polymer which facilitates the activation function there are at least two additional toehold domains: \texttt{te0} and \texttt{te1}.  We base the recognition of the inputs, as well as other two-domain strands in the system, on the long domains.  There are two long domains which remain the same regardless of the number of channels (\texttt{b}, \texttt{i}), and three which need to be defined when adding another input channel (\texttt{an}, \texttt{hn}, \texttt{fsin}).  Therefore, the system with $N=3$ input channels requires 11 long domain definitions ($2+3N$).

\subsubsection{Simulating the \neuron{}}

\begin{figure}[H]
\centering
\subfloat[fbex_l][Example of frequency bias learning ($m$=1).]{\includegraphics[width=0.495\textwidth]{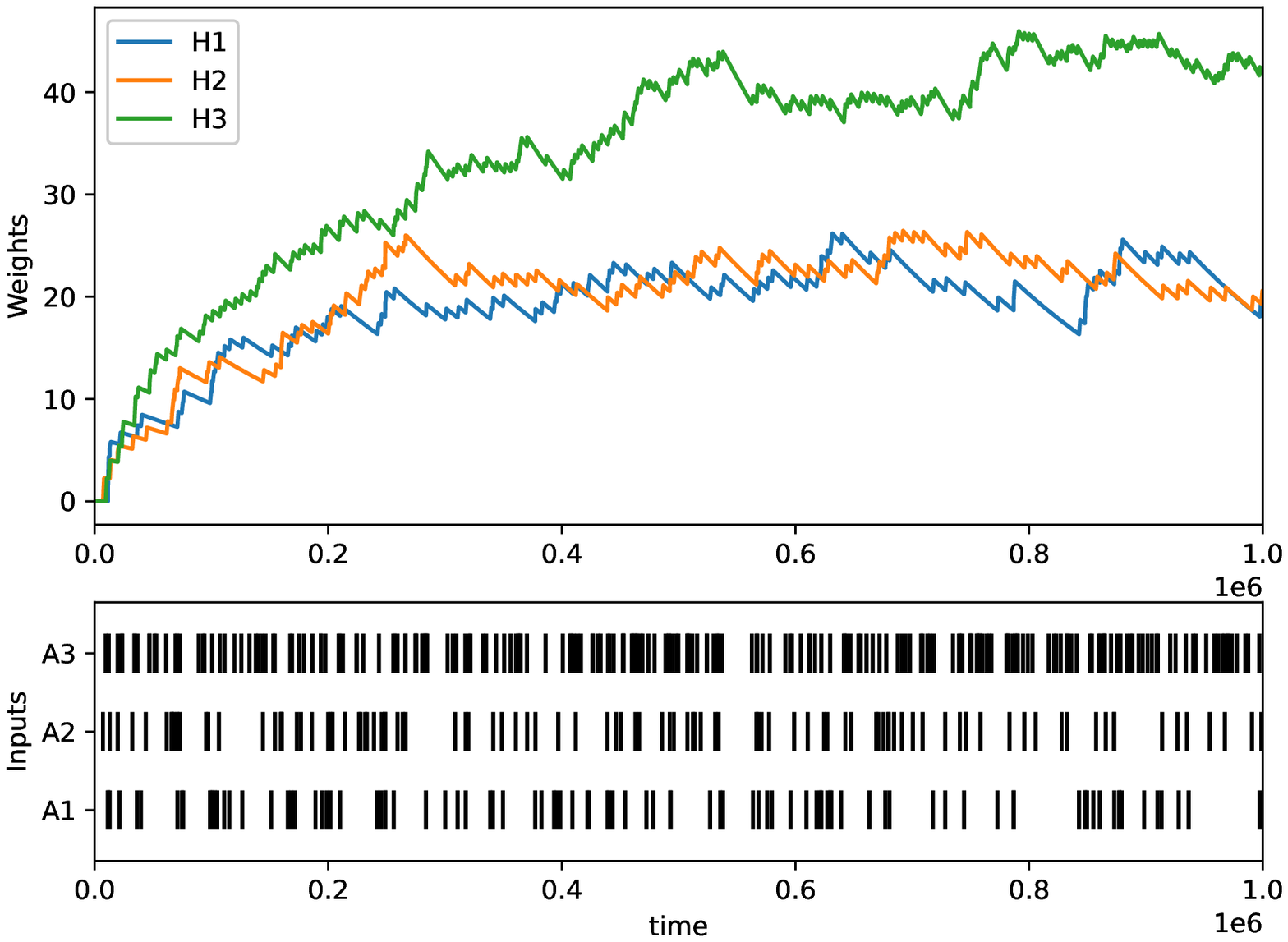}\label{fbex_l}}
\subfloat[tcex_l][Example of temporal correlation learning ($m$=3).]{\includegraphics[width=0.495\textwidth]{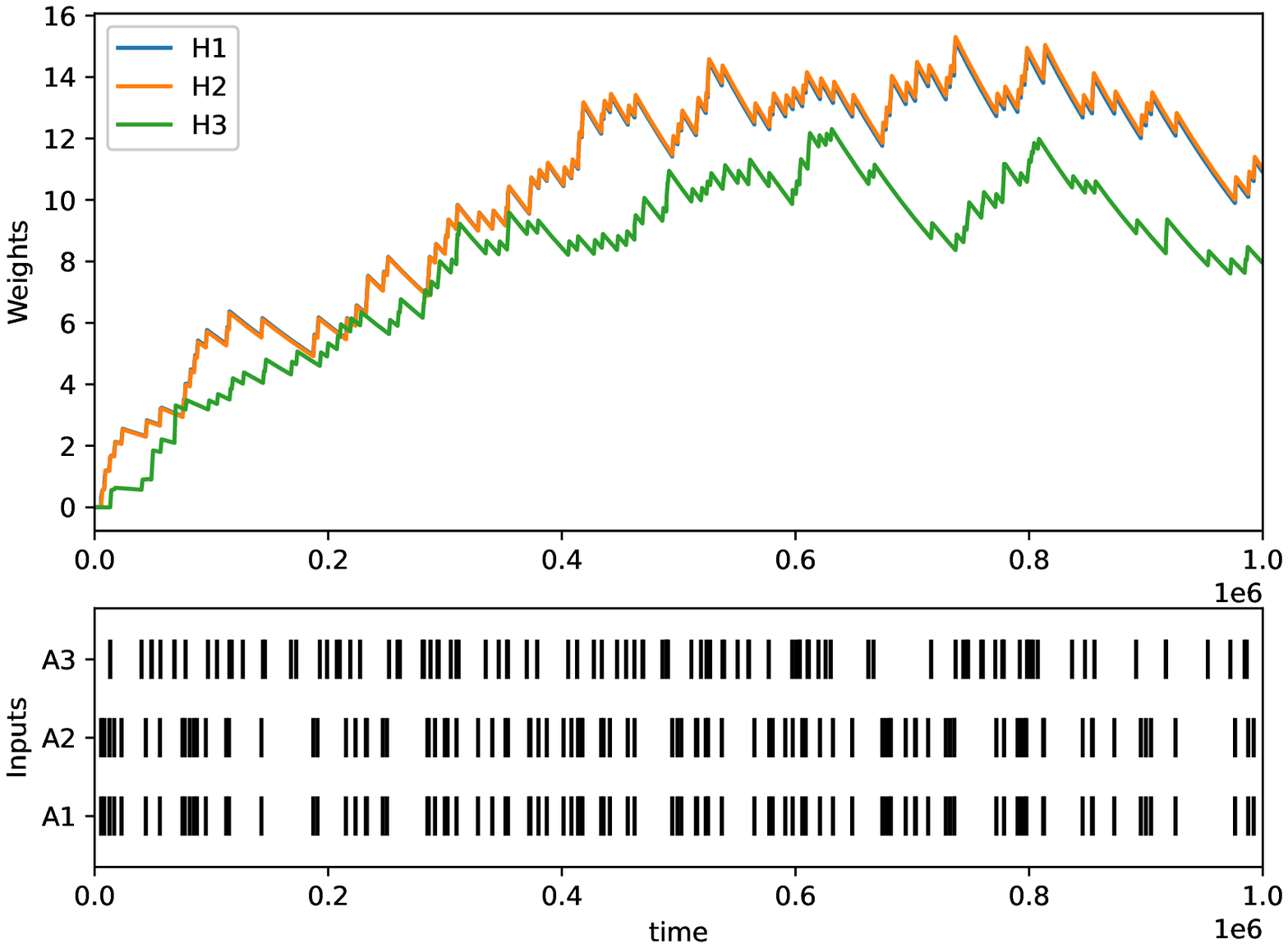}\label{tcex_l}}
\caption{
Examples of learning episodes in d-CN for
\protect\subref{fbex_l} frequency bias task,
 and \protect\subref{tcex_l} temporal correlation task.
 For statistical data about the weight distributions obtained over multiple runs see fig. \ref{TC_FB_results}.
}
\label{TC_FB_results_MAIN} 
\end{figure}

When simulating the \neuron{}, we initialise the system  with different amounts of gate complexes and helper strands needed for the computation by both Join and Fork gates depending on their function.  Signal modulation fuel molecules are initiated at 25000 $\mu$M, signal integration at 50000 $\mu$M, and weight accumulation at 10000 $\mu$M.  We also initialise the fuel molecules necessary for the signal integration mechanism $Fsi_{n}$ with 50000 $\mu$M. 
Lastly, in all of the experiments we choose to set the bolus size, i.e. the amount of $A_{n}$ species injected to the system at each spike, to $\beta=10$ $\mu$M. 
In order to model  decay  of $H_{n}$ species we introduce garbage collection molecules
\texttt{\string{th\string^\string*\string}\string[hn\string]}, which sequester and inactivate the molecular species $H_{n}$.  We inject 12 $\mu$M and 0.1 $\mu$M of these species to the system periodically every 1000 s.
\par
We have been careful to use strand displacement reaction rates that are within the range that has been measured experimentally  \citep{toehold_ex_zhang}.  In order to reproduce the desired dynamical behaviours, the binding rates associated with the \texttt{ta}, \texttt{th} toeholds have been set to lower values than the other toeholds; see \tabref{tab:strand_table_dsd_new} for details on the parameters.
\par
To determine whether the  \neuron{}  is capable of \emph{learning}, we carried out a range of simulations using Visual DSD.  We found that both \emph{infinite} and \emph{detailed} mode compilation could produce the intended dynamical behaviours.  Similarly, we found that these behaviours could be produced in both simulations at low copy numbers (using Gillespie's stochastic simulation algorithm) and in the fluid limit (deterministic rate equations).  Accordingly, we show infinite mode deterministic simulations in the main article, and other simulations in the SI (\figref{TC_FB_results_other}).

\begin{figure}[h]
  \centering
  \subfloat[mix_bias][FB + TC]{\includegraphics[width=0.495\textwidth]{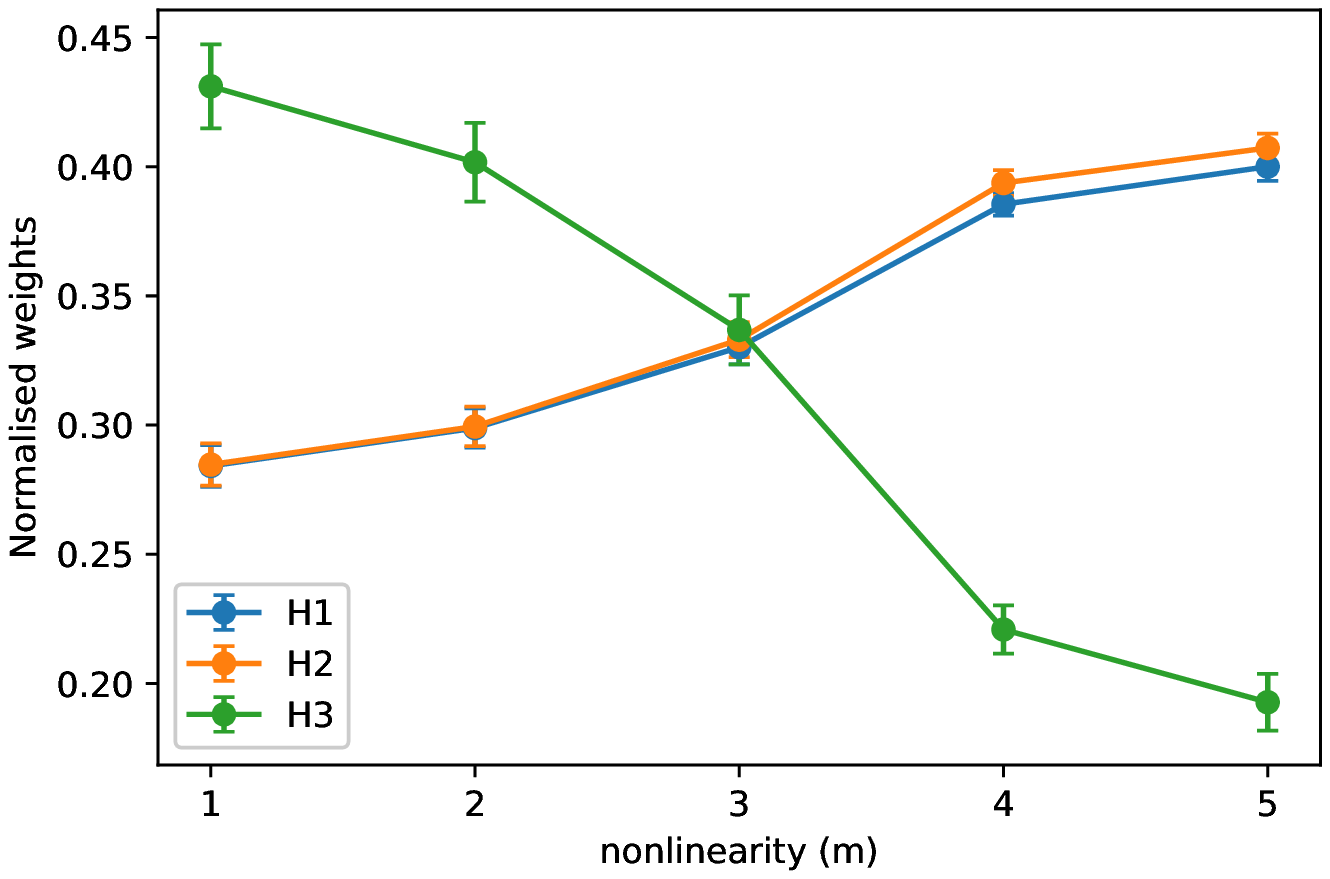}\label{mix_bias}}
  \subfloat[tc_bias][TC]{\includegraphics[width=0.495\textwidth]{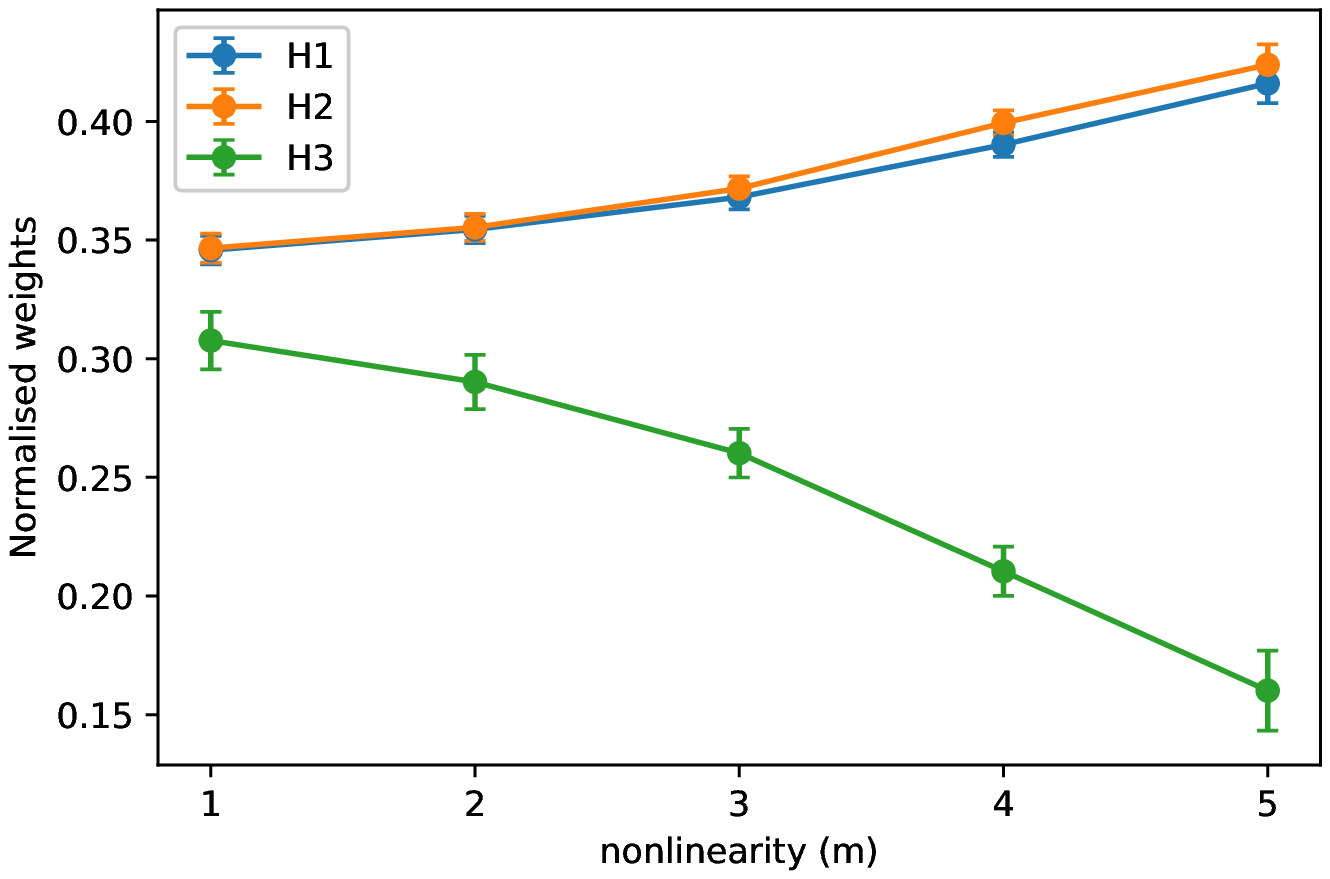}\label{tc_bias}}
  \caption{ Normalised steady state weights as a function of the length of the activation function polymer. As the polymer is extended the activation function becomes steeper and therefore requires a correlation of at least two signals to trigger learning. Therefore, the DNA neuron becomes better at recognising temporal correlations when $m$ is high. 
  }
  \label{bias_dCN} 
\end{figure}

To check whether the \neuron{} behaves as expected, we test its ability to distinguish the two types of biases on tasks where $A_2$ is temporally correlated with $A_1$, and further analyse how this depends on non-linearity/polymer length (\figref{bias_dCN}). First, we consider a scenario where $A_3$ is both uncorrelated with $A_1$/$A_2$ and additionally has a spiking frequency twice as high as the other input channels (0.0002 Hz; \figref{mix_bias}). Consistent with the CN, the \neuron{} is sensitive to frequency bias when the non-linearity is low, corresponding to the weights of $A_3$ are high for $m=1$. Vice versa, in the case of high non-linearity, the \neuron{} recognises the temporal correlations, corresponding to the weights of $A_1$ and $A_2$ being high.  
When removing the frequency bias of $A_3$, the system still differentiates between uncorrelated and correlated inputs, but the ability to distinguish the two types of signal increases with $m$ (see \figref{tc_bias}).
  
\begin{figure}[h]
  \centering
  \includegraphics[width=0.8\textwidth]{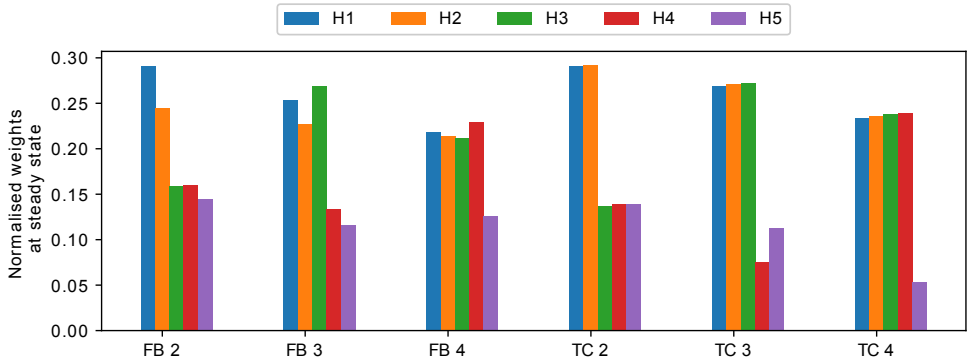}\label{tasks_fig_DSD}
  \caption{
Same as fig. \ref{bars_CN}, but for the \neuron{}. Data was only collected after the weights reached the steady state (after 800000 time units).
  }
  \label{bars_dCN}
\end{figure}

We also compared the ability of the \neuron{} directly with the CN. We found that the \neuron{} is able to detect both FB and TC biases (\figref{bars_dCN}). However, in the TC task the indication of the temporal order of the input signals is subtle, in the sense that the steady state weights of the correlated channels are almost the same, with only a slight difference indicating temporal order. 

\section{Discussion}

To the best of our knowledge, the  CN is the first fully autonomous  chemical model  of a Hebbian spiking neuron. While it is unlikely that the basic model can be engineered as is, it has some features that make the system interesting from a fundamental point of view. 

One of the attractive features of the (basic) CN neuron is that it is micro-reversible  and  therefore  thermodynamically plausible.  This makes it a useful theoretical tool to probe the thermodynamics of learning. While a thorough analysis of the energy requirements of the system is beyond the scope of this article, we note that the physical plausibility of the model has highlighted resource requirements of computation. In particular, we found that increasing the non-linearity comes at an additional cost in resource. The CN suffers from starvation of $B$ molecules as $m$ increases.  For a sufficiently high number of $m$, this leads to a breakdown of the mechanisms and the system loses its ability to detect coincidences, as illustrated  in \figref{iod_volume}.   This ``starvation'' effect can be alleviated by increasing the bolus size (while keeping the threshold fixed; \figref{iod_volume}).  In a biological context, the increase of the bolus size comes at a direct synthesis cost, if the molecules that make up the bolus need to be made by the cell. Yet, even if we assume that the particles are, somehow, pre-existing,  injecting a bolus requires  chemical work, which  is proportional to the number of particles, i.e. the bolus size. Hence, there is a fundamental thermodynamic cost involved in computing the non-linearity. We are not aware of any formal proofs that show that computing non-linearities necessarily requires an increased energy requirement. It therefore remains an open question whether or not this is a feature of the particular model choices, or the manifestation of a deeper constraint.  
Interestingly, the FB task, which does not rely on non-linearities, can be solved with much simpler, and thermodynamically cheaper, designs, e.g.~ a simple decaying particle. 

While the basic CN does not lend itself to a direct implementation in biochemistry, we presented a compartmentalised interpretation of the system that is biologically more plausible. It interprets different input species, and indeed the internal state molecule $B$ as one and the same species, but contained in different compartments. This makes the system feasible, in principle. Although creating many compartments with the required dynamics may remain challenging, significant progress has been made in recent years towards programming molecular systems in protocells \citep{lyu2020protocells}.

Interestingly, there are structural similarities between the c-CN and the {\em lac} system in \ecoli\citep{myscireppaper}. The essence of the {\em lac} system is that it only switches on the lactose metabolism (the equivalent to the weight molecules in the compartment) when it is stimulated by lactose in the environment (i.e. $B$). The principle of operation of the {\em lac} system is similar to the c-CN, except that \ecoli does of course not export lactose to the environment.  Taking this analogy seriously, it would be interesting to consider whether catabolite repression, which is a moderately complex decision process, can be mapped to a simple neural network. 

Amongst the three versions of the chemical neuron that we presented, we found that all could reproduce the same qualitative behaviours (Figs. \ref{asso} and \ref{straightline}). However, given that all three of them are different designs, each version required its own parametrisation, which had to be found by manual exploration in each case. It is thus not possible to reproduce the behaviour of one model with another one exactly. Qualitatively, however, we found the same behaviours in all models. The only major difference was on the TC task. Unlike the other two versions of the chemical neuron, the  \neuron{} did not clearly highlight  the temporal order of input signals (\figref{bars_dCN}). While the \neuron{} indicates  a strong difference between the correlated and non-correlated species, the weight difference between the correlated channels which should indicate the temporal order is marginal. Whether this can be improved with a better parametrisation, or whether this points to a fundamental limitation of the model must remain an open question. 

From an engineering perspective, the \neuron{} is certainly the easiest to realise experimentally.  DNA circuits are much less prone to crosstalk than more standard biochemical reaction networks. Synthesising DNA molecules is now a routine procedure. There are, however, several elements of the \neuron{} design that will require careful consideration before an implementation can be done. For any practical use, one would need to interface the DNA computer with the {\em in vivo} target systems.  How to do this in a general way remains an open question, but there have been a number of previous systems that indicate possible pathways \citep{connect1,connect2,connect3,connect4}. 
\par
More specifically for the \neuron{}, there are a number of experimental challenges that need to be addressed.
In order to ensure that the kinetics of the \neuron{} are conserved throughout the  learning and testing phases, we require the activation species $B$ to decay.  To remove the $B$ species, we employ simple helper complexes, which are periodically replenished during the simulation.
These complexes are capable of making $B$ and $H_{n}$ species unreactive, thereby removing them from the system. 
In order to achieve better reproducibility of the results the experimental realisation of this would necessitate a relatively frequent or continuous supply of these DNA complexes.
While difficult to achieve experimentally, there are known techniques to overcome the need for frequent replenishment, including the use of \emph{buffered} gates \citep{DNA_circuits} or timer circuits \citep{timer_circuits}.
\par
Scaling the system to more input channels requires an additional short domain sequences per additional channel.  Prima facia the scaling up of the \neuron{} is therefore limited by the availability of orthogonal short domain sequences.
A redesign based on localised design principles could be a feasible solution if the number of toeholds becomes a problem.
Here, instead of using a different set of long and short domains, distinct channels could be implemented through physical separation of the species \citep{Chatterjee110965}.

Describing the model as a neuron encourages the question of building networks capable of complex computational tasks. 
A major impediment for building networks of \neuron{} could be the immediate injection of $A$ species to the neurons in the next layers of the network. 
This would necessitate an inclusion of a different activation function, or a mechanism which would allow for the signal propagation.
Incorporating a buffered gate design \citep{DNA_circuits} could allow for a programmed release of a certain number of input species, once the activation signal is produced.
Nevertheless, we leave the question of constructing functional neural networks in DNA for future research.

\subsubsection{Supporting Information}

Supporting information includes the following items: 
\begin{itemize}
\item detailed description of the reaction rate constants necessary to realise CN and c-CN,
\item lists of nucleotide sequences and binding rates for d-CN,
\item weight distributions for FB 2 and TC 2 tasks as a function of $m$,
\item statistical data about the weight distributions for d-CN,
\item examples of d-CN training in other simulation modes and relearning of input statistics, 
\item study of signal modulation mechanism in the d-CN,
\item examination of strategies for garbage collection and stability of the learnt solutions as a function of bolus size and abundance of gate molecules,
\item analysis of performance of the CN and d-CN models on the FB and TC task,
\item Visual DSD code for d-CN.
\end{itemize}

\subsection{Author Contributions}
\label{sec:contributions}
Conceived the research: DC, ND, JF. Conducted the research: JF. Wrote the paper: DC, ND, JF.

\subsubsection{Conflicts of Interest}
None

\subsubsection{Funding}
None

\bibliography{biblio} 

\providecommand{\latin}[1]{#1}
\makeatletter
\providecommand{\doi}
  {\begingroup\let\do\@makeother\dospecials
  \catcode`\{=1 \catcode`\}=2 \doi@aux}
\providecommand{\doi@aux}[1]{\endgroup\texttt{#1}}
\makeatother
\providecommand*\mcitethebibliography{\thebibliography}
\csname @ifundefined\endcsname{endmcitethebibliography}
  {\let\endmcitethebibliography\endthebibliography}{}
\begin{mcitethebibliography}{56}
\providecommand*\natexlab[1]{#1}
\providecommand*\mciteSetBstSublistMode[1]{}
\providecommand*\mciteSetBstMaxWidthForm[2]{}
\providecommand*\mciteBstWouldAddEndPuncttrue
  {\def\EndOfBibitem{\unskip.}}
\providecommand*\mciteBstWouldAddEndPunctfalse
  {\let\EndOfBibitem\relax}
\providecommand*\mciteSetBstMidEndSepPunct[3]{}
\providecommand*\mciteSetBstSublistLabelBeginEnd[3]{}
\providecommand*\EndOfBibitem{}
\mciteSetBstSublistMode{f}
\mciteSetBstMaxWidthForm{subitem}{(\alph{mcitesubitemcount})}
\mciteSetBstSublistLabelBeginEnd
  {\mcitemaxwidthsubitemform\space}
  {\relax}
  {\relax}

\bibitem[Govern and {ten Wolde}(2014)Govern, and {ten Wolde}]{government1}
Govern,~C.; {ten Wolde},~P. Energy dissipation and noise correlations in
  biochemical sensing. \emph{Physical Review Letters} \textbf{2014},
  \emph{113}, 258102\relax
\mciteBstWouldAddEndPuncttrue
\mciteSetBstMidEndSepPunct{\mcitedefaultmidpunct}
{\mcitedefaultendpunct}{\mcitedefaultseppunct}\relax
\EndOfBibitem
\bibitem[Govern and {ten Wolde}(2014)Govern, and {ten Wolde}]{government2}
Govern,~C.; {ten Wolde},~P. Optimal resource allocation in cellular sensing
  systems. \emph{PNAS} \textbf{2014}, \emph{111}, 17486--17491\relax
\mciteBstWouldAddEndPuncttrue
\mciteSetBstMidEndSepPunct{\mcitedefaultmidpunct}
{\mcitedefaultendpunct}{\mcitedefaultseppunct}\relax
\EndOfBibitem
\bibitem[Alon(2019)]{alon2019introduction}
Alon,~U. \emph{An Introduction to Systems Biology: Design Principles of
  Biological Circuits}; Chapman \& Hall/CRC Computational Biology Series; CRC
  Press LLC, 2019\relax
\mciteBstWouldAddEndPuncttrue
\mciteSetBstMidEndSepPunct{\mcitedefaultmidpunct}
{\mcitedefaultendpunct}{\mcitedefaultseppunct}\relax
\EndOfBibitem
\bibitem[Yi \latin{et~al.}(2000)Yi, Huang, Simon, and Doyle]{Yi_2000}
Yi,~T.; Huang,~Y.; Simon,~M.; Doyle,~J. Robust perfect adaptation in bacterial
  chemotaxis through integral feedback control. \emph{Proceedings of the
  National Academy of Sciences} \textbf{2000}, \emph{97}, 4649--4653\relax
\mciteBstWouldAddEndPuncttrue
\mciteSetBstMidEndSepPunct{\mcitedefaultmidpunct}
{\mcitedefaultendpunct}{\mcitedefaultseppunct}\relax
\EndOfBibitem
\bibitem[Hoffer \latin{et~al.}(2001)Hoffer, Westerhoff, Hellingwerf, Postma,
  and Tommassen]{Hoffer_2001}
Hoffer,~S.~M.; Westerhoff,~H.~V.; Hellingwerf,~K.~J.; Postma,~P.~W.;
  Tommassen,~J. Autoamplification of a two-component regulatory system results
  in "learning" behavior. \emph{Journal of bacteriology} \textbf{2001},
  \emph{183}, 4914—4917\relax
\mciteBstWouldAddEndPuncttrue
\mciteSetBstMidEndSepPunct{\mcitedefaultmidpunct}
{\mcitedefaultendpunct}{\mcitedefaultseppunct}\relax
\EndOfBibitem
\bibitem[Chu(2018)]{my1}
Chu,~D. Performance limits and trade-offs in entropy-driven biochemical
  computers. \emph{Journal of Theoretical Biology} \textbf{2018}, \emph{443},
  1--9\relax
\mciteBstWouldAddEndPuncttrue
\mciteSetBstMidEndSepPunct{\mcitedefaultmidpunct}
{\mcitedefaultendpunct}{\mcitedefaultseppunct}\relax
\EndOfBibitem
\bibitem[Chu(2017)]{my2}
Chu,~D. Limited by sensing - A minimal stochastic model of the lag-phase during
  diauxic growth. \emph{Journal of Theoretical Biology} \textbf{2017},
  \emph{414}, 137--146\relax
\mciteBstWouldAddEndPuncttrue
\mciteSetBstMidEndSepPunct{\mcitedefaultmidpunct}
{\mcitedefaultendpunct}{\mcitedefaultseppunct}\relax
\EndOfBibitem
\bibitem[Chu and Barnes(2016)Chu, and Barnes]{my3}
Chu,~D.; Barnes,~D. The lag-phase during diauxic growth is a trade-off between
  fast adaptation and high growth rate. \emph{Scientific Reports}
  \textbf{2016}, \emph{6}, 25191\relax
\mciteBstWouldAddEndPuncttrue
\mciteSetBstMidEndSepPunct{\mcitedefaultmidpunct}
{\mcitedefaultendpunct}{\mcitedefaultseppunct}\relax
\EndOfBibitem
\bibitem[Aggarwal(2018)]{aggravate}
Aggarwal,~C. \emph{Neural networks and deep learning : a textbook}; Springer:
  Cham, Switzerland, 2018\relax
\mciteBstWouldAddEndPuncttrue
\mciteSetBstMidEndSepPunct{\mcitedefaultmidpunct}
{\mcitedefaultendpunct}{\mcitedefaultseppunct}\relax
\EndOfBibitem
\bibitem[Sengupta \latin{et~al.}(2019)Sengupta, Ye, Wang, Liu, and
  Roy]{deep_snn}
Sengupta,~A.; Ye,~Y.; Wang,~R.; Liu,~C.; Roy,~K. Going Deeper in Spiking Neural
  Networks: VGG and Residual Architectures. \emph{Frontiers in Neuroscience}
  \textbf{2019}, \emph{13}, 95\relax
\mciteBstWouldAddEndPuncttrue
\mciteSetBstMidEndSepPunct{\mcitedefaultmidpunct}
{\mcitedefaultendpunct}{\mcitedefaultseppunct}\relax
\EndOfBibitem
\bibitem[Afshar \latin{et~al.}(2020)Afshar, Ralph, Xu, Tapson, Schaik, and
  Cohen]{afshar_snn}
Afshar,~S.; Ralph,~N.; Xu,~Y.; Tapson,~J.; Schaik,~A.~v.; Cohen,~G. Event-Based
  Feature Extraction Using Adaptive Selection Thresholds. \emph{Sensors}
  \textbf{2020}, \emph{20}\relax
\mciteBstWouldAddEndPuncttrue
\mciteSetBstMidEndSepPunct{\mcitedefaultmidpunct}
{\mcitedefaultendpunct}{\mcitedefaultseppunct}\relax
\EndOfBibitem
\bibitem[Fil and Chu(2020)Fil, and Chu]{Fil_2020}
Fil,~J.; Chu,~D. Minimal Spiking Neuron for Solving Multilabel Classification
  Tasks. \emph{Neural Computation} \textbf{2020}, \emph{32}, 1408--1429\relax
\mciteBstWouldAddEndPuncttrue
\mciteSetBstMidEndSepPunct{\mcitedefaultmidpunct}
{\mcitedefaultendpunct}{\mcitedefaultseppunct}\relax
\EndOfBibitem
\bibitem[G{\"{u}}tig(2016)]{Gutig2016}
G{\"{u}}tig,~R. {Spiking neurons can discover predictive features by
  aggregate-label learning}. \emph{Science} \textbf{2016}, \emph{351},
  aab4113--aab4113\relax
\mciteBstWouldAddEndPuncttrue
\mciteSetBstMidEndSepPunct{\mcitedefaultmidpunct}
{\mcitedefaultendpunct}{\mcitedefaultseppunct}\relax
\EndOfBibitem
\bibitem[Oja(1982)]{oja_1982}
Oja,~E. Simplified neuron model as a principal component analyzer.
  \emph{Journal of Mathematical Biology} \textbf{1982}, \emph{15},
  267--273\relax
\mciteBstWouldAddEndPuncttrue
\mciteSetBstMidEndSepPunct{\mcitedefaultmidpunct}
{\mcitedefaultendpunct}{\mcitedefaultseppunct}\relax
\EndOfBibitem
\bibitem[Diehl and Cook(2015)Diehl, and Cook]{diehl_digits}
Diehl,~P.; Cook,~M. Unsupervised learning of digit recognition using
  spike-timing-dependent plasticity. \emph{Frontiers in Computational
  Neuroscience} \textbf{2015}, \emph{9}, 99\relax
\mciteBstWouldAddEndPuncttrue
\mciteSetBstMidEndSepPunct{\mcitedefaultmidpunct}
{\mcitedefaultendpunct}{\mcitedefaultseppunct}\relax
\EndOfBibitem
\bibitem[Afshar \latin{et~al.}(2019)Afshar, Hamilton, Tapson, van Schaik, and
  Cohen]{van_schaik_planes}
Afshar,~S.; Hamilton,~R.~J.; Tapson,~J.; van Schaik,~A.; Cohen,~G.
  Investigation of Event-Based Surfaces for High-Speed Detection, Unsupervised
  Feature Extraction, and Object Recognition. \emph{Frontiers in Neuroscience}
  \textbf{2019}, \emph{12}, 1047\relax
\mciteBstWouldAddEndPuncttrue
\mciteSetBstMidEndSepPunct{\mcitedefaultmidpunct}
{\mcitedefaultendpunct}{\mcitedefaultseppunct}\relax
\EndOfBibitem
\bibitem[Okamoto \latin{et~al.}(1988)Okamoto, Sakai, and Hayashi]{Okamoto_1988}
Okamoto,~M.; Sakai,~T.; Hayashi,~K. Biochemical Switching Device Realizing
  McCulloch-Pitts Type Equation. \emph{Biol. Cybern.} \textbf{1988}, \emph{58},
  296–299\relax
\mciteBstWouldAddEndPuncttrue
\mciteSetBstMidEndSepPunct{\mcitedefaultmidpunct}
{\mcitedefaultendpunct}{\mcitedefaultseppunct}\relax
\EndOfBibitem
\bibitem[Hjelmfelt \latin{et~al.}(1991)Hjelmfelt, Weinberger, and
  Ross]{Hjelmfelt_1991}
Hjelmfelt,~A.; Weinberger,~E.~D.; Ross,~J. Chemical implementation of neural
  networks and Turing machines. \emph{Proceedings of the National Academy of
  Sciences} \textbf{1991}, \emph{88}, 10983--10987\relax
\mciteBstWouldAddEndPuncttrue
\mciteSetBstMidEndSepPunct{\mcitedefaultmidpunct}
{\mcitedefaultendpunct}{\mcitedefaultseppunct}\relax
\EndOfBibitem
\bibitem[Banda \latin{et~al.}(2014)Banda, Teuscher, and
  Stefanovic]{chemperceptron}
Banda,~P.; Teuscher,~C.; Stefanovic,~D. Training an asymmetric signal
  perceptron through reinforcement in an artificial chemistry. \emph{Journal of
  The Royal Society Interface} \textbf{2014}, \emph{11}, 20131100\relax
\mciteBstWouldAddEndPuncttrue
\mciteSetBstMidEndSepPunct{\mcitedefaultmidpunct}
{\mcitedefaultendpunct}{\mcitedefaultseppunct}\relax
\EndOfBibitem
\bibitem[Blount \latin{et~al.}(2017)Blount, Banda, Teuscher, and
  Stefanovic]{chemperceptron2}
Blount,~D.; Banda,~P.; Teuscher,~C.; Stefanovic,~D. Feedforward Chemical Neural
  Network: An In Silico Chemical System That Learns xor. \emph{Artificial Life}
  \textbf{2017}, \emph{23}, 295--317\relax
\mciteBstWouldAddEndPuncttrue
\mciteSetBstMidEndSepPunct{\mcitedefaultmidpunct}
{\mcitedefaultendpunct}{\mcitedefaultseppunct}\relax
\EndOfBibitem
\bibitem[Shirakawa and Sato(2013)Shirakawa, and Sato]{assmodel}
Shirakawa,~T.; Sato,~H. Construction of a Molecular Learning Network.
  \emph{Journal of Advanced Computational Intelligence and Intelligent
  Informatics} \textbf{2013}, \emph{17}, 913--918\relax
\mciteBstWouldAddEndPuncttrue
\mciteSetBstMidEndSepPunct{\mcitedefaultmidpunct}
{\mcitedefaultendpunct}{\mcitedefaultseppunct}\relax
\EndOfBibitem
\bibitem[Nesbeth \latin{et~al.}(2016)Nesbeth, Zaikin, Saka, Romano, Giuraniuc,
  Kanakov, and Laptyeva]{assmodel2}
Nesbeth,~D.; Zaikin,~A.; Saka,~Y.; Romano,~M.; Giuraniuc,~C.; Kanakov,~O.;
  Laptyeva,~T. Synthetic biology routes to bio-artificial intelligence.
  \emph{Essays in Biochemistry} \textbf{2016}, \emph{60}, 381--391\relax
\mciteBstWouldAddEndPuncttrue
\mciteSetBstMidEndSepPunct{\mcitedefaultmidpunct}
{\mcitedefaultendpunct}{\mcitedefaultseppunct}\relax
\EndOfBibitem
\bibitem[Chen and Xu(2015)Chen, and Xu]{biochemimp}
Chen,~M.; Xu,~J. Construction of a genetic conditional learning system in
  Escherichia coli. \emph{Science China Information Sciences} \textbf{2015},
  \emph{58}, 1--6\relax
\mciteBstWouldAddEndPuncttrue
\mciteSetBstMidEndSepPunct{\mcitedefaultmidpunct}
{\mcitedefaultendpunct}{\mcitedefaultseppunct}\relax
\EndOfBibitem
\bibitem[Racovita and Jaramillo(2020)Racovita, and Jaramillo]{forcement}
Racovita,~A.; Jaramillo,~A. Reinforcement learning in synthetic gene circuits.
  \emph{Biochemical Society Transactions} \textbf{2020}, \emph{48},
  1637--1643\relax
\mciteBstWouldAddEndPuncttrue
\mciteSetBstMidEndSepPunct{\mcitedefaultmidpunct}
{\mcitedefaultendpunct}{\mcitedefaultseppunct}\relax
\EndOfBibitem
\bibitem[Fernando \latin{et~al.}(2009)Fernando, Liekens, Bingle, Beck, Lenser,
  Stekel, and Rowe]{Fernando2009}
Fernando,~C.; Liekens,~A.; Bingle,~L.; Beck,~C.; Lenser,~T.; Stekel,~D.;
  Rowe,~J. {Molecular circuits for associative learning in single-celled
  organisms}. \emph{Journal of The Royal Society Interface} \textbf{2009},
  \emph{6}, 463--469\relax
\mciteBstWouldAddEndPuncttrue
\mciteSetBstMidEndSepPunct{\mcitedefaultmidpunct}
{\mcitedefaultendpunct}{\mcitedefaultseppunct}\relax
\EndOfBibitem
\bibitem[McGregor \latin{et~al.}(2012)McGregor, Vasas, Husbands, and
  Fernando]{mcgregor_cn}
McGregor,~S.; Vasas,~V.; Husbands,~P.; Fernando,~C. Evolution of Associative
  Learning in Chemical Networks. \emph{PLOS Computational Biology}
  \textbf{2012}, \emph{8}, 1--19\relax
\mciteBstWouldAddEndPuncttrue
\mciteSetBstMidEndSepPunct{\mcitedefaultmidpunct}
{\mcitedefaultendpunct}{\mcitedefaultseppunct}\relax
\EndOfBibitem
\bibitem[Macia \latin{et~al.}(2017)Macia, Vidiella, and
  Sol{\'{e}}]{solebiochemimp}
Macia,~J.; Vidiella,~B.; Sol{\'{e}},~R. Synthetic associative learning in
  engineered multicellular consortia. \emph{Journal of The Royal Society
  Interface} \textbf{2017}, \emph{14}, 20170158\relax
\mciteBstWouldAddEndPuncttrue
\mciteSetBstMidEndSepPunct{\mcitedefaultmidpunct}
{\mcitedefaultendpunct}{\mcitedefaultseppunct}\relax
\EndOfBibitem
\bibitem[Lakin \latin{et~al.}(2011)Lakin, Youssef, Polo, Emmott, and
  Phillips]{visual_dsd}
Lakin,~M.; Youssef,~S.; Polo,~F.; Emmott,~S.; Phillips,~A. Visual DSD: a design
  and analysis tool for DNA strand displacement systems. \emph{Bioinformatics
  (Oxford, England)} \textbf{2011}, \emph{27}, 3211--3213\relax
\mciteBstWouldAddEndPuncttrue
\mciteSetBstMidEndSepPunct{\mcitedefaultmidpunct}
{\mcitedefaultendpunct}{\mcitedefaultseppunct}\relax
\EndOfBibitem
\bibitem[Amir \latin{et~al.}(2014)Amir, Ben-Ishay, Levner, Ittah, Abu-Horowitz,
  and Bachelet]{amir2014universal}
Amir,~Y.; Ben-Ishay,~E.; Levner,~D.; Ittah,~S.; Abu-Horowitz,~A.; Bachelet,~I.
  Universal computing by DNA origami robots in a living animal. \emph{Nature
  nanotechnology} \textbf{2014}, \emph{9}, 353--357\relax
\mciteBstWouldAddEndPuncttrue
\mciteSetBstMidEndSepPunct{\mcitedefaultmidpunct}
{\mcitedefaultendpunct}{\mcitedefaultseppunct}\relax
\EndOfBibitem
\bibitem[Seelig \latin{et~al.}(2006)Seelig, Soloveichik, Zhang, and
  Winfree]{Seelig1585}
Seelig,~G.; Soloveichik,~D.; Zhang,~D.; Winfree,~E. Enzyme-Free Nucleic Acid
  Logic Circuits. \emph{Science} \textbf{2006}, \emph{314}, 1585--1588\relax
\mciteBstWouldAddEndPuncttrue
\mciteSetBstMidEndSepPunct{\mcitedefaultmidpunct}
{\mcitedefaultendpunct}{\mcitedefaultseppunct}\relax
\EndOfBibitem
\bibitem[Soloveichik \latin{et~al.}(2010)Soloveichik, Seelig, and
  Winfree]{soloveichik2010}
Soloveichik,~D.; Seelig,~G.; Winfree,~E. DNA as a universal substrate for
  chemical kinetics. \emph{Proceedings of the National Academy of Sciences}
  \textbf{2010}, \emph{107}, 5393--5398\relax
\mciteBstWouldAddEndPuncttrue
\mciteSetBstMidEndSepPunct{\mcitedefaultmidpunct}
{\mcitedefaultendpunct}{\mcitedefaultseppunct}\relax
\EndOfBibitem
\bibitem[Chen \latin{et~al.}(2013)Chen, Dalchau, Srinivas, Phillips, Cardelli,
  Soloveichik, and Seelig]{chen_dalchau2013}
Chen,~Y.-J.; Dalchau,~N.; Srinivas,~N.; Phillips,~A.; Cardelli,~L.;
  Soloveichik,~D.; Seelig,~G. Programmable chemical controllers made from DNA.
  \emph{Nature Nanotechnology} \textbf{2013}, \emph{8}, 755–762\relax
\mciteBstWouldAddEndPuncttrue
\mciteSetBstMidEndSepPunct{\mcitedefaultmidpunct}
{\mcitedefaultendpunct}{\mcitedefaultseppunct}\relax
\EndOfBibitem
\bibitem[Yurke \latin{et~al.}(2000)Yurke, Turberfield, Mills, Simmel, and
  Neumann]{Yurke_2000}
Yurke,~B.; Turberfield,~A.; Mills,~A.; Simmel,~F.; Neumann,~J. A DNA-fuelled
  molecular machine made of DNA. \emph{Nature} \textbf{2000}, \emph{406},
  605--608\relax
\mciteBstWouldAddEndPuncttrue
\mciteSetBstMidEndSepPunct{\mcitedefaultmidpunct}
{\mcitedefaultendpunct}{\mcitedefaultseppunct}\relax
\EndOfBibitem
\bibitem[Fontana(2006)]{Fontana2006}
Fontana,~W. Pulling Strings. \emph{Science} \textbf{2006}, \emph{314},
  1552--1553\relax
\mciteBstWouldAddEndPuncttrue
\mciteSetBstMidEndSepPunct{\mcitedefaultmidpunct}
{\mcitedefaultendpunct}{\mcitedefaultseppunct}\relax
\EndOfBibitem
\bibitem[Badelt \latin{et~al.}(2020)Badelt, Grun, Sarma, Wolfe, Shin, and
  Winfree]{Badelt_2020}
Badelt,~S.; Grun,~C.; Sarma,~K.~V.; Wolfe,~B.; Shin,~S.~W.; Winfree,~E. A
  domain-level DNA strand displacement reaction enumerator allowing arbitrary
  non-pseudoknotted secondary structures. \emph{Journal of The Royal Society
  Interface} \textbf{2020}, \emph{17}, 20190866\relax
\mciteBstWouldAddEndPuncttrue
\mciteSetBstMidEndSepPunct{\mcitedefaultmidpunct}
{\mcitedefaultendpunct}{\mcitedefaultseppunct}\relax
\EndOfBibitem
\bibitem[Cardelli(2010)]{Cardelli}
Cardelli,~L. Two-Domain {DNA} Strand Displacement. Proceedings Sixth Workshop
  on Developments in Computational Models: Causality, Computation, and Physics,
  {DCM} 2010, Edinburgh, Scotland, 9-10th July 2010. 2010; pp 47--61\relax
\mciteBstWouldAddEndPuncttrue
\mciteSetBstMidEndSepPunct{\mcitedefaultmidpunct}
{\mcitedefaultendpunct}{\mcitedefaultseppunct}\relax
\EndOfBibitem
\bibitem[R. \latin{et~al.}(2015)R., Arbona, Lad, Mendoza, Aimé, and
  Elezgaray]{cross_talk}
R.,~M.; Arbona,~J.; Lad,~A.; Mendoza,~O.; Aimé,~J.; Elezgaray,~J. Connecting
  localized DNA strand displacement reactions. \emph{Nanoscale} \textbf{2015},
  \emph{7}, 12970--12978\relax
\mciteBstWouldAddEndPuncttrue
\mciteSetBstMidEndSepPunct{\mcitedefaultmidpunct}
{\mcitedefaultendpunct}{\mcitedefaultseppunct}\relax
\EndOfBibitem
\bibitem[Qian and Winfree(2011)Qian, and Winfree]{qian2011scaling}
Qian,~L.; Winfree,~E. Scaling up digital circuit computation with DNA strand
  displacement cascades. \emph{Science} \textbf{2011}, \emph{332},
  1196--1201\relax
\mciteBstWouldAddEndPuncttrue
\mciteSetBstMidEndSepPunct{\mcitedefaultmidpunct}
{\mcitedefaultendpunct}{\mcitedefaultseppunct}\relax
\EndOfBibitem
\bibitem[Dalchau \latin{et~al.}(2018)Dalchau, Sz{\'e}p, Hernansaiz-Ballesteros,
  Barnes, Cardelli, Phillips, and Csik{\'a}sz-Nagy]{biol_switches}
Dalchau,~N.; Sz{\'e}p,~G.; Hernansaiz-Ballesteros,~R.; Barnes,~C.~P.;
  Cardelli,~L.; Phillips,~A.; Csik{\'a}sz-Nagy,~A. Computing with biological
  switches and clocks. \emph{Natural Computing} \textbf{2018}, \emph{17},
  761--779\relax
\mciteBstWouldAddEndPuncttrue
\mciteSetBstMidEndSepPunct{\mcitedefaultmidpunct}
{\mcitedefaultendpunct}{\mcitedefaultseppunct}\relax
\EndOfBibitem
\bibitem[Lakin \latin{et~al.}(2012)Lakin, Youssef, Cardelli, and
  Phillips]{DNA_circuits}
Lakin,~M.; Youssef,~S.; Cardelli,~L.; Phillips,~A. Abstractions for DNA circuit
  design. \emph{Journal of The Royal Society Interface} \textbf{2012},
  \emph{9}, 470--486\relax
\mciteBstWouldAddEndPuncttrue
\mciteSetBstMidEndSepPunct{\mcitedefaultmidpunct}
{\mcitedefaultendpunct}{\mcitedefaultseppunct}\relax
\EndOfBibitem
\bibitem[Qian \latin{et~al.}(2011)Qian, Winfree, and Bruck]{DNA_NN_cascades}
Qian,~L.; Winfree,~E.; Bruck,~J. Neural network computation with DNA strand
  displacement cascades. \emph{Nature} \textbf{2011}, \emph{475},
  368--372\relax
\mciteBstWouldAddEndPuncttrue
\mciteSetBstMidEndSepPunct{\mcitedefaultmidpunct}
{\mcitedefaultendpunct}{\mcitedefaultseppunct}\relax
\EndOfBibitem
\bibitem[Genot \latin{et~al.}(2013)Genot, Fujii, and
  Rondelez]{Genot2013ScalingDD}
Genot,~A.; Fujii,~T.; Rondelez,~Y. Scaling down DNA circuits with competitive
  neural networks. \emph{Journal of The Royal Society Interface} \textbf{2013},
  \emph{10}\relax
\mciteBstWouldAddEndPuncttrue
\mciteSetBstMidEndSepPunct{\mcitedefaultmidpunct}
{\mcitedefaultendpunct}{\mcitedefaultseppunct}\relax
\EndOfBibitem
\bibitem[Cherry and Qian(2018)Cherry, and Qian]{DNA_WTA}
Cherry,~K.~M.; Qian,~L. Scaling up molecular pattern recognition with DNA-based
  winner-take-all neural networks. \emph{Nature} \textbf{2018}, \emph{559},
  370--376\relax
\mciteBstWouldAddEndPuncttrue
\mciteSetBstMidEndSepPunct{\mcitedefaultmidpunct}
{\mcitedefaultendpunct}{\mcitedefaultseppunct}\relax
\EndOfBibitem
\bibitem[LeCun and Cortes(2010)LeCun, and Cortes]{mnist}
LeCun,~Y.; Cortes,~C. {MNIST} handwritten digit database. \textbf{2010}, \relax
\mciteBstWouldAddEndPunctfalse
\mciteSetBstMidEndSepPunct{\mcitedefaultmidpunct}
{}{\mcitedefaultseppunct}\relax
\EndOfBibitem
\bibitem[Lakin and Stefanovic(2016)Lakin, and Stefanovic]{supervised}
Lakin,~M.; Stefanovic,~D. Supervised Learning in Adaptive DNA Strand
  Displacement Networks. \emph{ACS Synthetic Biology} \textbf{2016}, \emph{5},
  885--897\relax
\mciteBstWouldAddEndPuncttrue
\mciteSetBstMidEndSepPunct{\mcitedefaultmidpunct}
{\mcitedefaultendpunct}{\mcitedefaultseppunct}\relax
\EndOfBibitem
\bibitem[Chu \latin{et~al.}(2009)Chu, Zabet, and Mitavskiy]{Chu_2009}
Chu,~D.; Zabet,~N.; Mitavskiy,~B. Models of transcription factor binding:
  Sensitivity of activation functions to model assumptions. \emph{Journal of
  Theoretical Biology} \textbf{2009}, \emph{257}, 419 -- 429\relax
\mciteBstWouldAddEndPuncttrue
\mciteSetBstMidEndSepPunct{\mcitedefaultmidpunct}
{\mcitedefaultendpunct}{\mcitedefaultseppunct}\relax
\EndOfBibitem
\bibitem[Zhang and Winfree(2009)Zhang, and Winfree]{toehold_ex_zhang}
Zhang,~D.; Winfree,~E. Control of DNA Strand Displacement Kinetics Using
  Toehold Exchange. \emph{Journal of the American Chemical Society}
  \textbf{2009}, \emph{131}, 17303--17314\relax
\mciteBstWouldAddEndPuncttrue
\mciteSetBstMidEndSepPunct{\mcitedefaultmidpunct}
{\mcitedefaultendpunct}{\mcitedefaultseppunct}\relax
\EndOfBibitem
\bibitem[Lyu \latin{et~al.}(2020)Lyu, Peng, Liu, Kuai, Mo, Han, Li, and
  Tan]{lyu2020protocells}
Lyu,~Y.; Peng,~R.; Liu,~H.; Kuai,~H.; Mo,~L.; Han,~D.; Li,~J.; Tan,~W.
  Protocells programmed through artificial reaction networks. \emph{Chemical
  Science} \textbf{2020}, \emph{11}, 631--642\relax
\mciteBstWouldAddEndPuncttrue
\mciteSetBstMidEndSepPunct{\mcitedefaultmidpunct}
{\mcitedefaultendpunct}{\mcitedefaultseppunct}\relax
\EndOfBibitem
\bibitem[Chu and Barnes(2016)Chu, and Barnes]{myscireppaper}
Chu,~D.; Barnes,~D. The lag-phase during diauxic growth is a trade-off between
  fast adaptation and high growth rate. \emph{Scientific Reports}
  \textbf{2016}, \emph{6}, 25191\relax
\mciteBstWouldAddEndPuncttrue
\mciteSetBstMidEndSepPunct{\mcitedefaultmidpunct}
{\mcitedefaultendpunct}{\mcitedefaultseppunct}\relax
\EndOfBibitem
\bibitem[Groves \latin{et~al.}(2015)Groves, Chen, Zurla, Pochekailov,
  Kirschman, Santangelo, and Seelig]{connect1}
Groves,~B.; Chen,~Y.; Zurla,~C.; Pochekailov,~S.; Kirschman,~J.;
  Santangelo,~P.; Seelig,~G. Computing in mammalian cells with nucleic acid
  strand exchange. \emph{Nature Nanotechnology} \textbf{2015}, \emph{11},
  287--294\relax
\mciteBstWouldAddEndPuncttrue
\mciteSetBstMidEndSepPunct{\mcitedefaultmidpunct}
{\mcitedefaultendpunct}{\mcitedefaultseppunct}\relax
\EndOfBibitem
\bibitem[Oesinghaus and Simmel(2019)Oesinghaus, and Simmel]{connect2}
Oesinghaus,~L.; Simmel,~F. Switching the activity of Cas12a using guide {RNA}
  strand displacement circuits. \emph{Nature Communications} \textbf{2019},
  \emph{10}\relax
\mciteBstWouldAddEndPuncttrue
\mciteSetBstMidEndSepPunct{\mcitedefaultmidpunct}
{\mcitedefaultendpunct}{\mcitedefaultseppunct}\relax
\EndOfBibitem
\bibitem[J. and Ellington(2014)J., and Ellington]{connect3}
J.,~C.; Ellington,~A. Diagnostic Applications of Nucleic Acid Circuits.
  \emph{Accounts of Chemical Research} \textbf{2014}, \emph{47},
  1825--1835\relax
\mciteBstWouldAddEndPuncttrue
\mciteSetBstMidEndSepPunct{\mcitedefaultmidpunct}
{\mcitedefaultendpunct}{\mcitedefaultseppunct}\relax
\EndOfBibitem
\bibitem[Douglas \latin{et~al.}(2012)Douglas, Bachelet, and Church]{connect4}
Douglas,~S.~M.; Bachelet,~I.; Church,~G.~M. A Logic-Gated Nanorobot for
  Targeted Transport of Molecular Payloads. \emph{Science} \textbf{2012},
  \emph{335}, 831--834\relax
\mciteBstWouldAddEndPuncttrue
\mciteSetBstMidEndSepPunct{\mcitedefaultmidpunct}
{\mcitedefaultendpunct}{\mcitedefaultseppunct}\relax
\EndOfBibitem
\bibitem[Fern \latin{et~al.}(2017)Fern, Scalise, Cangialosi, Howie, Potters,
  and Schulman]{timer_circuits}
Fern,~J.; Scalise,~D.; Cangialosi,~A.; Howie,~D.; Potters,~L.; Schulman,~R. DNA
  Strand-Displacement Timer Circuits. \emph{ACS Synthetic Biology}
  \textbf{2017}, \emph{6}, 190--193\relax
\mciteBstWouldAddEndPuncttrue
\mciteSetBstMidEndSepPunct{\mcitedefaultmidpunct}
{\mcitedefaultendpunct}{\mcitedefaultseppunct}\relax
\EndOfBibitem
\bibitem[Chatterjee \latin{et~al.}(2017)Chatterjee, Dalchau, Muscat, Phillips,
  and Seelig]{Chatterjee110965}
Chatterjee,~G.; Dalchau,~N.; Muscat,~R.; Phillips,~A.; Seelig,~G. A spatially
  localized architecture for fast and modular DNA computing. \emph{Nature
  Nanotechnology} \textbf{2017}, \emph{12}, 920\relax
\mciteBstWouldAddEndPuncttrue
\mciteSetBstMidEndSepPunct{\mcitedefaultmidpunct}
{\mcitedefaultendpunct}{\mcitedefaultseppunct}\relax
\EndOfBibitem
\end{mcitethebibliography}
\end{document}


\appendix

\section{Supplementary Tables}

\subsection{CN parameters}

\begin{table}[H]
\centering
\caption{List of reaction rate constants in the CN model (s$^{-1}$).}
\label{params_table}
\begin{tabular}{ll}
\toprule
Function & Parameters \\ 
\midrule
\multirow{2}{*}{Input}              
& $k_{IA}=10$, $k_{AI}=0.000001$  \\
& $k_{AB}=0.1$, $k_{BA}=0.000001$  \\ 
\midrule
\multirow{2}{*}{Activation function }
& $k^{+}=1$, $k^{-}=5$ \\
& $k^{-}_{last}=0.5$  \\ 
\midrule
\multirow{2}{*}{Learning}
& $k_{AE}=0.05$, $k_{EA}=0.000001$, $k_{EH}=100$, $k_{HE}=0.000001$   \\
& $k_{AH}=0.001$, $k_{HA}=0.000001$, $k_{HB}=100$, $k_{BH}=0.000001$ \\ 
\midrule
\multirow{2}{*}{Leak}
& $k_{H\varnothing}=0.0003$  \\
& $k_{B\varnothing}=0.1$ \\ 
\bottomrule
\end{tabular}
\end{table}

\subsection{c-CN parameters}

\begin{table}[H]
\centering
\caption[List of reaction rate constants in a cellular interpretation of the chemical neuron.]{List of reaction rate constants in a c-CN model (s$^{-1}$).}
\label{params_table2}
\begin{tabular}{ccccc}
\toprule
Function & Reaction rates \\
\midrule
\multirow{2}{*}{Input}
 & $k_{IA}=10$, $k_{AI}=0.000001$  \\
&  $k_{AH}=0.03$, $k_{HA}=0.000001$, $k_{HB}=100$, $k_{BH}=0.000001$  \\
\midrule 
\multirow{2}{*}{Activation function }
& $k^{+}=1$, $k^{-}=5$ \\
& $k^{-}_{last}=0.5$  \\ 
\midrule
\multirow{5}{*}{Weight accumulation}
& $k_{AE}=0.2$, $k_{EA}=0.000001$, $k_{A*E}=0.2$, $k_{EA*}=0.000001$,\\
& $k_{A*A}=0.05$, $k_{AA*}=0.000001$ \\
& $k_{A*h}=1$, $k_{hA*}=0.1$ \\
& $k_{\textrm{leak}}=0.0001$ \\
& $k_{h}=1$ \\ 
\midrule
\multirow{2}{*}{Leak}
& $k_{H\varnothing}=0.0003$ \\
& $k_{B\varnothing}=0.1$   \\
\bottomrule
\end{tabular}
\end{table}

\subsection{List of nucleotide sequences and binding rates for d-CN}
\label{sec:dCN_details}

\begin{table}[H]
\centering
\caption[List of toehold domains and their respective binding and unbinding rates for the d-CN model.]{List of toehold domains and their respective binding and unbinding rates in $\mu$M s$^{-1}$ for the simulations carried out in the infinite compilation mode.}
\label{tab:strand_table_dsd_new}
\begin{tabular}{cccc} 
\toprule
Signal Species & Toehold & Bind & Unbind \\
\midrule
$A_{n}$ & \texttt{ta} & 1 & 10 \\ 
$H_{n}$ & \texttt{th} & 0.001 & 10 \\
$B$ & \texttt{tb} & 1 & 10 \\ 
$E_{m}$ & \texttt{tem}& 5 & 10 \\ 
$Fsi$ & \texttt{tfsi} & 1 & 10 \\
$Faf$ & \texttt{tfaf} & 1 & 10 \\ 
\midrule
$Ism$ & \texttt{tism} & 1 & 10 \\
$Iaf$ & \texttt{tiaf} & 1 & 10 \\
$Isi$ & \texttt{tisi} & 1 & 10 \\
\midrule
$Iwa_{n}$ & \texttt{itwan} & 1 & 10 \\
\bottomrule
\end{tabular}
\end{table}

\begin{table}[H]
\centering
\caption{List of the two-domain DNA strands and their respective nucleotide sequences in the d-CN activation function.}
\label{nucleotide_table_dsd_new}
\begin{tabular}{cc} 
\toprule
Strand & Sequence \\
\midrule
	\texttt{<ta\textasciicircum\ a1>}  & $GACA + CCCTAAACTTATCTAAACAT$ \\ 
	\texttt{<ta\textasciicircum\ a2>}  & $GACA + CCCATTTCAAATCAAAACTT$ \\ 
	\texttt{<ta\textasciicircum\ a3>}  & $GACA + CCCATTACTAATCAATTCAA$ \\ 
\midrule
	\texttt{<th\textasciicircum\ h1>}  & $CTCAG + CCCTTTTCTAAACTAAACAA$ \\ 
	\texttt{<th\textasciicircum\ h2>}  & $CTCAG + CCCTTATCATATCAATACAA$ \\ 
	\texttt{<th\textasciicircum\ h3>}  & $CTCAG + CCCTTAACTTAACAAATCTA$ \\ 
\midrule
	\texttt{<tb\textasciicircum\ b>}  & $ACTACAC + CCCAAAACAAAACAAAACAA$ \\   
	\texttt{<te0\textasciicircum\ b>}  & $CATCG + CCCAAAACAAAACAAAACAA$ \\ 
	\texttt{<te1\textasciicircum\ b>}  & $TACCAA + CCCTTATCATATCAATACAA$ \\ 
	\texttt{<te2\textasciicircum\ b>}  & $GTCA + CCCTTATCATATCAATACAA$ \\ 
	\texttt{<te3\textasciicircum\ b>}  & $GCTA + CCCTTATCATATCAATACAA$ \\ 
	\texttt{<te4\textasciicircum\ b>}  & $TATTCC + CCCTTATCATATCAATACAA$ \\ 
	\texttt{<te5\textasciicircum\ b>}  & $CACACA + CCCTTATCATATCAATACAA$ \\ 
\midrule
	\texttt{<tfsi\textasciicircum\ fsi1>}  & $ACCT + CCCTATTCAATTCAAATCAA$ \\ 
	\texttt{<tfsi\textasciicircum\ fsi2>}  & $ACCT + CCCTATACTATACAATACTA$ \\ 
	\texttt{<tfsi\textasciicircum\ fsi3>}  & $ACCT + CCCTAATCTAATCATAACTA$ \\ 
\bottomrule
	\texttt{<tisi\textasciicircum>}  & $TAGCCA$ \\ 
	\texttt{<tism\textasciicircum>}  & $CCCT$ \\ 
	\texttt{<tiwa1\textasciicircum>}  & $CTCAATC$ \\ 
	\texttt{<tiwa2\textasciicircum>}  & $CCTACG$ \\ 
	\texttt{<tiwa3\textasciicircum>}  & $TCTCCA$ \\ 
\bottomrule
	\texttt{<i>}  & $CCCTTTACATTACATAACAA$ \\ 
\bottomrule
\end{tabular}
\end{table}

\clearpage

\section{Supplementary Figures}

\subsection{Weight distributions for FB 2 and TC 2 tasks.}
\label{a1}

\begin{figure}[H]
\centering
\subfloat[fb2][FB 2]{\includegraphics[width=\textwidth]{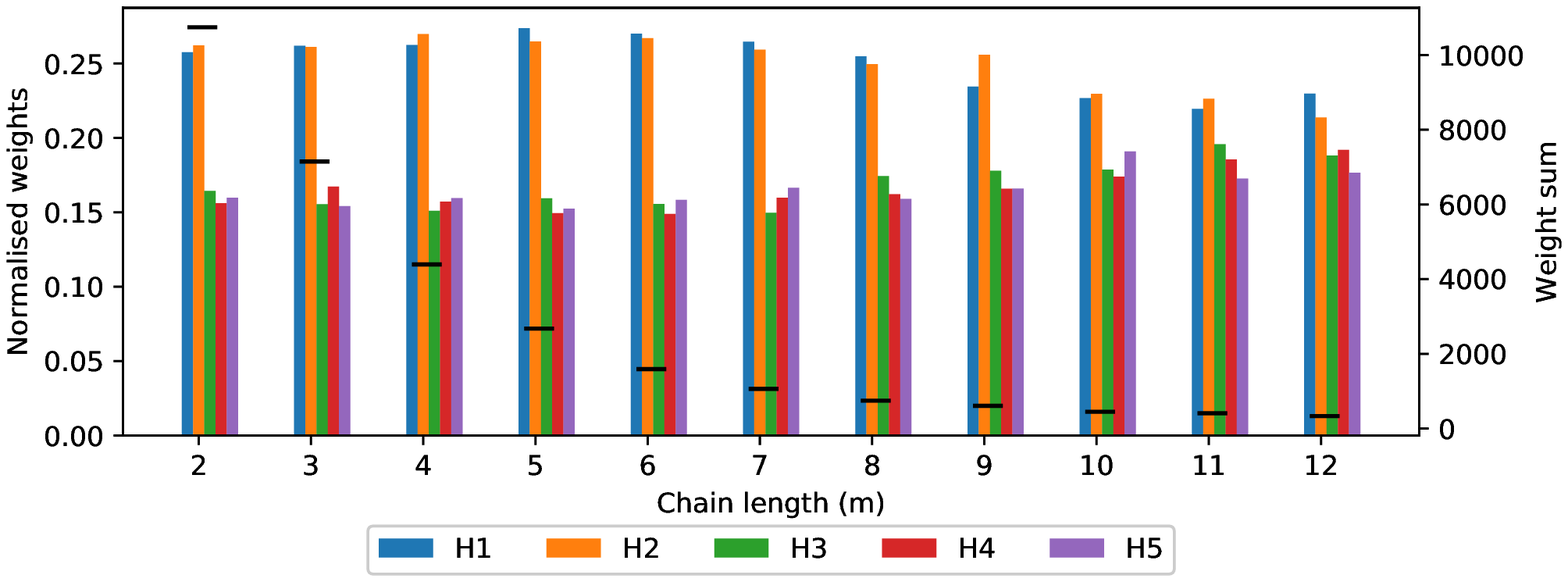}\label{fb2}}\\
\subfloat[tc2][TC 2]{\includegraphics[width=\textwidth]{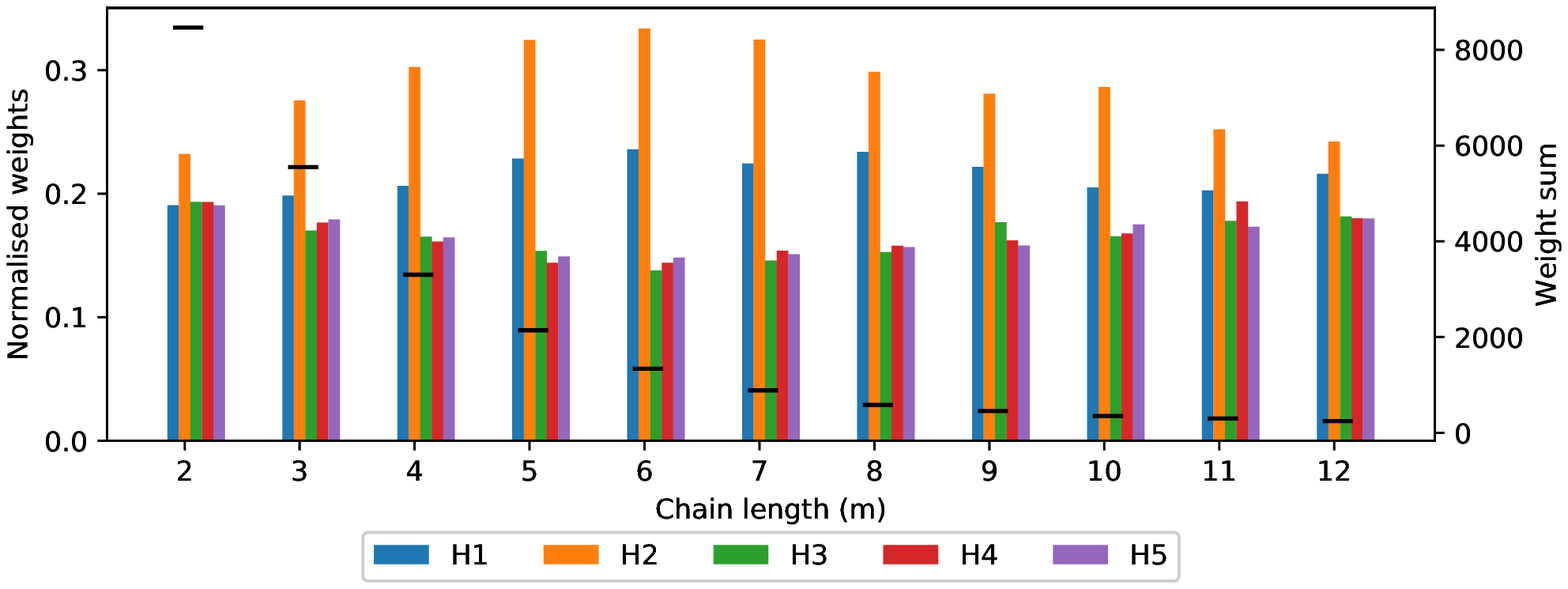}\label{tc2}}
\caption{Normalised weights for \protect\subref{fb2} FB 2 and \protect\subref{tc2} TC 2 tasks as a function of chain length $m$. 
The y-axis on the right side of the figures describes the sum of weights across all input channels, which is a normalisation constant for the corresponding sets of weights. 
It can be noticed that the accumulation of weights becomes more difficult as the chain length increases. 
However, it's worth noting that both low and high $m$ result in low variance of the weight representations.}
\label{CN_weights} 
\end{figure}

\subsection{Examples of learning episodes}
\label{learning_episodes}
\begin{figure}[H]
\centering
\subfloat[fb_l][Average weights for frequency bias ($m=3$).]{\includegraphics[width=0.495\textwidth]{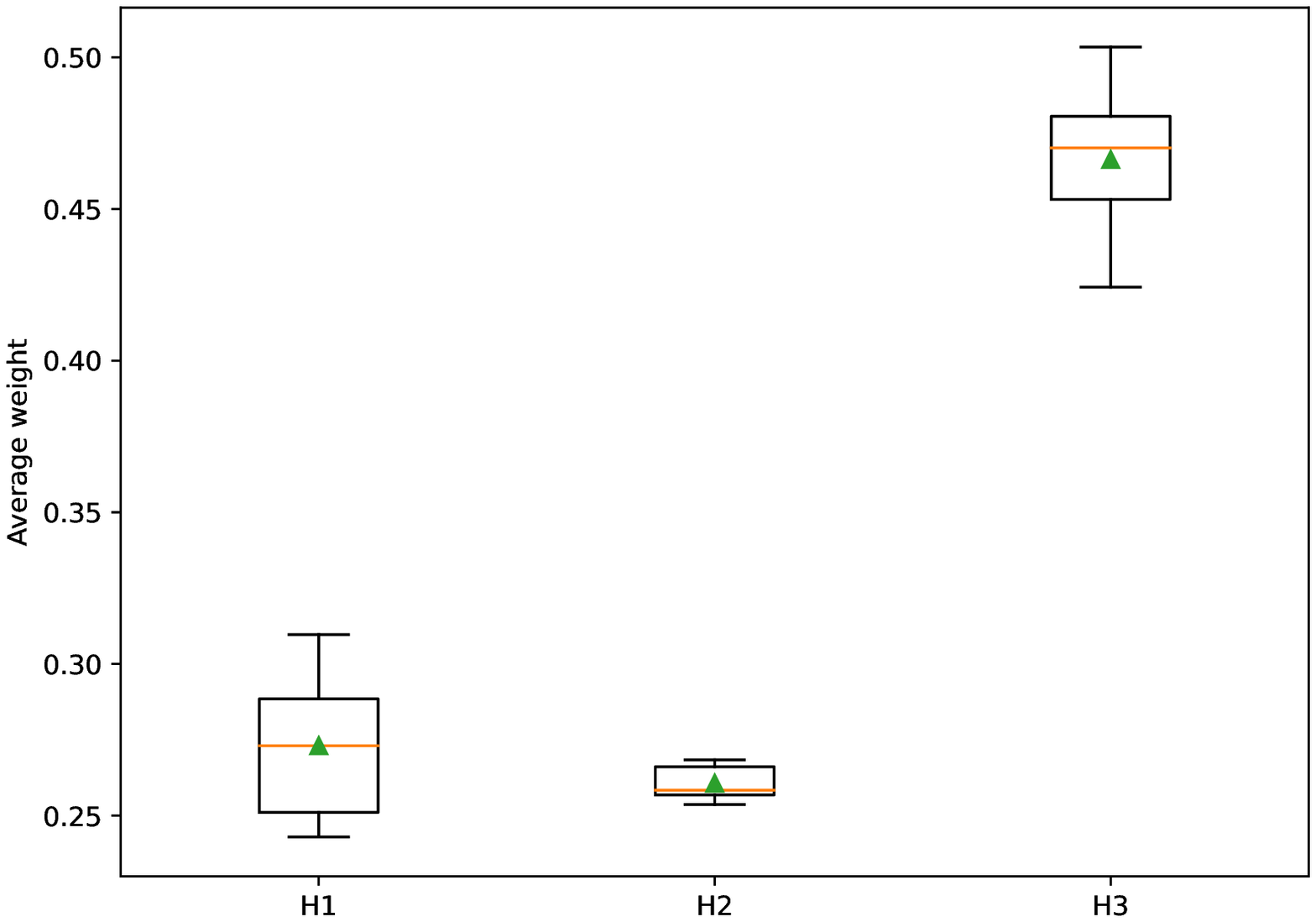}\label{fb_l}}
\subfloat[tc_l][Average weights for temporal correlation ($m=1$).]{\includegraphics[width=0.495\textwidth]{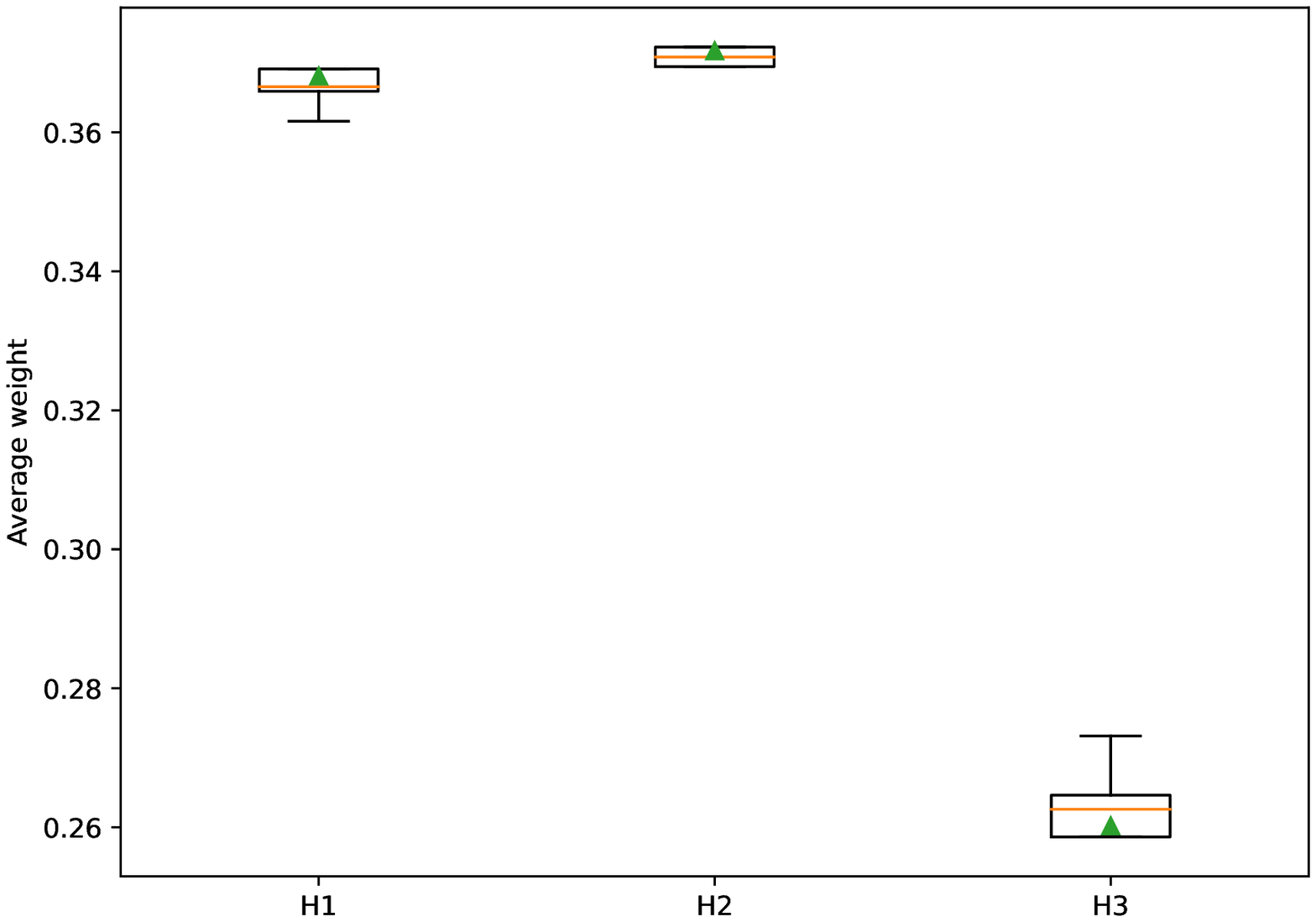}\label{tc_l}}
\caption{
Normalised weights in the steady state for input data with
\protect\subref{fb_l} frequency bias, and 
\protect\subref{tc_l} temporal correlation.
The values represent averages over 200000s of stimulation, and were only collected once the system reached a homeostatic state (after 800000s).
We then repeat each experiment 10 times, and calculate the statistics. 
We define the temporal correlation as pairing of inputs from channels $A_1$ and $A_2$.
The inputs from $A_1$ always precede that of the $A_2$ with a fixed temporal distance $\delta$.
We can see that the steady state abundances of channel specific $H_n$ molecules reflect the temporal order of the inputs provided. 
Moreover, the input channel that spiked in an asynchronous way ($A_3$) accumulated lower weights than the temporally correlated channels.
In the frequency bias experiment we assume $A_1$ to come at frequency twice as high as that of the two other input channels. 
The system learns by accumulating steady state abundances of weight molecules $H_n$ for each channel.
After training, the weights of the channels with a high input frequency will be high, whereas the ones less likely to spike are on a similar low level. 
}
\label{TC_FB_results} 
\end{figure}

\subsection{Simulation of learning tasks}
\label{a6}

\begin{figure}[H]
\centering
\subfloat[tcex][Example of temporal correlation learning.]{\includegraphics[width=0.5\textwidth]{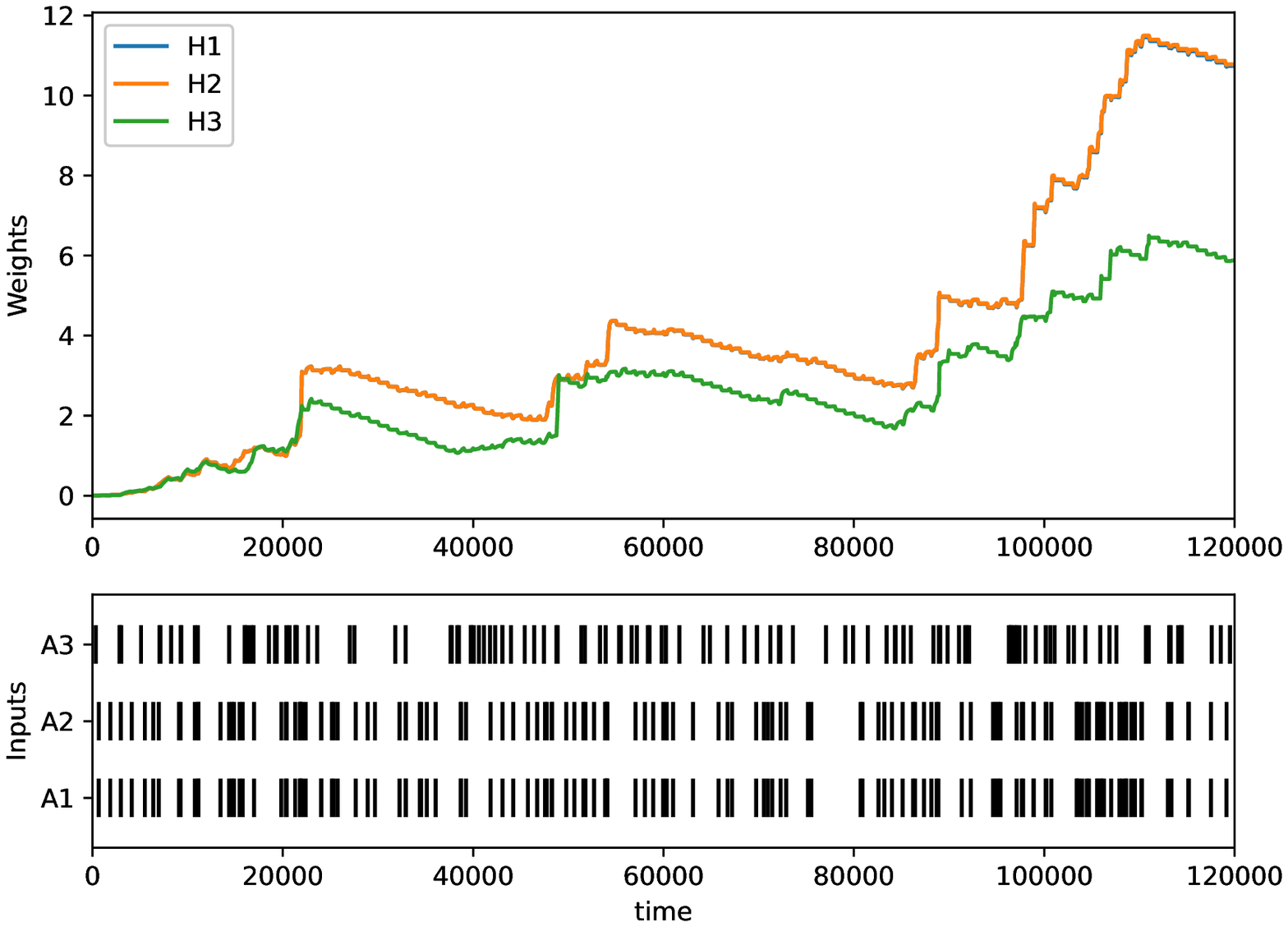}\label{tcex_l_other}} 
\subfloat[fbex][Example of frequency bias learning.]{\includegraphics[width=0.5\textwidth]{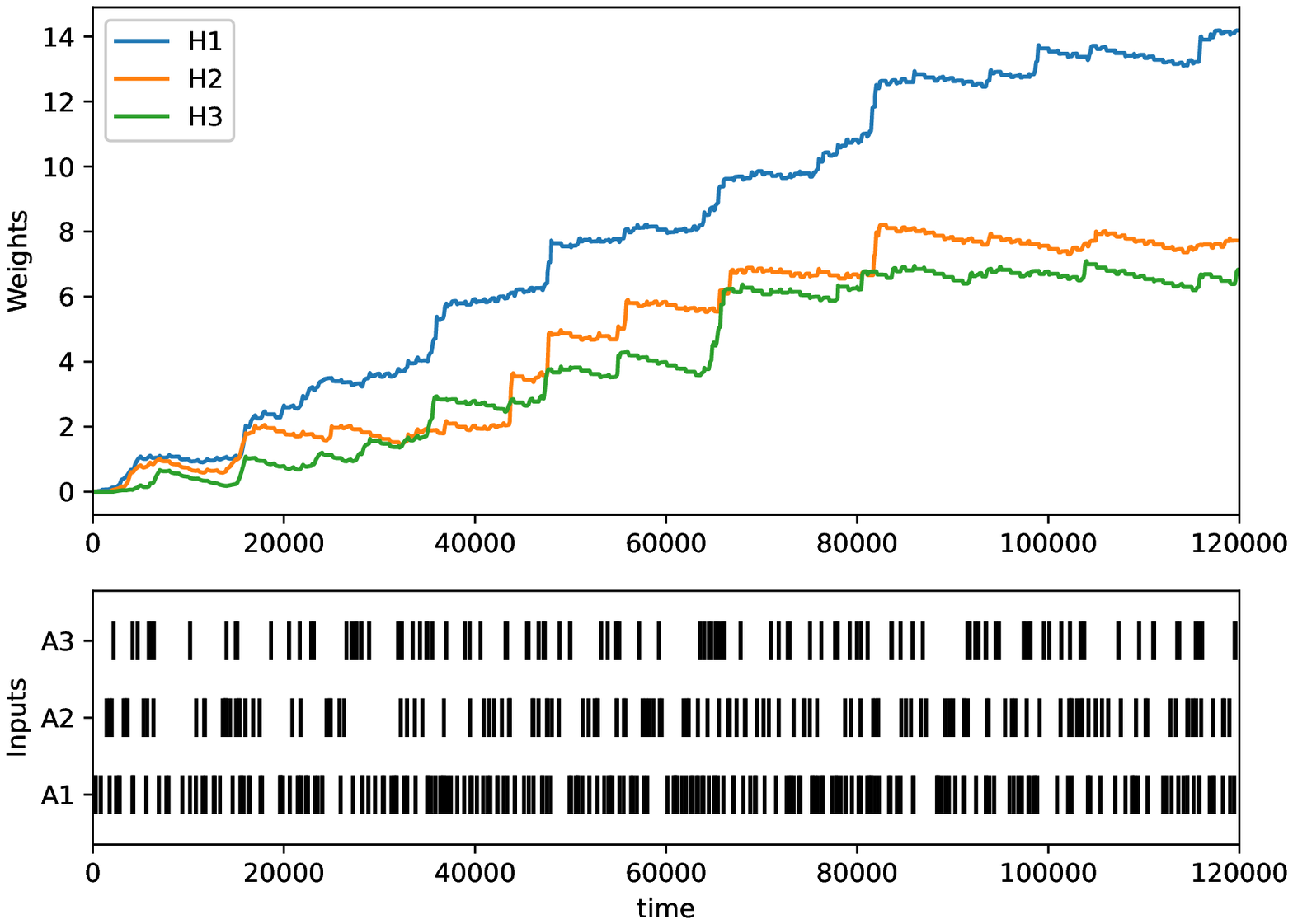}\label{fbex_l_other}}
\caption{
Example learning episode for
\protect\subref{tcex_l_other}
temporal correlation task, and 
\protect\subref{fbex_l_other} frequency bias task simulated using a detailed compilation mode of Visual DSD. The detailed mode is the most realistic setting, which assumes that binding, unbinding and branch migration have finite rates. 
}
\label{TC_FB_results_other} 
\end{figure}

\subsection{Learning new input statistics}
\label{a7}

\begin{figure}[H]
\centering
\subfloat[p1p2fb][Example of frequency bias learning with input statistics changing during the simulation]{\includegraphics[width=\textwidth]{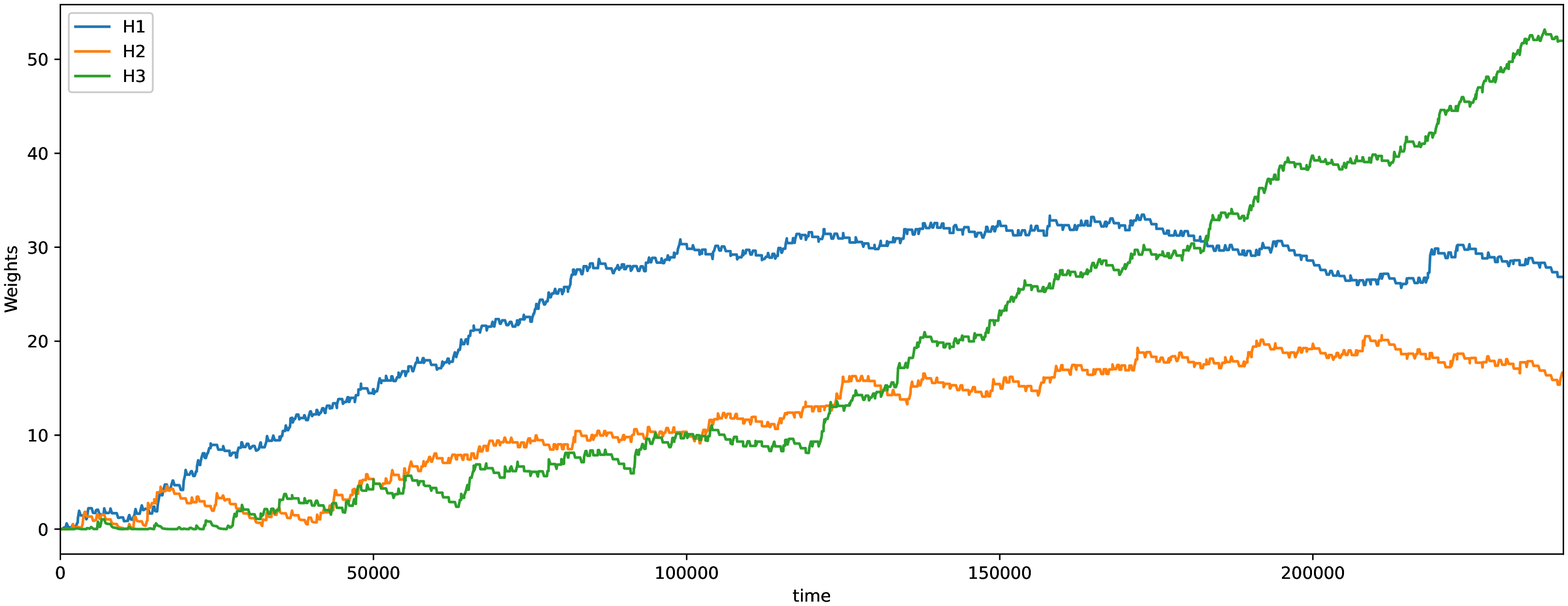}\label{p1p2fb}}\\
\subfloat[p1fb][Normalised weights at t=120000.]{\includegraphics[width=0.495\textwidth]{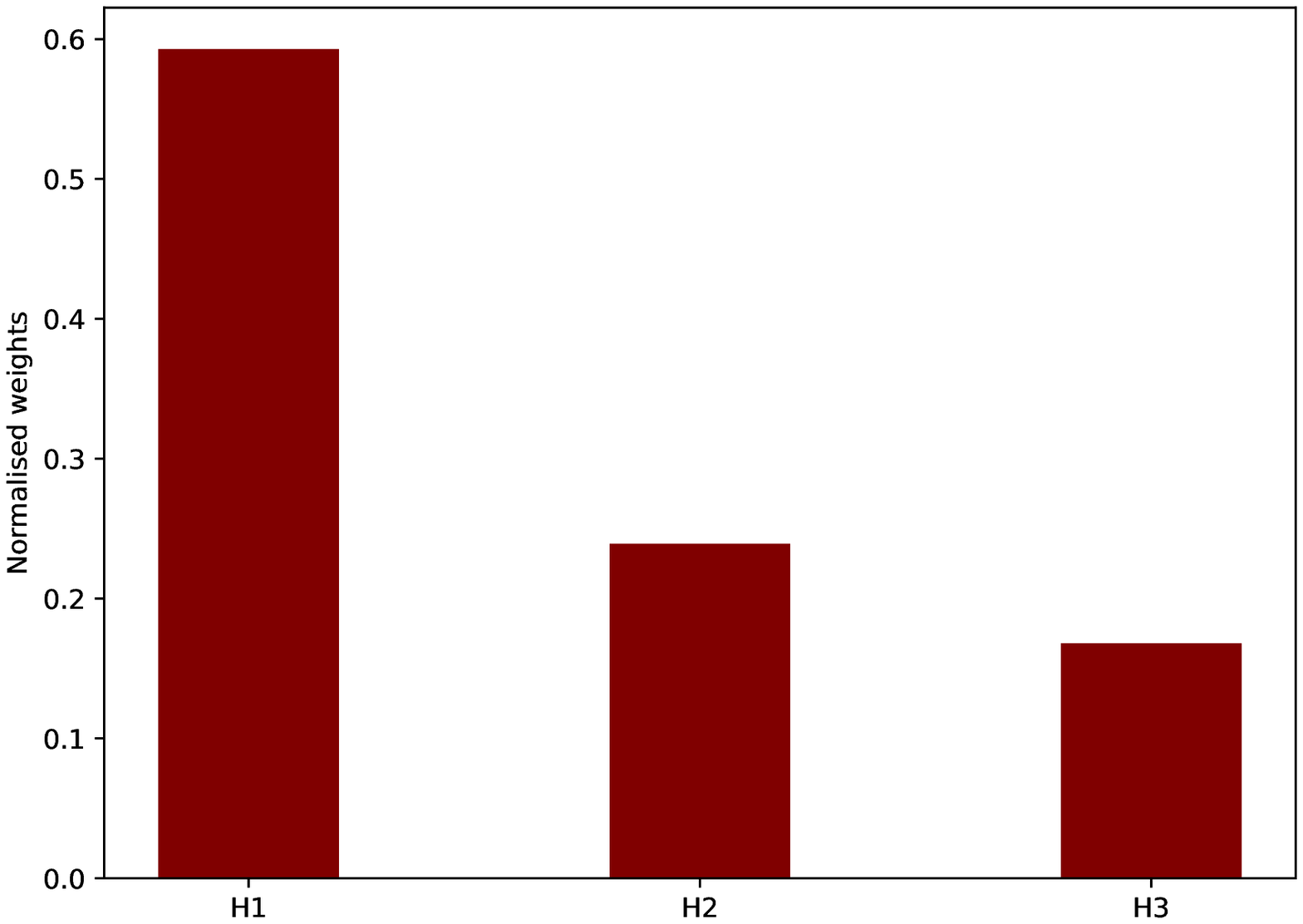}\label{p1fb}}
\subfloat[[p2fb][Normalised weights at t=240000.]{\includegraphics[width=0.495\textwidth]{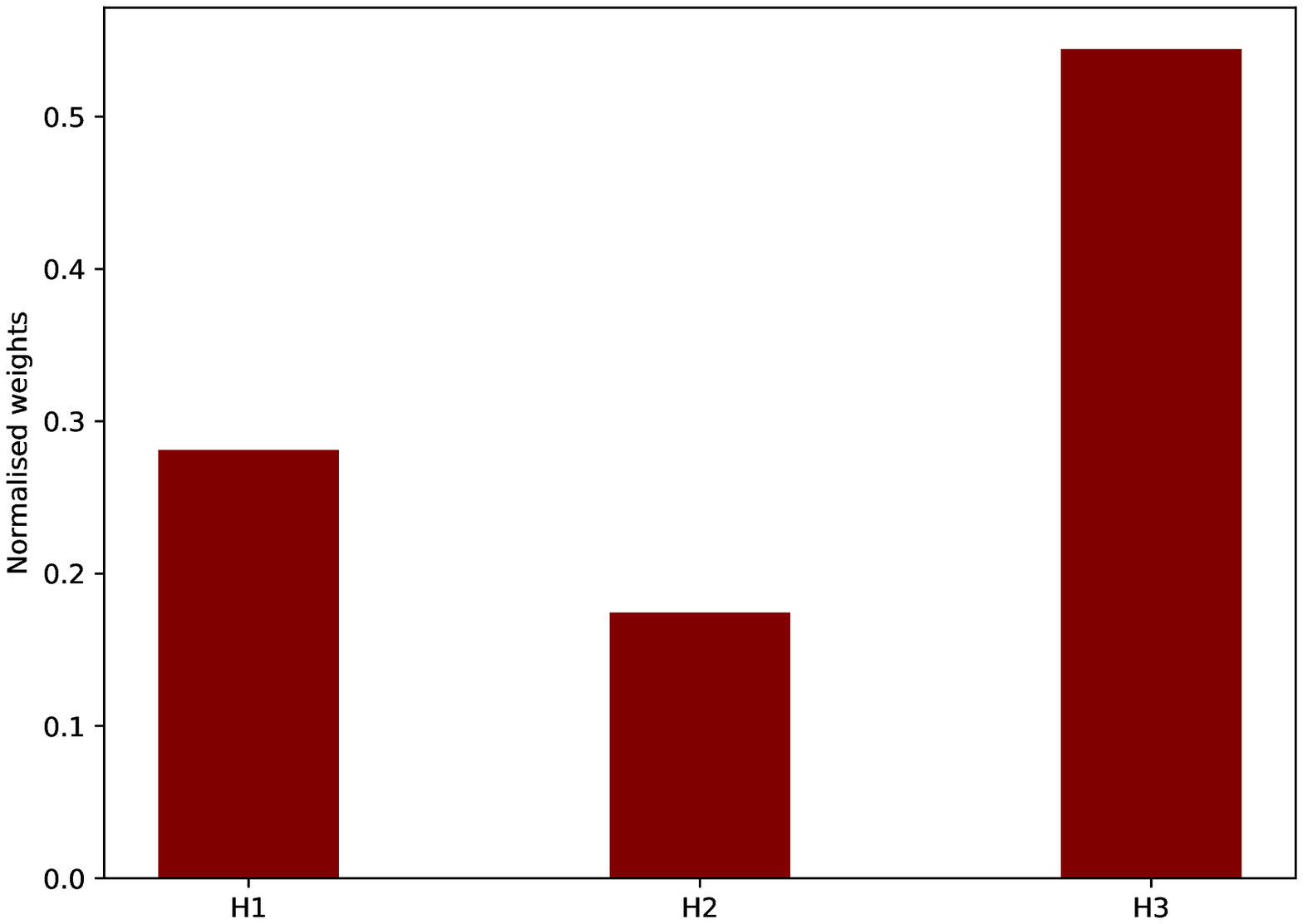}\label{[p2fb}}
\caption{We examine a case when the statistical bias of the inputs changes during the simulation. The simulation starts with inputs having a statistical bias towards inputs from channel A1, which arrive twice as frequently compared to the other channels. At t=120000 the statistics change and A3 becomes the statistically over-represented input channel. The neuron's weights reflect this change and the weight associated with the new biased input channel increases. }
\label{FB_relearn} 
\end{figure}

\begin{figure}[H]
\centering
\subfloat[p1p2tc][Example of temporal correlation learning with input statistics changing during the simulation]{\includegraphics[width=\textwidth]{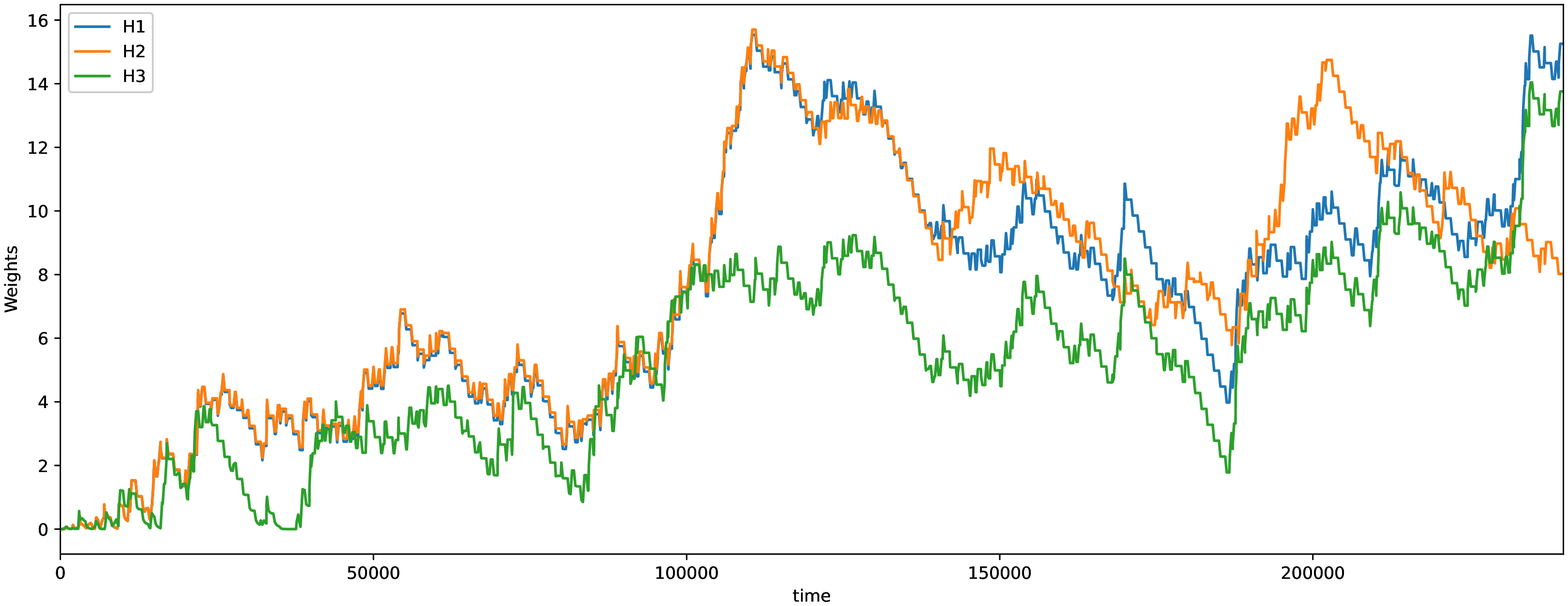}\label{p1p2tc}}\\
\subfloat[p1tc][Normalised weights at t=120000.]{\includegraphics[width=0.495\textwidth]{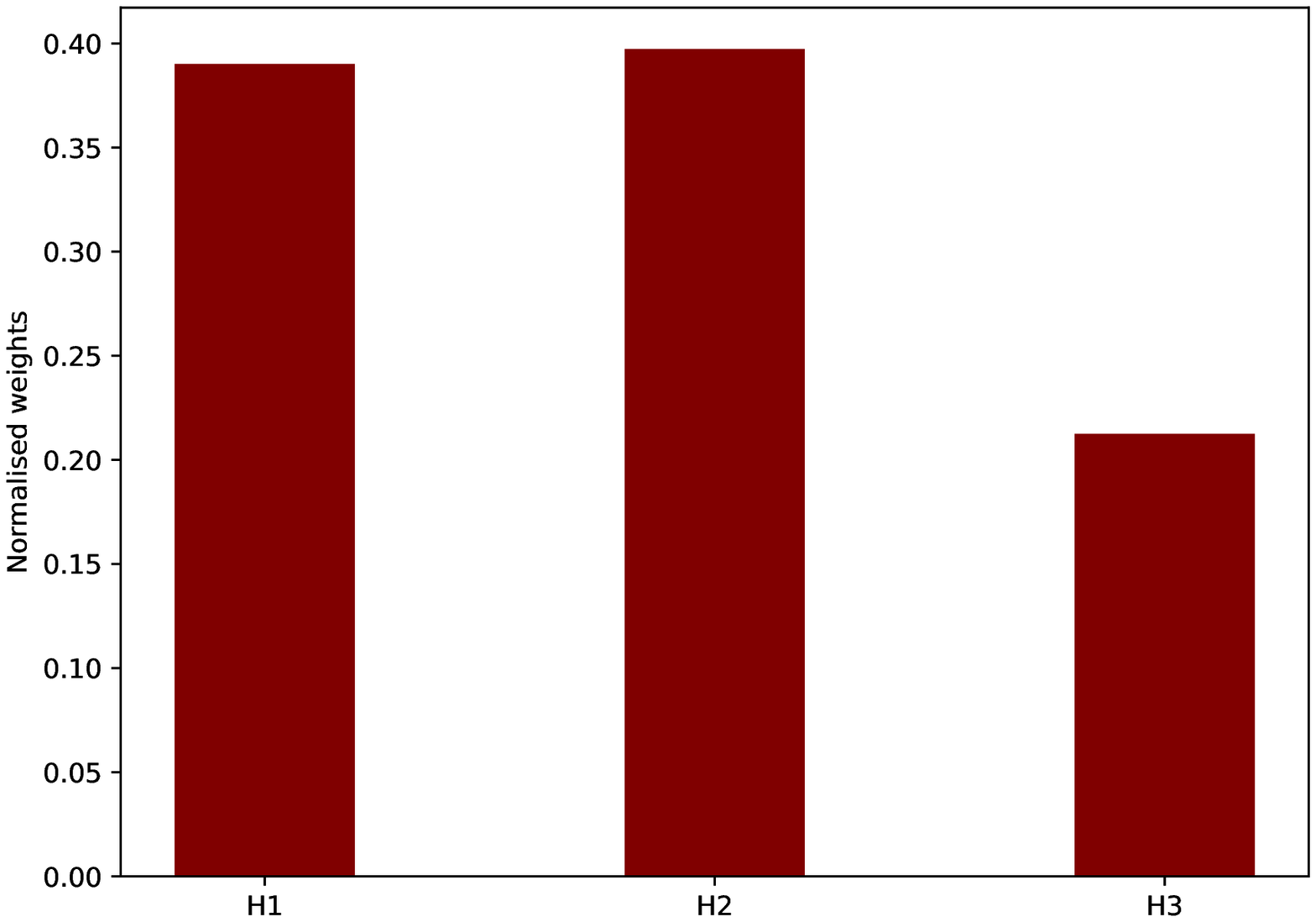}\label{p1tc}}
\subfloat[[p2tc][Normalised weights at t=240000.]{\includegraphics[width=0.495\textwidth]{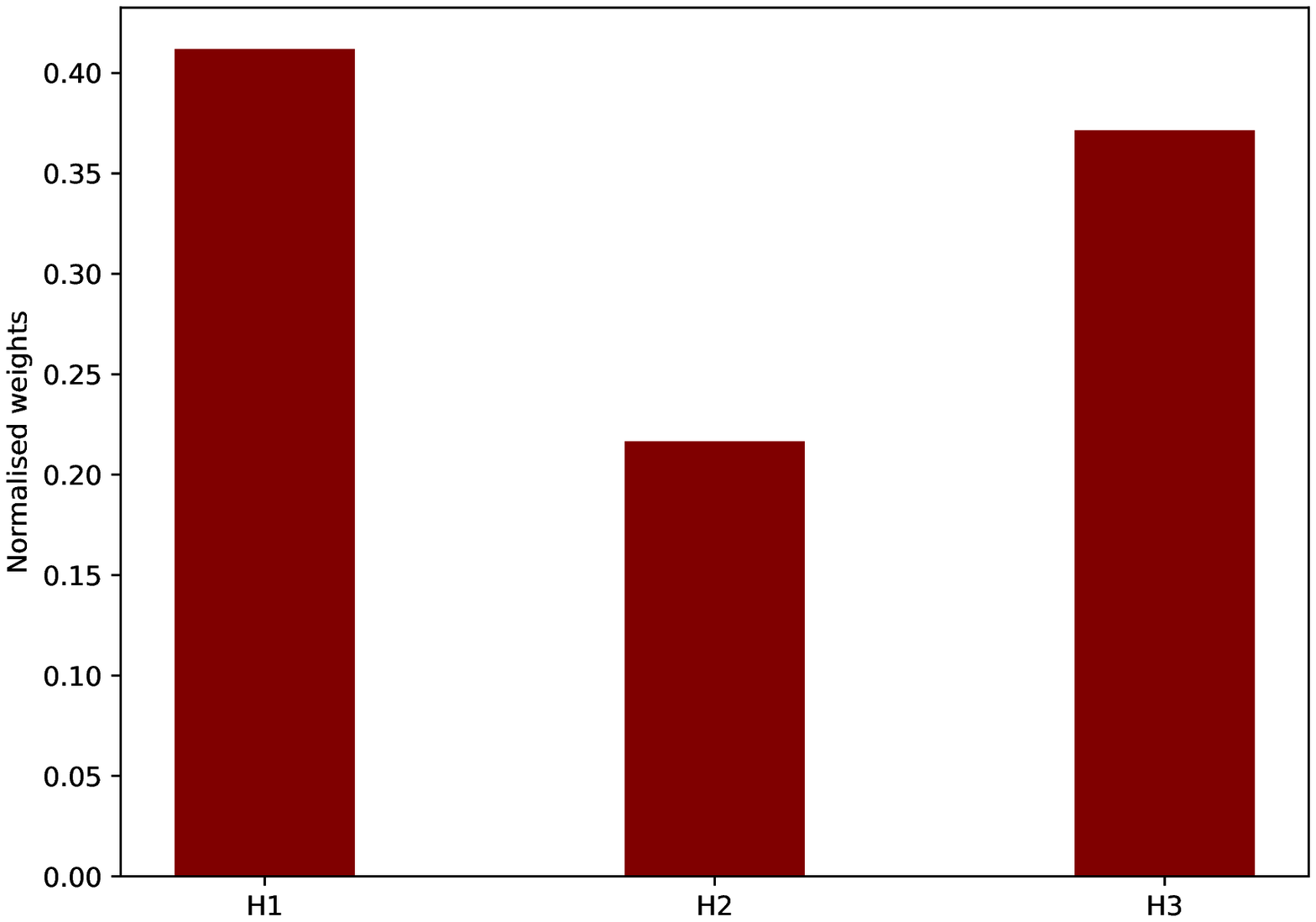}\label{[p2tc}}
\caption{We examine a case when the statistical bias of the inputs changes during the simulation. The simulation starts with inputs having a temporal correlation between inputs A1 and A2, where the inputs from A2 always arrive 1s after that of A1. After 120000s, the weight associated with the second input channel H2 is slightly higher than H1, and both of them are significantly higher than H3.
At t=120000 the statistics change and we introduce a correlation between A1 and A3, where the inputs from A1 always arrive 1s after that of A3. The neuron's weights reflect this change at t=240000. Now, the weight associated with the first input channel H1 is slightly higher than H3, and both of them are significantly higher than H2. }
\label{TC_relearn} 
\end{figure}

\subsection{Signal modulation in the DNA neuron}

\begin{figure}[H]
  \centering
  \subfloat[sm_b_l][Internal state ($B$) as a function of $H_{n}$.]{
      \includegraphics[width=0.48\textwidth]{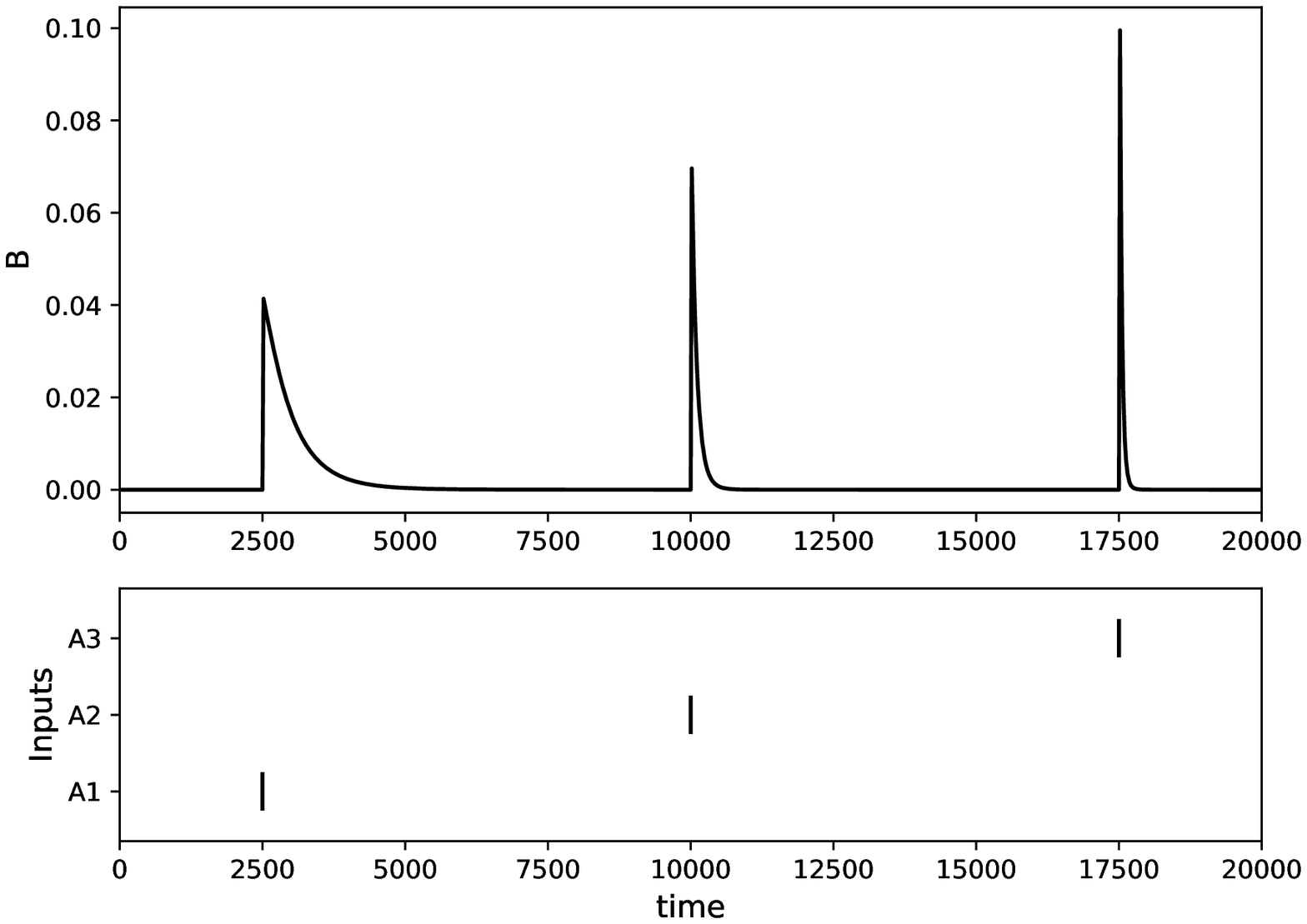}
      \label{sm_b_l}}
  \subfloat[sm_ebh_l][Learning signal ($\mathcal E$) as a function of $H_{n}$.]{
      \includegraphics[width=0.48\textwidth]{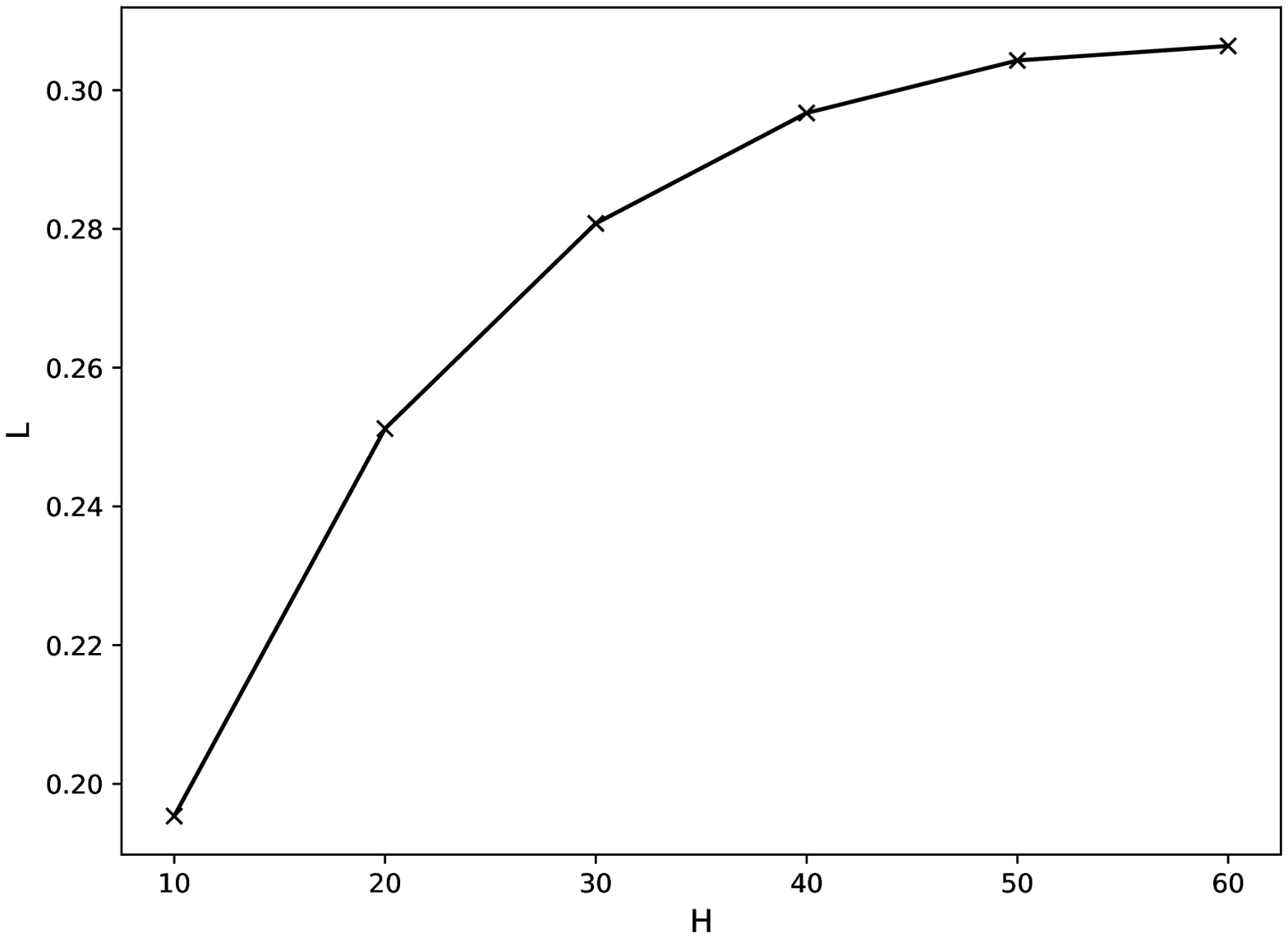}
        \label{sm_ebh_l}}
  \caption{
  Signal modulation in \neuron  ($m=1$).  
  \protect\subref{sm_b_l} We inject 5 $\mu$M of $A_{n}$ molecules to the system at $t=100$, $300$, and $500$.  Each channel has a different amount of $H$ molecules associated with it: $H_1=10$, $H_2=30$, and $H_3=50$.  This has the effect that the conversion of $A_1$ is slower than that of  $A_2$, which is in turn slower than the conversion of $A_3$.  
  \protect\subref{sm_ebh_l} The influence of inputs on the concentration of $L$ species as a function of its weight ($H$). The \neuron shows a different amount of activation given inputs from differently weighted channels. Here, we assume a constant decay of $H_{n}$ and $B$ species with rate constants $k_{H}=0.00002$ and $k_{B}=2$.
  }
  \label{SM}
\end{figure}

\subsection{Strategies of garbage collection}
\label{a4}

\begin{figure}[H]
\centering
\subfloat[h3][Normalised $H_3$.]{\includegraphics[width=0.5\textwidth]{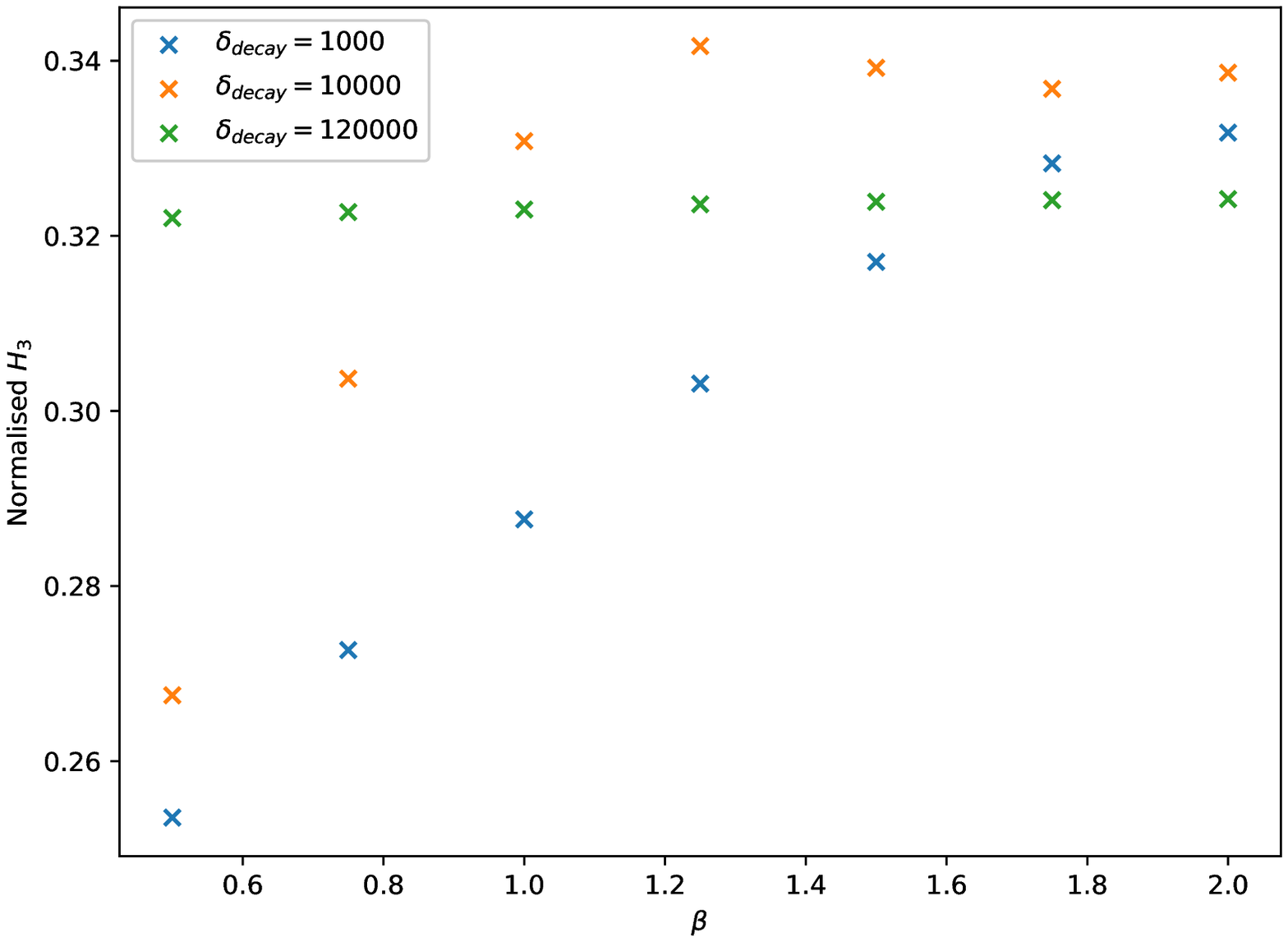}\label{h3_decay}} 
\subfloat[iod][Index of dispersion.]{\includegraphics[width=0.5\textwidth]{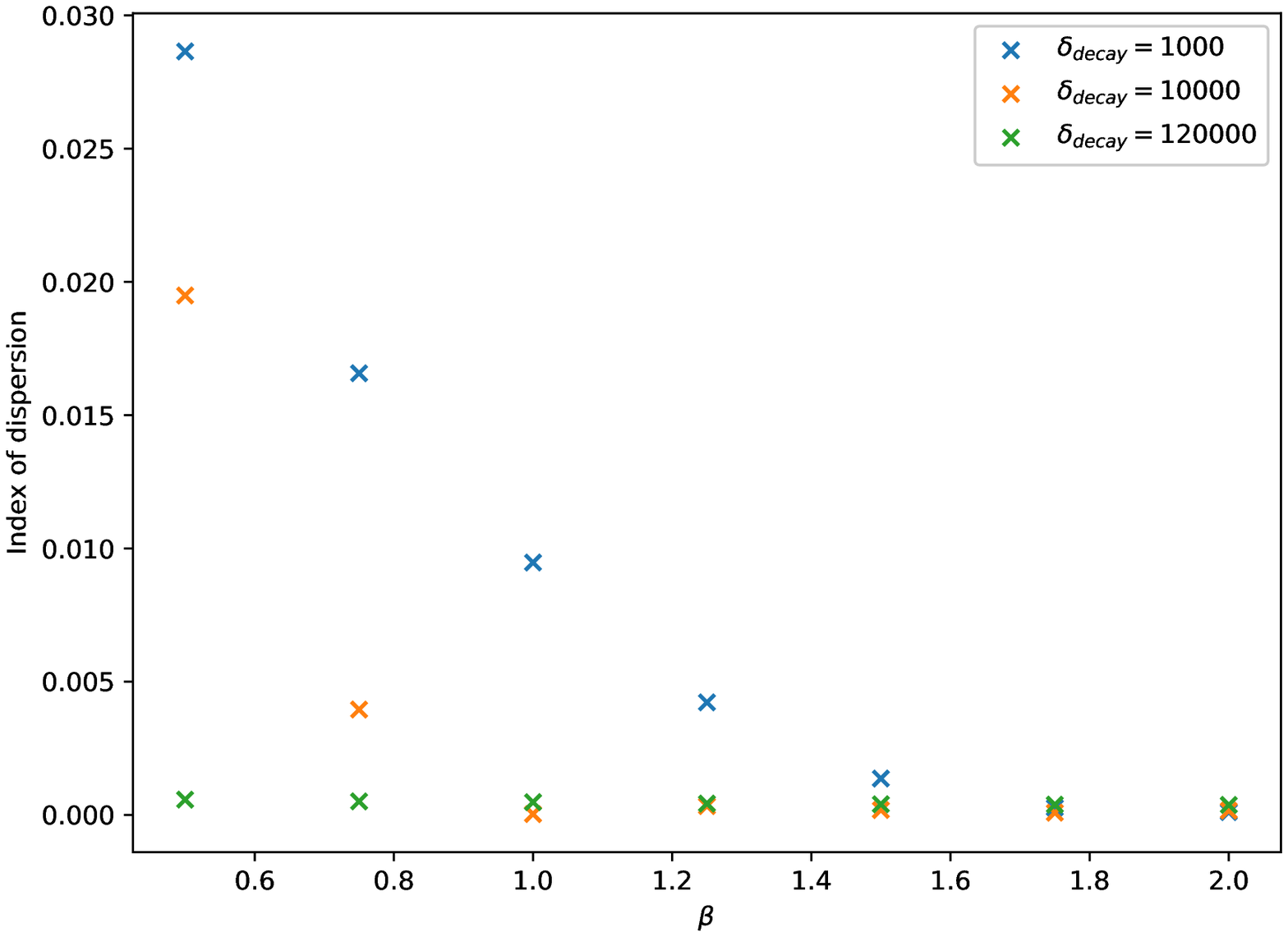}\label{iod_decay}}
\caption{We examine the performance of different strategies of supplying garbage collector molecules to the system as a function of bolus size $\beta$.
We vary the temporal distance between subsequent injections of these species ($\delta_\mathrm{decay}$).
We measure the diversity of the weight set using index of dispersion, i.e. the standard deviation divided by the mean of the weights.
As expected, the more frequent but smaller injections (more similar to a decay with a constant rate) are more conducive to learning. 
The extreme case of $\delta_\mathrm{decay}=120000$ demonstrates that the system fails to learn if the garbage collection complexes are provided only once at the beginning of the simulation.}
\label{E_removal} 
\end{figure}
\subsection{Stability of the d-CN}

\begin{figure}[H]
\centering
\subfloat[h3][Normalised $H_3$.]{\includegraphics[width=0.495\textwidth]{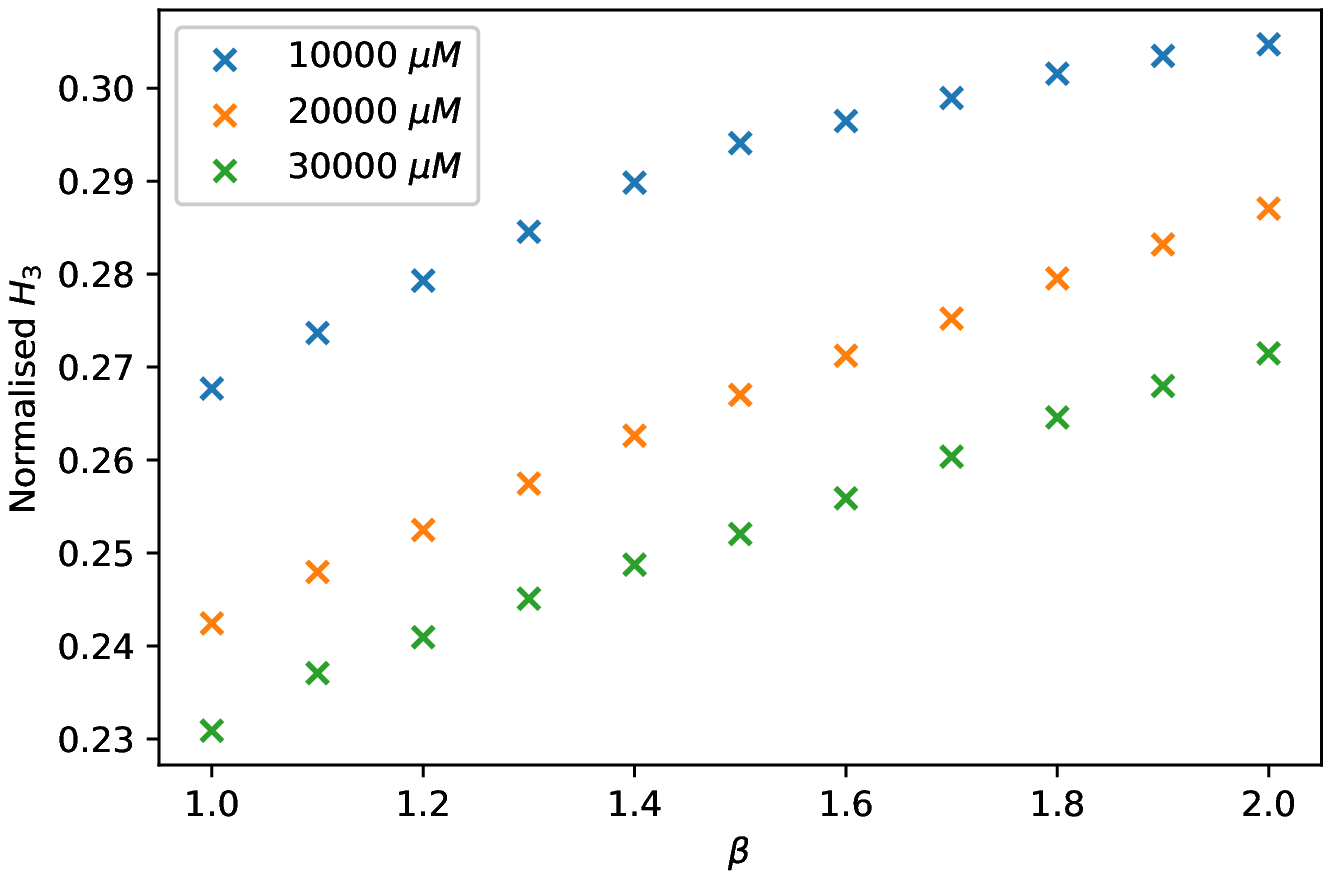}\label{h3_l}}
\subfloat[iod][Index of dispersion.]{\includegraphics[width=0.495\textwidth]{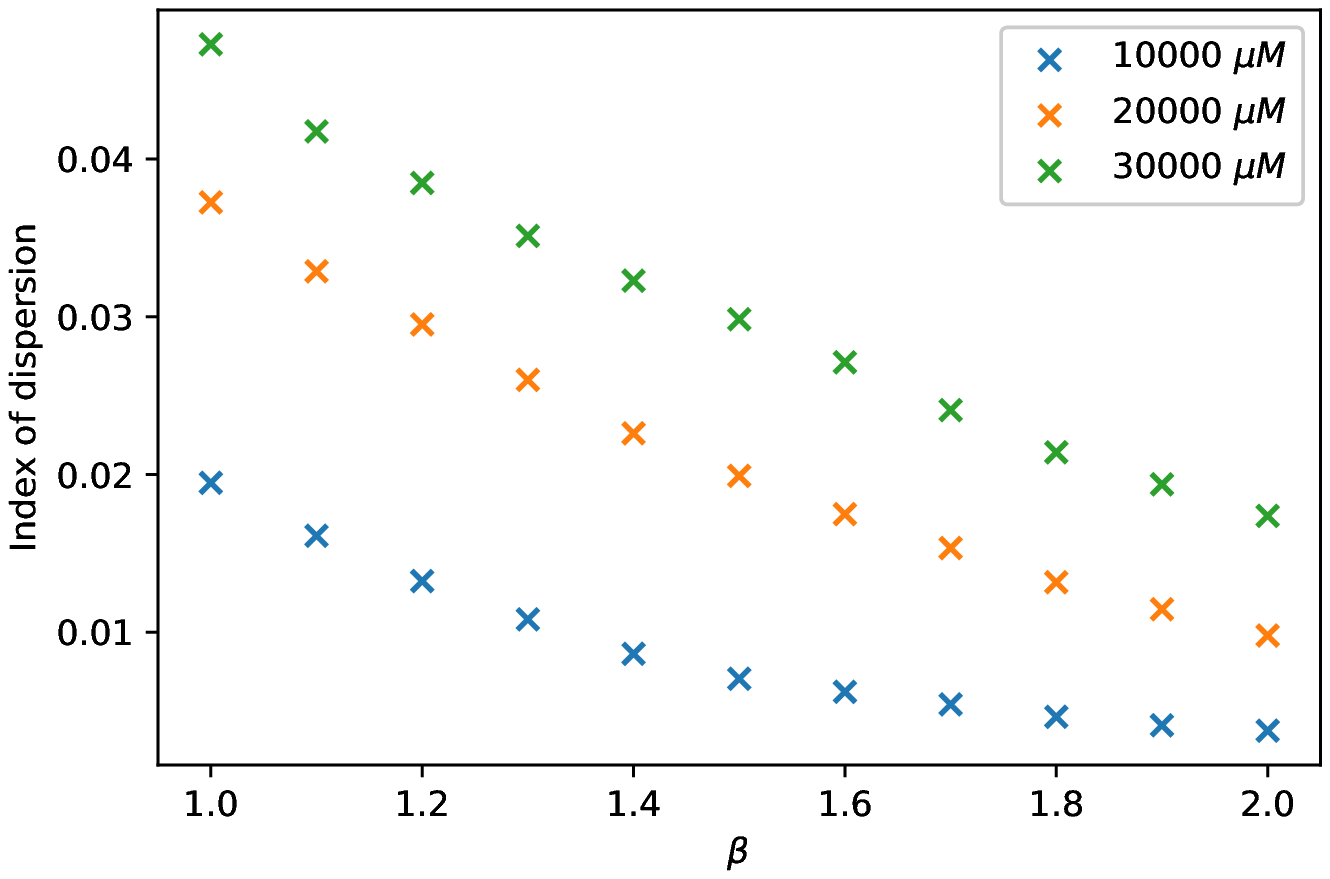}\label{iod_l}}
\caption{ 
\protect\subref{h3_l} The normalised value of the uncorrelated weight ($H_3$), and \protect\subref{iod_l} index of dispersion of the weights at the steady state as a function of the bolus size $\beta$.  The index of dispersion is a measure of diversity of the steady state weights, hence indicates how well the \neuron distinguished between input channels.  This means that an input stream with no bias would result in an index of dispersion $\approx0$.
}
\label{TC_results_bolus} 
\end{figure}

Here, we investigate the ability of the systems to distinguish temporarily correlated inputs as a function of the bolus size $\beta$.
\figref{TC_results_bolus} shows that the increase of $\beta$ results in less diverse weight representations. 
We use the  index of dispersion, i.e. the standard deviation divided by the mean of the weights, as a measure of diversity in the weight set.
As the amount of $A_n$ molecules injected at each spike increases, the system's performance declines as a result of resource starvation.
Each input spike results in a complete release of $E$ molecules, regardless of the abundance of $B$ molecules.
This results in the steady state weight of the uncorrelated input ($H_3$) approaching the weights of the two other inputs. As a consequence, the system is no longer able to detect temporal correlations.  Moreover, we vary the abundance of gate fuel molecules available at the beginning of each simulation.
As the amount of available gate complexes increases the ability of the system to distinguish temporally correlated inputs also increases.
Therefore, we show that the performance of temporal correlation detection can be increased at a cost of more fuel molecules, and therefore longer simulation time.

\section{Performance of the CN and d-CN on the FB and TC task}
\label{fdsec}

\begin{figure}[ht]
\centering
\subfloat[][]{\includegraphics[width=0.48\textwidth]{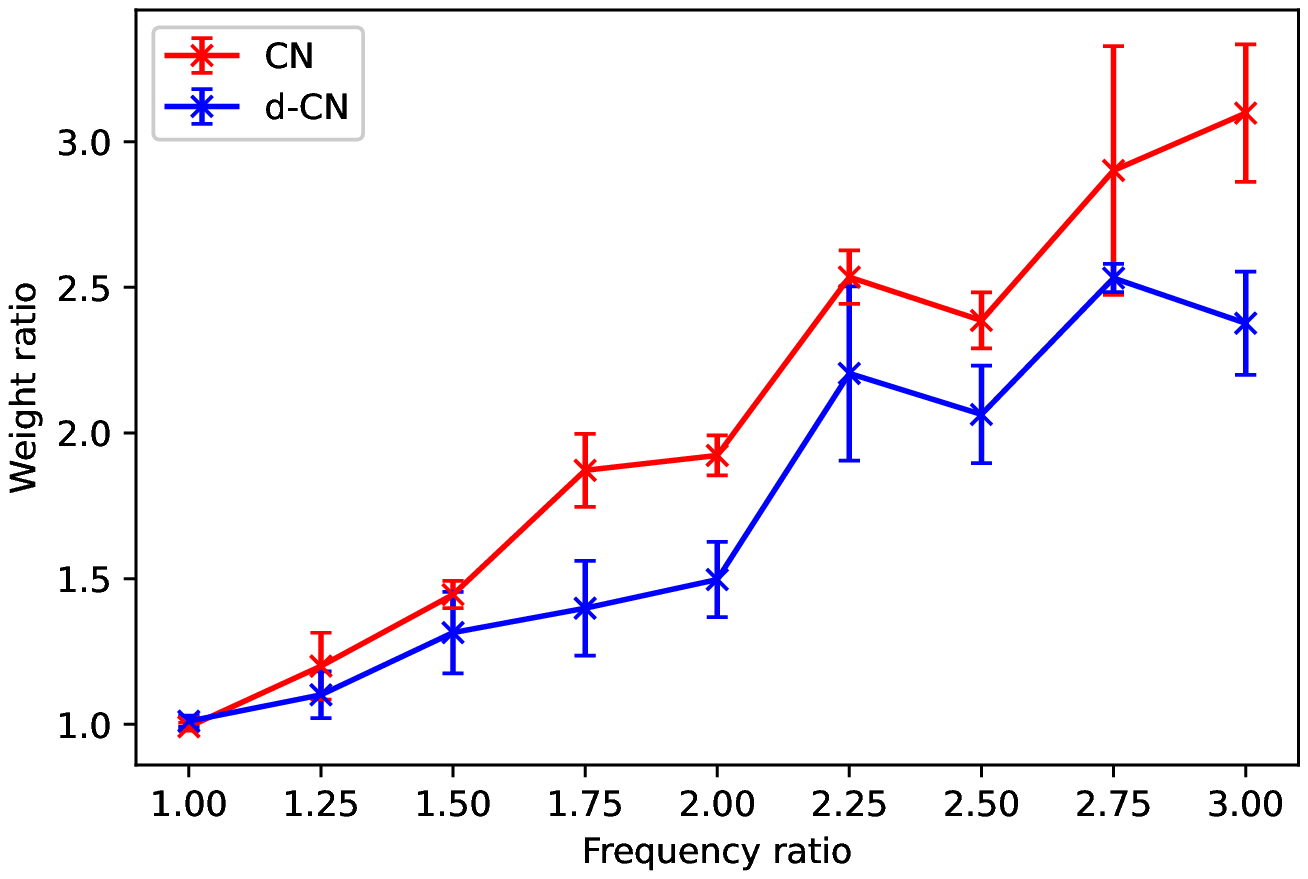}\label{straightlinefb}}
\subfloat[][]{\includegraphics[width=0.48\textwidth]{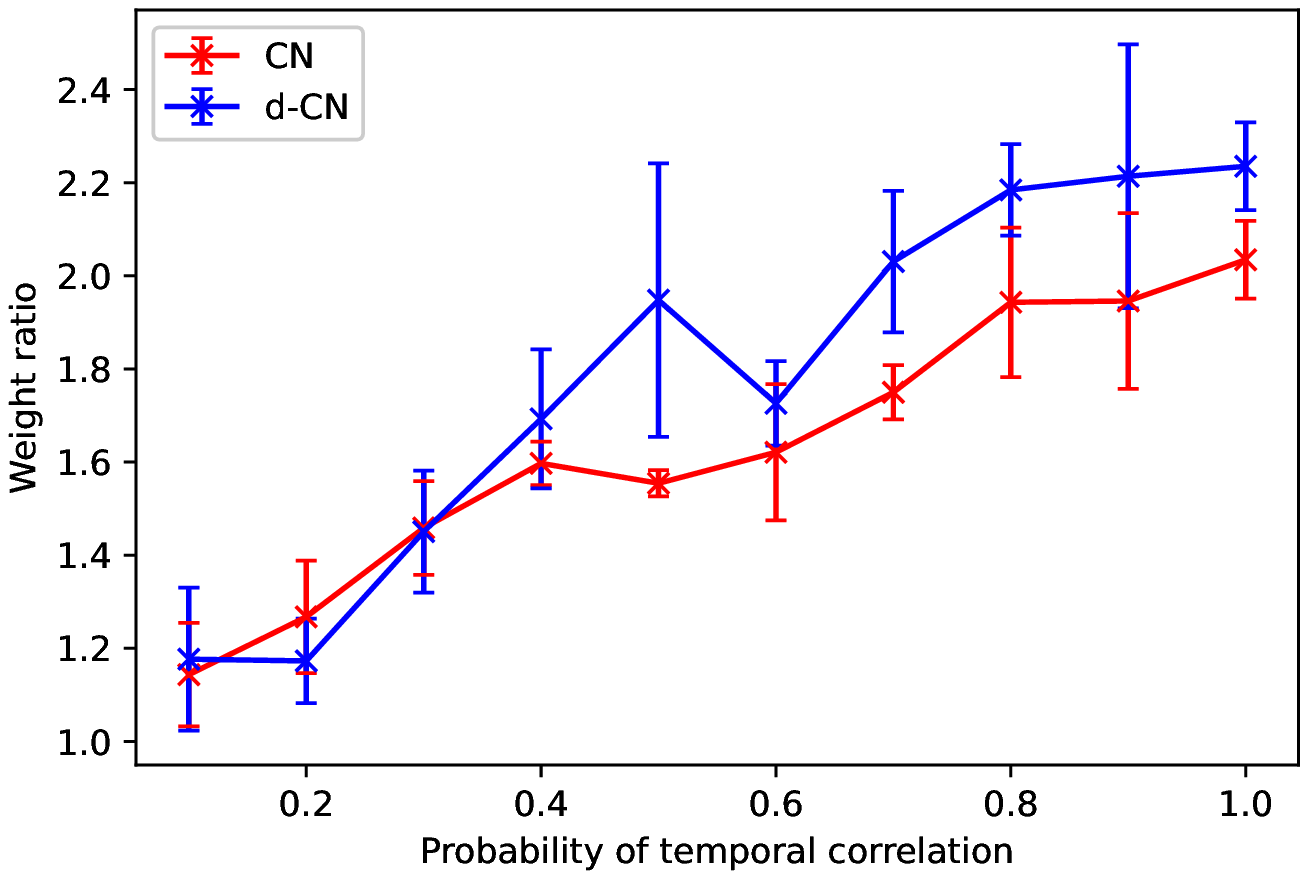}\label{straightlinetc}}
\caption{The response of the neuron, as measured by the ratio of the first and the second channel weight,  as a function of the signal strength. \protect\subref{straightlinefb} We kept the frequency of the first channel fixed at 1 Hz  and  decreased the  frequency of the second channel. The non-linearity was set to 1.   \protect\subref{straightlinetc} Both input channels have the same frequency of 1 Hz, but we varied the probability that an output spike of the second channel follows an input to channel 1.   The nonlinearity was set to 5.}
\label{straightline}
\end{figure}

When using the CN to detect a FB, the difference between the steady state weight abundances will reflect the difference of the frequencies with which the input channels fired, although the exact relationship between the  two is not immediately clear.   In order to understand this better, we considered a CN with three input channels. We then varied  the frequency of channel 1 while keeping the input frequency to channel 2 fixed and  recorded the ratio $w_1/w_2$ as a function of  $f_1/f_2$. We found that the weight ratio was proportional to the frequency ratio (\figref{straightlinefb}). While it remains unclear to what extent this qualitative result generalises to more complicated cases, it is apparent that CN is able to detect very small biases, albeit with a correspondingly small output signal strength. 
\par
We performed an equivalent analysis with the TC task. We varied the probability of  an input spike in channel 1 being followed by an input spike in channel 2, while keeping the total frequency of all input channels constant. So, for example, a probability of $0.5$ means that on average every second input spike of channel 1 is followed by an input spike of channel 2 after a delay of $\delta$ and half of the input spikes of channel 2 occur at random times. Again, we find that  even for small probabilities, there is a reliable difference in weights between the first and the second channel (\figref{straightlinetc}).

\clearpage

\section{Visual DSD code for d-CN model}
\label{a5}

\begin{lstlisting}[style=dsd]
directive simulation {final=1000000; plots=[B(); L(); H1(); H2(); H3(); A1(); A2(); A3()]}   
directive simulator deterministic
directive deterministic {stiff=true}
directive units{concentration=uM}
directive compilation infinite
directive parameters [bOUT=2; Ein=5; AFgateIn=10; backIn=10; rateb=1; tedeg=1; t1b_deg =0.1; NovelB1=10; Ain =  10;  SMFin = 0.5; WAFin = 0.5; SIFin = 0.5; AFFin=1; NdegE = 0; NdegH = 0.05; Hs = 0; kdegB = 0.1; kdegE = 0.1; kdegH = 0.000005]

//// Toehold domain reactivity
dom ta = {bind=1; unbind=10; colour="green"}   // Inputs
dom th = {bind=0.001; unbind=10; colour="orange"}  // Weights
dom tb = {bind=1; unbind=10; colour="#00fbff"}    // State
dom te0 = {bind=5; unbind=10; colour="#eb34e8"} // Activation
dom te1 = {bind=5; unbind=10; colour="#eb34e8"} // Activation
dom te2 = {bind=5; unbind=10; colour="#eb34e8"} // Activation
dom te3 = {bind=5; unbind=10; colour="#eb34e8"} // Activation
dom te4 = {bind=5; unbind=10; colour="#eb34e8"} // Activation
dom te5 = {bind=5; unbind=10; colour="#eb34e8"} // Activation
dom tfsi = {bind=1; unbind=10; colour="black"}    // Fuel for SI

//// Translator toeholds reactivity
dom tiwa1 = {bind=1; unbind=10; colour="#ffe000"}  // WA1
dom tiwa2 = {bind=1; unbind=10; colour="#ffe000"}  // WA2
dom tiwa3 = {bind=1; unbind=10; colour="#ffe000"}  // WA3
dom tiwa4 = {bind=1; unbind=10; colour="#ffe000"}  // WA4
dom tism = {bind=1; unbind=10; colour="blue"}	  // SM
dom tisi = {bind=1; unbind=10; colour="red"}	  // SI

//// Activation
def E0() = <te0^ b>
def E1() = <te1^ b>
def E2() = <te2^ b>
def E3() = <te3^ b>
def E4() = <te4^ b>

//def L() = <b te1^ b>  // Learning signal (m=1)
//def L() = <b te2^ b>  // Learning signal (m=2)
//def L() = <b te3^ b>  // Learning signal (m=3)
//def L() = <b te4^ b>  // Learning signal (m=4)
def L() = <b te5^ b>    // Learning signal (m=5)

def nE0() = <b te0^>
def nE1() = <b te1^>
def nE2() = <b te2^>
def nE3() = <b te3^>
def nE4() = <b te4^>

def nB() = <b tb^>

//// Inputs
def A1() = <ta^ a1>
def A2() = <ta^ a2>
def A3() = <ta^ a3>
def A4() = <ta^ a4>

//// Weights
def H1() = <th^ h1>
def H2() = <th^ h2>
def H3() = <th^ h3>
def H4() = <th^ h4>

//// State
def B() = <tb^ b>

//// Fuels
//// Join gate (R1 + R2 <-> T)
def Join(ta, a, tb, b, tr) = {ta^*}[a tb^]:[b tr^]:[i] 
//// Fork gate (T <-> P1 + P2)
def Fork(ta, a, tb, b, tr) = [i]:[ta^ a]:[tb^ b]{tr^*}
def Fork_WA(ta, a, tb, b, tr) = [i]:[ta^ a]:<b>[tb^ b]{tr^*}

//// Decay modules
def degE() = {te^*}[e]                                 // E removal
def degB() = {tb^*}[b]                                 // B removal
def degH1() = {th^*}[h1]                               // H1 removal
def degH2() = {th^*}[h2]                               // H2 removal
def degH3() = {th^*}[h3]                               // H3 removal
def degH4() = {th^*}[h4]                               // H4 removal

//// Weight accumulation: An + Ex <-> Ex + Hn (OLD) mx
def WA_fuel_mx(an, hn, tiwan, fuel, time) = 
( fuel Join(ta, an, te5, b, tiwan)  @ time
| fuel Fork_WA(th, hn, te5, b, tiwan) @ time
| fuel <hn te5^>  @ time
| fuel <i th^> @ time
| fuel <tiwan^ i> @ time
)

//// Signal modulation: An + Hn <-> Hn + B
def SM_fuel(an, hn, fuel, time) = 
( fuel Join(ta, an, th, hn, tism) @ time
| fuel Fork(tb, b, th, hn, tism) @ time
| fuel <b th^> @ time
| fuel <i tb^> @ time
| fuel <tism^ i> @ time		
)

//// Signal integration: An + In <-> In + B - NEW
def SI_fuel_new(an, in, fuel, time) = 
( fuel Join(ta, an, tfsi, in, tisi) @ time
| fuel Fork(tb, b, tfsi, in, tisi) @ time
| fuel <b tfsi^> @ time
| fuel <tfsi^ in> @ time
| fuel <i tb^> @ time
| fuel <tisi^ i> @ time		
)

//// Signal integration: I + An <-> An + B - OLD
def SI_fuel_old(in, an, fuel, time) = 
( fuel Join(tfsi, in, ta, an, tisi) @ time
| fuel  Fork(tb, b, ta, an, tisi) @ time
| fuel <b ta^> @ time
| fuel <tfsi^ in> @ time
| fuel <i tb^> @ time
| fuel <tisi^ i> @ time
)

//// Activation function:  B + Faf <-> Faf + E
def AF_newE_5(fuel_gate, time) = 
( fuel_gate {tb^*}[b te0^]:[b tb^]:[b te1^]:[b tb^]:[b te2^]:[b tb^]:[b te3^]:[b tb^]:[b te4^]:[b te5^]<b> @ time
)

//// Activation function:  B + Faf <-> Faf + E
def AF_newE_4(fuel_gate, time) = 
( fuel_gate {tb^*}[b te0^]:[b tb^]:[b te1^]:[b tb^]:[b te2^]:[b tb^]:[b te3^]:[b te4^]<b> @ time
)

//// Activation function:  B + Faf <-> Faf + E
def AF_newE_3(fuel_gate, time) = 
( fuel_gate {tb^*}[b te0^]:[b tb^]:[b te1^]:[b tb^]:[b te2^]:[b te3^]<b> @ time
)

//// Activation function:  B + Faf <-> Faf + E
def AF_newE_2(fuel_gate, time) = 
( fuel_gate {tb^*}[b te0^]:[b tb^]:[b te1^]:[b te2^]<b> @ time
)

//// Activation function:  B + Faf <-> Faf + E
def AF_newE_1(fuel_gate, time) = 
( fuel_gate {tb^*}[b te0^]:[b te1^]<b> @ time
)

//// Short example of randomly generated input with frequency bias
( 0 B()  | B() ->{bOUT}    
| 0 H1() | H1() ->{kdegH}
| 0 H2() | H2() ->{kdegH} 
| 0 H3() | H3() ->{kdegH} 

| Ein E0() 
| Ein E1() 
| Ein E2()
| Ein E3() 
| Ein E4()

| backIn nE0()
| backIn nE1()
| backIn nE2()
| backIn nE3()
| backIn nE4()

| backIn nB()
| 0 L()
| AF_newE_5(AFgateIn, 0) 

| SM_fuel(a1, h1, 25000, 0) 
| SI_fuel_old(i1, a1, 80000, 0)
| WA_fuel_mx(a1, h1, tiwa1, 10000, 0) 

| SM_fuel(a2, h2, 25000, 0) 
| SI_fuel_old(i2, a2, 80000, 0)
| WA_fuel_mx(a2, h2, tiwa2, 10000, 0) 

| SM_fuel(a3, h3, 25000, 0) 
| SI_fuel_old(i3, a3, 80000, 0)
| WA_fuel_mx(a3, h3, tiwa3, 10000, 0) 

| Ain A1() @ 1376 
| Ain A1() @ 1794 
| Ain A3() @ 5713 
| Ain A3() @ 5963 
| Ain A1() @ 9357 
// etc.
)
\end{lstlisting}